\documentclass[twocolumn]{article}
\pdfoutput=1
\usepackage{graphicx,lipsum,textcomp}
\usepackage{lineno,hyperref}
\usepackage{amsmath,array}
\usepackage{booktabs}
\usepackage[section]{placeins}
\modulolinenumbers[5]
\makeatletter
\newcommand{\eqnum}{\refstepcounter{equation}\textup{\tagform@{\theequation}}}
\makeatother
\newcolumntype{L}[1]{>{\raggedright\let\newline\\\arraybackslash\hspace{0pt}}m{#1}}
\newcolumntype{C}[1]{>{\centering\let\newline\\\arraybackslash\hspace{0pt}}m{#1}}
\newcolumntype{R}[1]{>{\raggedleft\let\newline\\\arraybackslash\hspace{0pt}}m{#1}}
\usepackage{siunitx}
\usepackage[pass]{geometry} 
\usepackage[font=small,labelfont=bf]{caption}
\usepackage{framed} 
\usepackage{multicol} 

\usepackage{nomencl} 
\usepackage{color,soul} 
\usepackage{xcolor}
\newcommand{\mathcolorbox}[2]{\colorbox{#1}{$\displaystyle #2$}}  
\usepackage{ifthen}
\usepackage{setspace} 

\renewcommand{\nomgroup}[1]{%
\ifthenelse{\equal{#1}{A}}{\item[\textbf{ }]}{%
\ifthenelse{\equal{#1}{G}}{\item[\textbf{ }]}{%
\ifthenelse{\equal{#1}{S}}{\item[\textbf{Subscripts}]}{}}}
}
    
\makenomenclature
\renewcommand*\nompreamble{\begin{multicols}{2}}
\renewcommand*\nompostamble{\end{multicols}}
\newcommand\SA{S\!A}

\usepackage{amssymb}

\usepackage[sort&compress]{natbib}
\setcitestyle{numbers,square}
\newcommand\blfootnote[1]{%
  \begingroup
  \renewcommand\thefootnote{}\footnote{#1}%
  \addtocounter{footnote}{-1}%
  \endgroup
}

\usepackage{authblk}
\title{On the drag of freely falling non-spherical particles}
\author[1,2]{Gholamhossein Bagheri}
\author[1]{Costanza Bonadonna}
\affil[1]{Department of Earth Sciences, University of Geneva, Rue des Maraîchers 13, 1205 Gen{\`e}ve, Switzerland}
\affil[2]{Max Planck Institute for Dynamics and Self-Organization, Am Fassberg 17, 37077 G{\"o}ttingen, Germany}

\begin{document}

\twocolumn[
  \begin{@twocolumnfalse}
  \vspace{-7em}
\maketitle
\begin{abstract}
We present a new general model for the prediction of the drag coefficient of non-spherical solid particles of regular and irregular shapes falling in gas or liquid valid for sub-critical particle Reynolds numbers (i.e.  $Re < 3 \times 10^5$). Results are obtained from experimental measurements on 300 regular and irregular particles in the air and analytical solutions for ellipsoids. Depending on their size, irregular particles are accurately characterized with a 3D laser scanner or SEM micro-CT method. The experiments are carried out in settling columns with height of 0.45 to 3.60 \SI{}{\metre} and in a 4m-high vertical wind tunnel. In addition, $881$ additional experimental data points are also considered that are compiled from the literature for particles of regular shapes falling in liquids. New correlation is based on the particle Reynolds number and two new shape descriptors defined as a function of particle flatness, elongation and diameter. New shape descriptors are easy-to-measure and can be more easily characterized than sphericity. The new correlation has an average error of $\sim 10\%$, which is significantly lower than errors associated with existing correlations. Additional aspects of particle sedimentation is also investigated. First, it is found that particles falling in dense liquids, in particular at $Re>1000$, tend to fall with their maximum projection area perpendicular to their falling direction, whereas in gases their orientation is random. Second, effects of small-scale surface vesicularity and roughness on the drag coefficient of non-spherical particles found to be $<10\%$. Finally, the effect of particle orientation on the drag coefficient is discussed and additional correlations are presented to predict the end members of drag coefficient due to change in the particle orientation.
\vspace{1em}

\emph{keywords: Drag Coefficient; Terminal Velocity; Free Fall; Particle Shape; Non-Spherical;, Irregular }

\href{http://dx.doi.org/10.1016/j.powtec.2016.06.015}{doi:10.1016/j.powtec.2016.06.015}
\end{abstract}
  \end{@twocolumnfalse}
]
\blfootnote{Corresponding author: G. Bagheri}
\blfootnote{   Email: \href{gholamhossein.bagheri@ds.mpg.de}{gholamhossein.bagheri@ds.mpg.de}}
\blfootnote{   Tel.: +49-551-5176-317}
\blfootnote{\textcopyright 2016. This manuscript version is made available under}
\blfootnote{the CC-BY-NC-ND 4.0 license (\href{http://creativecommons.org/licenses/by-nc-nd/4.0/}{link})}

\hl{Highlighted parts indicate corrections with respect to the officially published version of the manuscript (Thanks to Anand).}

\section{Introduction}
Non-spherical particles are encountered in numerous fields of science and engineering, such as chemical engineering, civil engineering, mining engineering, physical sciences, biology and earth sciences \citep{Happel1983a,Blott2007}. The category of non-spherical particles, in general, includes both regular (e.g. ellipsoid, cube, cylinder) and irregular shapes (e.g. pharmaceutical powders, spore, pollen, coal particles, cosmic and atmospheric dust, sand, pebble, volcanic particles). Nonetheless, in many studies that deal with particulate flows, particles are assumed to be perfect spheres. This is mainly due to the fact that the shape characterization of irregular particles is a complex process and numerous shape descriptors have been developed in the past few decades to quantify various aspects, such as form, roundness, irregularity and sphericity \citep{Blott2007,Bagheri2014}. More importantly, the most accurate models for predicting the behavior of non-spherical particles in fluids are based on studies on regular particles \citep{Haider1989,Ganser1993,Chhabra1999a}, for which the characterization of the particle shape is not complex and can be obtained analytically. \\
Particles of arbitrary shapes when transported in a fluid experience forces and momentum on all three coordinate axes \citep{White1998}. In many applications the most important force acting on a particle is the one that is exerted in the opposite direction of particle motion, which is called the drag force $F_D$ and defined as:

\begin{equation}
  \mathbf{F_D}=- \frac{1}{2} \rho_f C_D A \vert \mathbf{u_p} - \mathbf{u_f} \vert (\mathbf{u_p} - \mathbf{u_f}) \label{eq:Fdrag}
\end{equation}

\nomenclature[AF]{$\mathbf{F_D}$}{drag force, see Eq. (\ref{eq:Fdrag}), [$\SI{}{\newton}$]}
\nomenclature[Gr]{$\rho_f$}{surrounding fluid density, [$\SI{}{\kilogram\per\cubic\metre}$]}
\nomenclature[AC]{$C_D$}{drag coefficient, see Eq. (\ref{eq:CD})}
\nomenclature[Au]{$\mathbf{u_p}$}{particle velocity, [$\SI{}{\metre\per\second}$]}
\nomenclature[Au]{$\mathbf{u_t}$}{particle terminal velocity, [$\SI{}{\metre\per\second}$]}
\nomenclature[St]{$t$}{terminal (velocity)}
\nomenclature[Ss]{$smooth$}{irregular particles wrapped in Parafilm\textsuperscript{\textregistered}}
\nomenclature[Ss]{$rough$}{irregular particles without Parafilm\textsuperscript{\textregistered}-wrap}
\nomenclature[Au]{$\mathbf{u_f}$}{extrapolated fluid velocity at the particle centroid, [$\SI{}{\metre\per\second}$]}
\nomenclature[AA]{$A$}{a reference area of the particle, $\pi \, d_{eq}^2 / 4, \, [\SI{}{\square\metre}$]}

where $\rho_f $ is the fluid density, $C_D$ is the drag coefficient of the particle, $A$ is a reference area related to the particle size (e.g. $\pi d^2/4$ for a sphere with diameter of  $d$), $\mathbf{u_p}$ is the particle velocity, $\mathbf{u_f}$ is the fluid velocity extrapolated to the particle centroid (i.e. unhindered velocity)\citep{Loth2008}. The terminal velocity of the particle, $\mathbf{u_t}$, (i.e. the highest falling velocity at which the particle acceleration reaches zero) can be simply obtained by replacing $\mathbf{-F_D}$ with the particle weight considering the buoyancy force. The most challenging parameter to be determined in Eq. (\ref{eq:Fdrag}) is the drag coefficient  $C_D$, which is dependent on many parameters including particle Reynolds number, shape, orientation, secondary motions, particle-to-fluid density ratio, fluid turbulence intensity and particle/fluid acceleration \citep{Isaacs1967, Stringham1969a, Clift1971, Marchildon1979, Haider1989, Ganser1993, Clift2005, Loth2008, Holzer2008, Bagheri2013a, Brosse2013}. However, parameters that have a first order influence on $C_D$ are particle Reynolds number, shape, particle-to-fluid density ratio and orientation \citep{Loth2008, Holzer2008, Brosse2013}. Here, particle Reynolds number, $Re$, for both spherical and non-spherical particles is defined as:
\begin{equation}
  Re=\frac{\rho_f d_{eq}  \vert \mathbf{u_p} - \mathbf{u_f} \vert }{\mu_f}   \label{eq:Re}
\end{equation}
\nomenclature[AR]{$Re$}{particle Reynolds number, $\rho_f , \vert \mathbf{u_p} - \mathbf{u_f} \vert d_{eq} / \mu_f$}
\nomenclature[Ad]{$d_{eq}$}{diameter of a sphere with the same volume as the particle, $\SI{}{\metre}$}
\nomenclature[Gm]{$\mu_f$}{fluid dynamic viscosity, [$\SI{}{\pascal\cdot\second}$]}

where $d_{eq} $ is the diameter of a sphere with the same volume as the particle and $\mu_f$ is the fluid dynamic viscosity.
Except at very low values of $Re$ ($\ll 1$), where an analytical solution exists for spheres based on Stokes' solution \citep{GeorgeGabriel1851} and for ellipsoids based on Oberbeck solution \citep{Oberbeck1876}, no general solution can be found for calculating the drag coefficient of particles of any shape \citep{Happel1983a, Clift2005, Loth2008}. At higher $Re$, even for spherical particles, where quantification of particle shape is not an issue, experimental measurements are the main source of information while numerical solutions and boundary layer theory can provide additional information \citep{Clift2005}. 

In the absence of a general solution, a large number of empirical correlations for predicting the drag coefficient of spherical and non-spherical particles are introduced that are associated with different ranges of validity and accuracy \citep{Wieselsberger1922, Albertson1953, Willmarth1964, Marchildon1964, Christiansen1965a, Jayaweera1965, Isaacs1967, Stringham1969a, Komar1978, Marchildon1979, Wilson1979, Baba1981, Leith1987, McKay1988, Haider1989, Ganser1993, Cheng1997, Gogus2001, Clift2005, Loth2008, Mando2010, Chow2011,Alfano2011c}. However, correlations available in the literature are associated with some drawbacks. First, data used in previous studies are mostly based on experiments with particles of regular shapes (e.g. cube, cylinder, disk). Available data for irregular particles lack of an accurate characterization of particle shape and size, or they do not cover a wide range of Reynolds numbers \citep{Albertson1953, Komar1978, Wilson1979, Baba1981, Cheng1997, Dellino2005}.

Second, most formulations are based on sphericity, a function of particle surface area, which, in the case of irregular particles, is one of the most challenging parameters to be determined and requires sophisticated instruments \citep{Bagheri2014}. Third, almost all the available data are based on experiments in liquids for which the particle-to-fluid density ratio, $\rho'=\rho_p / \rho_f$, is in the order of $~1-11$, whereas $\rho'$ for solid particles moving in gases can be up to the order of $10^3$. $\rho'$ is an important parameter that can influence particle drag coefficient, especially at high Reynolds numbers. Finally, the effect of surface roughness and vesicularity on the drag coefficient of irregular particles is not yet well understood.
 
\nomenclature[Gr]{$\rho'$}{particle-to-fluid density ratio, $\rho_p / \rho_f$} 
\nomenclature[Gr]{$\rho_p$}{particle density, [$\SI{}{\kilogram\per\cubic\metre}$]} 
 
In the present study, a comprehensive analytical and experimental investigation on the drag coefficient of non-spherical particles including regular and irregular shapes with $Re<3 \times 10^5$ is carried out. At $Re<0.1$ (i.e. Stokes' regime) the analytical solution of Oberbeck \citep{Oberbeck1876} is solved numerically for ellipsoids with various elongation and flatness ratios. At $0.1 \leq Re<1000$ (i.e. intermediate regime), the drag coefficient of $100$ highly irregular volcanic particles and $17$ regular shape particles (i.e. cylinders and cubes) are measured in air-filled settling columns of various heights ($0.45-3.6 m$). Finally, a vertical wind tunnel \citep{Bagheri2013a} is used to measure the drag coefficient of $116$ irregular volcanic particles and $61$ regular shape particles (i.e. ellipsoids, circular cylinder, disks, other geometrical shapes) at $1000 \leq Re<3 \times 10^5$ (i.e. Newton's regime). A total of $10^4$ analytical and $1285$ experimental data points measured in the air are obtained. In addition, $881$ experimental data points compiled from the literature for spherical and regular particles, most of which measured in liquids, are also considered \citep{Pettyjohn1948, Willmarth1964, Christiansen1965a, Isaacs1967, McKay1988}. 

The main objective of this study is to find the simplest and the best correlated shape descriptors that could be used to estimate the drag coefficient of both regular and irregular particles moving in liquids (based on published data) or gases (based on new results). In addition, types of particle secondary motion, the effect of particle orientation, the effect of particle-to-fluid density ratio $\rho'$ and surface roughness on the drag coefficient are discussed. Finally, a general drag coefficient model is presented that is valid for predicting the average and end members of drag coefficient of non-spherical particles freely moving in gases or liquids.

In the following sections, first we present a introduction on the aerodynamics of particles and associated parameters followed by a thorough review of the existing models for predicting the drag coefficient of spherical and non-spherical particles. Then methods and materials used in this study are described. Finally, results are presented and the impact of important parameters on the drag coefficient of non-spherical particles is discussed in detail.

\section{Aerodynamics studies: state-of-the-art}
\subsection{Aerodynamics of spherical particles}
Drag of non-spherical particles can be framed if first we analyze aerodynamics of spherical particles. Several analytical, numerical and experimental studies can be found that are focused on the aerodynamics and, in particular, on the drag of spheres  \citep{Clift2005}. In addition, the dependency of the drag of non-spherical particles on Reynolds number is, in general, very similar to that of spherical particles. 
\subsubsection{Flow development as a function of \texorpdfstring{$Re$}{TEXT}}
\begin{figure}
  \begin{center}
    \includegraphics[width=0.48\textwidth]{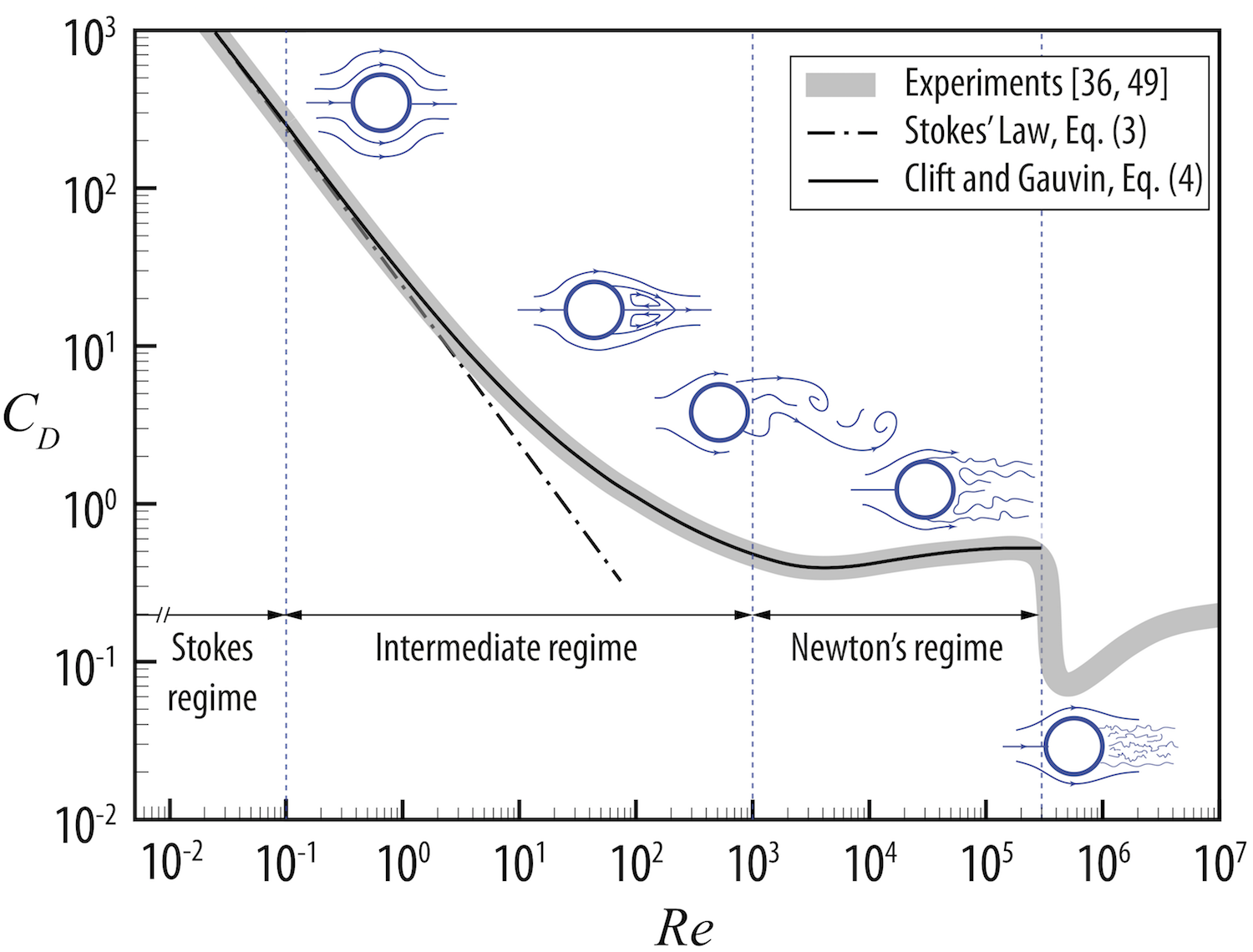}
  \end{center}
  \vspace{-2em}
  \caption{Dependency of $C_D$ on $Re$ for sphere. Streamlines around sphere at various $Re$ are also shown in the plot.}
  \label{SphereDrag}
\end{figure}

As it is shown in Fig. \ref{SphereDrag}, the fluid flow around spherical particles is strongly dependent on the particle Reynolds number. The flow at $Re \ll 1$ is called the \emph{Stokes' regime} (or \emph{creeping flow}), where the flow inertial terms are negligible with respect to viscous terms and flow remains attached to sphere with no wake behind \citep{White1998}. The flow remains attached up to $Re \approx 20$, which is the onset of flow separation. At $20<Re<130$ circular wakes behind a sphere grow but they remain steady and attached to the particle. As  $Re$ increases beyond 130 and up to 1000, vortex shedding begins and wakes behind the sphere gradually become instable and unsteady. At $1000<Re<3 \times 10^5$ wakes behind the sphere become fully turbulent while the boundary layer at the front of the sphere is laminar. This range of Reynolds number is called the \emph{Newton's regime} \citep{Clift2005}. $Re>3 \times 10^5$ corresponds to critical transition and supercritical regime where boundary layer and wake behind sphere are both turbulent. $Re=3 \times 10^5$ is called the \emph{critical Reynolds number}, at which the \emph{drag crisis} occurs and reduces the drag coefficient markedly \citep{Clift1971, Achenbach1972, Clift2005, Loth2008}. 

\subsubsection{Dependency of \texorpdfstring{$C_D$}{TEXT} on \texorpdfstring{$Re$}{TEXT}} 
The variation of the sphere drag coefficient at subcritical $Re$ can be studied in three different Reynolds regimes, namely the Stokes, intermediate and the Newton's regimes (Fig. \ref{SphereDrag}). Stokes \cite{GeorgeGabriel1851} showed that at $Re \ll 1$, where inertial terms in the Navier-Stokes equations are negligible, Navier-Stokes equations can be simplified to a linear differential equation, which can be solved analytically. Stokes' solution shows that the drag coefficient of a smooth solid spherical particle in standard conditions (i.e. moving with constant relative velocity in an undistributed, unbounded and incompressible flow) at $Re \ll 1$ is \citep{GeorgeGabriel1851, Clift2005}:
\begin{equation}
  C_D=\frac{24 }{Re}   \label{eq:CdStokes}
\end{equation}
Two thirds of this drag is due to viscous stresses (i.e. \emph{friction drag}) and one third to the pressure gradients (i.e. \emph{form drag} or \emph{pressure drag}). Sphere drag predicted by the Stokes' law at $Re=0.1$ is 2\% less than those obtained from more accurate solutions in which inertial terms are taken into account \citep{Happel1983a}. Thus, in this study $Re<0.1$ is assigned as the range for the Stokes' regime, where the Stokes' solution is associated with an error of $<2\%$ for spherical particles. 

In the intermediate regime ($0.1 \leq Re < 1000$), the sphere drag coefficient continues to decrease as $Re$ increases although the rate of decrease is lower than that at the Stokes' regime. Finally, the drag coefficient becomes almost constant in the Newton's regime ($1000 \leq Re <3 \times 10^5$) with a minimum of 0.38 at $5 \times 10^3$ and a maximum of 0.50 at $7 \times 10^4$ \citep{Clift1971}. Average of the drag coefficient for sphere in the Newton's regime is about 0.46. One of the most accurate correlations for predicting the drag coefficient of spherical particles at subcritical $Re$ is the model of Clift and Gauvin \cite{Clift1971}:
\begin{equation}
\begin{split}
  C_D=\frac{24 }{Re} \left( 1+0.15 Re^{0.687} \right)  \quad\quad\quad\quad \quad\quad\quad\quad\quad\quad\\ \quad\quad\quad\quad\quad\quad\quad +\frac{0.42}{\displaystyle 1+\frac{42500}{\displaystyle Re^{1.16}}}  \quad \mbox{for } Re<3 \times 10^5
  \end{split} \label{eq:CdClift}
\end{equation}
Eq. (\ref{eq:CdClift}) is valid for subcritical $Re$ and is within 6\% of experimental measurements (Fig. \ref{SphereDrag}) \citep{Clift2005}.

\subsection{Drag of non-spherical particles}
The dependency of the drag coefficient of non-spherical particles on the particle Reynolds number is very similar to that of spheres. In fact, for non-spherical particles, parameters other than the particle Reynolds number, such as particle shape, surface roughness, orientation and particle-to-fluid density ratio are the source of complexities in the determination of the drag coefficient. The impact of these parameters on the drag coefficient is dependent on the particle Reynolds number. To provide a clear background, the effect of these parameters on the drag coefficient is presented separately in the following sections. 

\subsubsection{Shape} \label{section:shape}
In general, at a given particle Reynolds number, the average of the drag coefficient of a falling non-spherical particle is higher than that of a sphere as a consequence of its non-spherical shape. As a result, the main challenge is to quantify the shape of particles through a shape descriptor that is well correlated with the drag coefficient. Shape descriptors are mathematical functions that require previous determination of dimensional variables of the particle, such as lengths, diameter, projection perimeter, surface area or volume \citep{Bagheri2014}. Ideally, the shape descriptor should be easy-to-measure for particles of both regular and irregular shapes. In studies related to transport and sedimentation of particles the most common shape descriptors are sphericity and \emph{form factors} (e.g. flatness, elongation and their combinations) \citep{Pettyjohn1948, McNown1950, Sneed1958, Christiansen1965a, Isaacs1967, Wilson1979, Baba1981, Leith1987, McKay1988, Haider1989, Ganser1993, Gogus2001, Loth2008, Chow2011}.
Sphericity $\psi$ is defined as the ratio of surface area of a sphere with the same volume as the particle to the actual surface area of the particle $\SA_{p}$ \citep{Wadell1933}:  
\begin{equation}
  \psi = \pi d_{eq}^2 / \SA_{p}  \label{eq:Sph}
\end{equation}

\nomenclature[Gp]{$\psi$}{sphericity, $\pi d_{eq}^2 / \SA_{p}$}
\nomenclature[AS]{$\SA_{p}$}{particle surface area, [$\SI{}{\square\metre}$]}

Sphericity is equal to 1 for spheres and decreases as particles become less spherical. As a result, for a fixed particle volume, the drag coefficient has an inverse correlation with the sphericity. 
The main disadvantage of sphericity is its dependency on the particle surface area. Although the surface area of a regular particle with smooth surface can be measured analytically, for irregular particles surface area can only be measured with sophisticated instruments, such as 3D scanners or gas adsorption. In addition, the measured surface area is a function of measurement accuracy and, in particular, it increases as the measurement resolution and accuracy increase \citep{Bagheri2014}. As a result, sphericity is not an absolute shape descriptor for irregular particles and should be reported with the measurement accuracy in order to be reproducible. Additionally, particles with different shapes can have the same sphericity. As an example, sphericity of a very elongated cylinder with height to diameter ratio of 20 ($h= 20\, d$) is equal to the sphericity of an extremely flat disk with height to diameter ratio of 0.1 ($h=0.1\, d$) (Table \ref{TabShape}). 

\begin{table*}[!h]
  \centering
  \caption{Sphericity and form dimensions of some geometrical shapes. Semi-axes lengths of the ellipsoid are $a$, $b$ and $c$, the edge length of cuboctahedron, octahedron, cube and tetrahedron is $a$, and the diameter and height of cylinders and disks are  $d$ and $h$, respectively. }
  \vspace{-1em}
\begin{tabular}{l c c ccc} \toprule
 Shape & $d_{eq}$ & $\psi$ & $L$& $I$& $S$
\\ \midrule
Ellipsoid ($a=2 \, b = 2 \, c$) & $2 \sqrt[3]{a\,b\,c}$& 0.791& $2 \, a$& $2 \, b$& $2 \, c$ \\
Cuboctahedron & $\sim 1.65 \,a$ &0.905& $2 \, a$& $\sqrt{2} \, a$& $\sqrt{2} \, a$\\
Octahedron & $\sim 0.97 \,a$ & 0.846&$\sqrt{2} \, a$& $a$ & $a$ \\
Cube& $\sim 1.24 \,a$  &  0.806&$\sqrt{3} \, a$& $\sqrt{2} \, a$& $a$ \\
Tetrahedron & $\sim 0.61 \,a$   &  0.670&$a$& $\sqrt{3/4} \, a$& $\sqrt{2/3} \, a$ \\
Cylinder ($h=20 \, d$) & $\sqrt[3]{3\,d^2\,h/2}$& 0.471 & $\sqrt{h^2+d^2}$ & $d$& $d$\\
Disk ($h=0.1 \, d$) & $\sqrt[3]{3\,d^2\,h/2}$& 0.471 & $\sqrt{h^2+d^2}$ & $d$& $d$
\\ \bottomrule
\label{TabShape}
\vspace{-2em}
\end{tabular} 
\end{table*}

\begin{figure}[!htb]
  \begin{center}
    \includegraphics[width=0.45\textwidth]{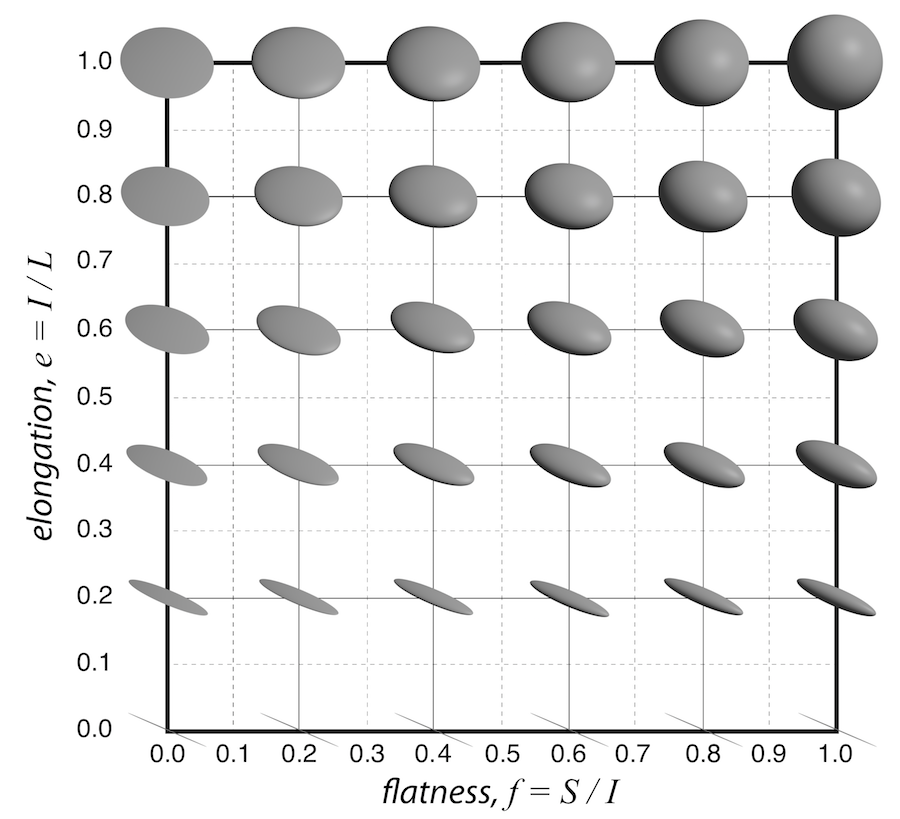}
      
  \end{center}
  \vspace{-2em}
  \caption[Effect of variation of flatness $f$ and elongation $e$ on the shape of ellipsoids.]{Effect of variation of flatness $f$ and elongation $e$ on the shape of ellipsoids. $L$, $I$ and $S$ are called from dimensions and defined as the longest, intermediate and shortest lengths of the particle, receptively.}
  \vspace{-3em}
  \label{EllipShape}
\end{figure}

Particle form factors are simpler to measure than sphericity, are less dependent on the measurement resolution and can better discriminate particles with different forms. In order to calculate form factors for a particle, its \emph{form dimensions} should be measured, which are defined and noted as $L$: longest, $I$: intermediate and $S$: shortest length of the particle \citep{Bagheri2014}. The most common form factors are flatness $f$ ($S/I$) and elongation $e$ ($I/L$). It should be noted that form factors, similar to sphericity, are a sub-category of shape descriptors and they are called form factors because they can provide information on the tri-dimensional characteristic of the particles (e.g. can quantify how flat or elongate a particle form is). Fig. \ref{EllipShape} shows how shapes of ellipsoids vary by changing their elongation and flatness ratios. The most common form factor related to drag of non-spherical particles is the Corey shape descriptor defined as $S/ \sqrt {L I} $ \citep{Albertson1953, Corey1963, Komar1978, Loth2008}, which is found to be highly correlated with the particle flatness \citep{Bagheri2014}. Interestingly, sphericity and Corey shape descriptor measured for irregular particles have a very weak correlation with each other \citep{Bagheri2014}, although they were both found to be correlated with the drag coefficient of non-spherical particles. 

\nomenclature[AL]{$L$}{longest length of the particle, [$\SI{}{\metre}$]}
\nomenclature[AI]{$I$}{intermediate length of the particle, [$\SI{}{\metre}$]}
\nomenclature[AS]{$S$}{shortest length of the particle, [$\SI{}{\metre}$]}
\nomenclature[Ae]{$e$}{elongation, $I/L$}
\nomenclature[Af]{$f$}{flatness, $S/I$}

The main shortcoming of form dimensions is their dependency on the operator judgments \citep{Bagheri2014}. Several methods exist for measuring particle form dimensions that are associated with different levels of simplicity and operator-dependent errors. Bagheri et al. \citep{Bagheri2014} reviewed some of these methods and presented a new method called the \emph{projection area protocol}. The projection area protocol is associated with the lowest operator-dependent errors and the measured form dimensions are better correlated with particle volume and surface area compared to other methods \citep{Bagheri2014}. Through the projection area protocol form dimensions are measured on two specific projections of the particle, namely the projections with maximum and minimum areas. $L$ and $I$ are defined as the largest and smallest dimensions measured on the maximum-area projection, and $S$ corresponds to the smallest dimension measured in the minimum-area projection. Note that through this procedure form dimensions do not need to be perpendicular with each other. In this way, $L$ and $S$ correspond to the largest and smallest lengths of the particle, respectively, and therefore, are less affected by operator-dependent errors \citep{Bagheri2014}. As an example, form dimensions of a cube with edge length of $a$ are $\sqrt{3 \, a}$, $ \sqrt{2 \, a}$ and $a$. Sphericity and form dimensions of some selected geometrical shapes measured through projection area protocol are shown in Table \ref{TabShape}.

\nomenclature[Aab]{$a,\, b,\,c$}{semi-axes lengths of ellipsoid, $a$ is also the edge length of cuboctahedron, octahedron, cube or tetrahedron, [$\SI{}{\metre}$]}
\nomenclature[Ad]{$d,\,h$}{diameter and height of cylinder or disk, [$\SI{}{\metre}$]}

\subsubsection{Surface roughness}
The drag coefficient in the Stokes' regime is relatively insensitive to surface roughness \citep{Loth2008}. This seems logical based on theorem of Hill and Power \citep{Hill1956}, which shows that in the Stokes' regime the drag exerted on a particle is bounded by the drag exerted on bodies that inscribe and circumscribe the particle. As a result, the surface roughness should not alter the drag coefficient significantly, since the drag coefficient of the rough particle can be narrowly constrained by the drag coefficient of two smooth bodies. On the other hand, in the Newton's regime surface roughness and small-scale vesicularity can significantly decrease the drag coefficient. This is due to the downwind shift of the boundary layer separation point, which results in the occurrence of the drag crisis (see above) at $Re$ lower than the critical $Re$. Experiments on spheres \citep{Achenbach2006} and cylinders \citep{Nakamura2006} clearly show how by increasing the surface roughness, the critical $Re$ shifts to lower values and triggers a premature transition to the drag crisis. Loth \cite{Loth2008} mentioned that irregular particles exhibit little or no drag crisis since such particles have a consistent bluff-body separation point throughout a wide range of $Re$. 
\begin{table*}
  \centering
  \caption{Most used correlations for estimating drag coefficient of non-spherical particles.}
  \vspace{-1em}
\begin{tabular}{L{35mm} l r} \toprule
Ref. & Formula & Eq.
\\ \midrule
Haider and Levenspiel \citep{Haider1989}& $C_D= \left(24/Re \right) \, \left( 1+C_1 \,  Re^{C_2}   \right) + C_3/ \left(1+C_4/Re\right)$ & \eqnum\label{Hider} \\  & $\quad \quad C_1= \exp \left(2.33-6.46 \, \psi +2.45 \, \psi^2\right)$
\\ &  $\quad \quad C_2=0.096+0.556 \, \psi $
\\ &  $\quad \quad C_3= \exp \left(4.90-13.89 \, \psi +18.42 \, \psi^2-10.26 \, \psi^3\right)$
\\ &  $\quad \quad C_4=\exp \left(1.47+12.26 \, \psi -20.73 \, \psi^2-15.89 \, \psi^3\right)$  \\ \midrule
 Ganser \cite{Ganser1993}  & $C_D=\left(24 k_S/Re \right) \, \left( 1+0.1118 \,  \left( Re \, k_N / k_S \right)^{0.6567}   \right) $  & \eqnum\label{GanserCd} \\
 & $\quad \quad \quad \quad+ 0.4305 \, k_N/ \left(1+3305/\left( Re \, k_N / k_S \right) \right)$ & \\
 Leith \cite{Leith1987} & $ \quad \quad k_S=1/ {3 \sqrt{\psi_{\perp}}} + 2/{3 \sqrt{\psi}}$ & \eqnum\label{LeithkS} \\
  Ganser \cite{Ganser1993} & $ \quad \quad k_S=1/ {3 } + 2/{3 \sqrt{\psi}}$ & \eqnum\label{GanserkS} \\
    Loth \cite{Loth2008} & $ \quad \quad k_S=\left( L \, I \, / S^2 \right )^{0.09} $ & \eqnum\label{eqLothkS} \\
  Ganser \cite{Ganser1993} & $\quad \quad k_N=10^{1.8148 \, \left(-\log \psi \right)^{0.5743}} $ & \eqnum\label{GanserkN} \\ \midrule
H{\"o}lzer and Sommerfeld\cite{Holzer2008}& $C_D=8/Re \sqrt{\psi_{\parallel}} + 16/ Re \sqrt{\psi}+3/\sqrt{Re} \, \psi^{3/4} $ & \eqnum\label{Holzer}
\\  & $\quad \quad \quad \quad +0.42 \times 10^{ 0.4 \left(- \log \psi \right)^{0.2}} \left( 1/\psi_{\perp} \right)$
\\ \bottomrule

\label{TabExistModel}

\end{tabular} 
\vspace{-3em}
\end{table*}
\subsubsection{Particle orientation and particle-to-fluid density ratio} \label{sec_intro_orientation}
Particle orientation is another parameter that can affect the drag coefficient of non-spherical particles. As a result, repeated experiments performed on a non-spherical particle of a given shape will show a spread in the measured drag coefficient due to the change in the particle orientation \citep{Happel1983a}. In the Stokes' regime, Cox \citep{Cox1965} showed that a freely falling spheroid with small eccentricity orients itself with the largest projection area normal to the direction of motion. Nevertheless, most particles with a certain well-defined symmetry properties (e.g. spheroidal, orthotropic, isometric, needle and plate particles) have no preferred orientations and fall without rotation in the Stokes' regime \citep{Pettyjohn1948, Albertson1953, Marchildon1964, Happel1983a, Clift2005, Loth2008}. If particles undergo Brownian motion, however, the particle orientation is changing randomly during descent. In such cases the most favorable estimation of the particle drag is an average value obtained from many random orientations \citep{Happel1983a, Clift2005}. Nonetheless, even when particles are not subjected to Brownian motion, an average of random orientations should be considered as the most relevant orientation for obtaining the average of the drag coefficient since in the Stokes' regime most particles do not have any preferred orientation and for a statistically representative run of experiments they can adopt any random orientation.  

As $Re$ increases up to $\approx 100$, particles tend to fall with the largest projection area normal to the direction of motion \citep{Marchildon1964, Komar1978, Clift2005}. Isometric particles show signs of oscillations and instability in the range $70<Re<300$  \citep{Pettyjohn1948}. Early studies on falling cylinders showed that the wake instability starts at $Re>50$ and angular oscillations and lateral deviations are observed at $Re>80-300$\citep{Marchildon1964, Jayaweera1965}. Disks exhibit a steady-falling regime with maximum projection normal to the falling direction at $Re<100$, and at $Re>100$ the falling pattern changes from oscillations to chaotic and tumbling \citep{Willmarth1964}. 

Finally, secondary motions become fully developed in the Newton's regime ($1000 \leq Re<3 \times 10^5$). In addition, in the Newton's regime particle-to-fluid density ratio $\rho'$ can significantly affect orientation and secondary motions of particles, and therefore, the drag coefficient \citep{Willmarth1964, Marchildon1964, Christiansen1965a, Isaacs1967, List1971, Tran-Cong2004, Chow2011, Bagheri2013a}. Studies on regular-shape particles show that as $\rho'$ increases, the secondary motion of particles increases too \citep{Marchildon1964, Christiansen1965a, Isaacs1967, Chow2011}. This leads to the reduction of the average projected area of the particle during falling and, hence, the drag coefficient reduces. However, most studies on falling particles are performed in the range $1<\rho'<15$, which is significantly lower than $\rho'$ for particles falling in the air that is $\mathcal{O} \left( 10^3\right)$. Thus, it is not yet well understood how $\rho'$ can influence the particle orientation at high $\rho'$. 

\subsubsection{Existing non-spherical drag coefficient models}
Table \ref{TabExistModel} shows the most common models for estimating drag coefficient of non-spherical particles. Here, models of Ganser \cite{Ganser1993} and Haider and Levenspiel \cite{Haider1989} are chosen since they were found to be the most accurate correlations for predicting the drag coefficient of non-spherical particles with average errors of 16.3\% and 23.5\%, respectively \citep{Chhabra1999a}. Model of Haider and Levenspiel \cite{Haider1989}, Eq. (\ref{Hider}), is the first generalized correlations for drag coefficient of regular shape particles, which is based on $Re$ and sphericity $\psi$. Haider and Levenspiel \cite{Haider1989} introduced Eq. (\ref{Hider}) based on experimental data on the drag coefficient of isometric particles and disks at $1<\rho'<15$.

\nomenclature[AC]{$C_{1-4}$}{empirical expressions used in Eq. (\ref{Hider})}
\nomenclature[Ak]{$k_S$}{Stokes' drag correction, $C_D / C_{D,\, s}$ (see Eqs. \ref{eq:kS})}
\nomenclature[Ak]{$k_N$}{Newton's drag correction, $C_D/C_{D,\, s}$ (see Eq. \ref{eq:kN})}
\nomenclature[AC]{$C_{D,\, s}$}{the drag coefficient of a sphere with same volume and Reynolds number as the particle}
\nomenclature[Sm2]{$min$}{minimum value}
\nomenclature[Sm1]{$max$}{maximum value}

Later, Ganser \cite{Ganser1993} proposed a simpler formulation, Eq. (\ref{GanserCd}), by using similarity and dimensional analyses. He showed that the drag coefficient of non-spherical particles can be predicted by $Re$ and two other shape-dependent parameters called Stokes' $k_S$ and Newton's $k_N$ drag corrections (Ganser \cite{Ganser1993} noted them as shape factors):
\begin{equation}
  k_S \equiv \frac {C_D}{C_{D,\, s}} = \frac{C_D}{24/Re}  \label{eq:kS}
\end{equation}
\begin{equation}
  k_N \equiv \frac {C_D}{C_{D,\, s}} = \frac{C_D}{0.463}   \label{eq:kN}
\end{equation}
where
\begin{equation}
  C_D=\frac{\vert\mathbf{F_D}\vert}{\frac{1}{2} \rho_f \left( d_{eq}/2 \right)^2 {\vert \mathbf{u_p} -\mathbf{u_f} \vert}^2}  \label{eq:CD}
\end{equation}
and $C_{D,\, s}$ is the drag coefficient of a sphere with same volume and Reynolds number as the particle. As the particle shape tends to a sphere, both $k_s$ and $k_N$ approach unity. Based on formulation of Ganser \cite{Ganser1993}, for a particle of a given shape the drag coefficient at any subcritical Reynolds number ($\approx Re< 3 \times 10^5$) can be predicted if $k_S$ and $k_N$ are known. Various correlations, Eqs. (\ref{LeithkS} -- \ref{GanserkN}), exist in the literature that estimate $k_s$ and $k_N$ as functions of sphericity $\psi$, the so called crosswise sphericity $\psi_{\perp}$ and particle form dimensions. Eq. (\ref{LeithkS}), i.e. $k_S=1/ {3 \sqrt{\psi_{\perp}}} + 2/{3 \sqrt{\psi}}$, suggested by Leith \cite{Leith1987} and used in the models of Ganser \cite{Ganser1993} and H\"{o}lzer and Sommerfeld \cite{Holzer2008}, is one of the most accepted model for estimating $k_S$ in the Stokes' regime that considers both shape and orientation. The crosswise sphericity $\psi_{\perp}$ is an orientation dependent parameter and is defined as 
\begin{equation}
  \resizebox{0.5\textwidth}{!}{
  $\psi_{\perp}=\frac{\mbox{projected area of the volume equivalent sphere}}{\mbox{projected area of the particle normal to the falling direction }}$
  }
  \label{eq:defCrossSph}
  \end{equation}
  
  \nomenclature[Gp]{$\psi_{\perp}$}{crosswise sphericity (see Eq. \ref{eq:defCrossSph})}
 
 \nomenclature[Gp]{$\psi_{\parallel}$}{lengthwise sphericity (see Eq. \ref{Holzer})}
 
Ganser \cite{Ganser1993} suggested to approximate $\psi_{\perp}$  to unity for isometric particles, Eq. (\ref{GanserkS}), but he did not discussed how it would change for non-isometric particles. 
Models of Ganser \cite{Ganser1993} and Haider and Levenspiel \cite{Haider1989} are general models that can predict average drag coefficient of particles falling at $1<\rho'<15$. As a result, they cannot be used to predict the drag coefficient of non-spherical particles in a specific orientation. In order to do so, more complex models similar to the one introduced by H\"{o}lzer and Sommerfeld \cite{Holzer2008}, Eq. (\ref{Holzer}), is needed, in which the particle orientation is also taken into account. In fact, H\"{o}lzer and Sommerfeld \cite{Holzer2008} used three different shape/orientation descriptors, namely particle sphericity $\psi$, crosswise sphericity $\psi_{\perp}$ and lengthwise sphericity $\psi_{\parallel}$. Amongst these parameters, the lengthwise sphericity $\psi_{\parallel}$ is the most complicated parameter to be obtained, which is defined as the ratio between the cross-sectional area of the volume equivalent sphere and the difference between half the surface area and the mean projected longitudinal cross-sectional area of the considered particle \citep{Holzer2008}. Given that the evaluation of $\psi_{\parallel}$  is very complex, H\"{o}lzer and Sommerfeld \cite{Holzer2008} suggested to replace it with $\psi_{\perp}$ with the cost of slight reduction in accuracy. In any case, calculation of both $\psi_{\perp}$ and $\psi_{\parallel}$ needs orientation of the particle to be known, and therefore, Eq. (\ref{Holzer}) is more suitable for Lagrangian computations where the particle orientation along the trajectory is also computed.

\section{Materials}
\begin{figure}[!htb]
  \begin{center}
    \vspace{1em}
    \includegraphics[width=0.47\textwidth]{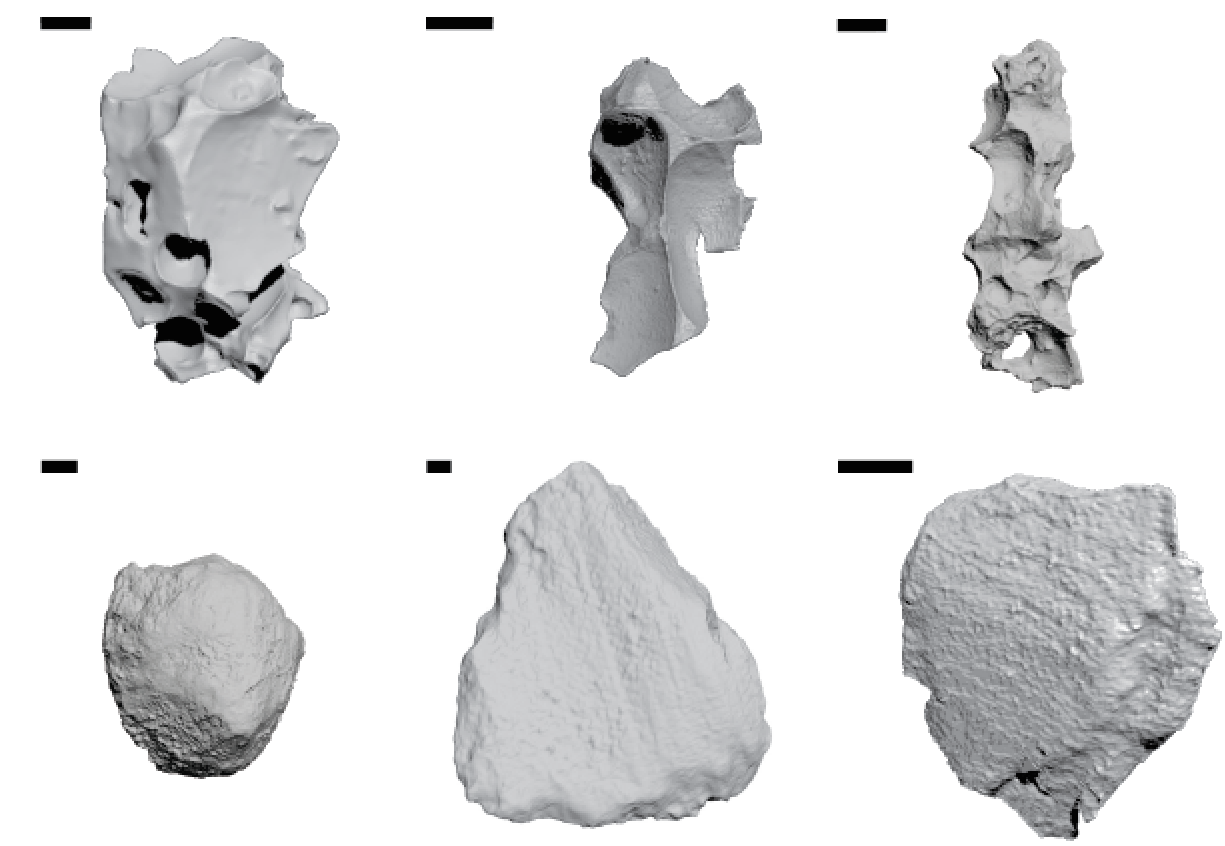}
  \end{center}
  \vspace{-2em}
  \caption[A selection of volcanic particles of Sample Set I tested in the settling columns.]{A selection of volcanic particles of Sample Set I tested in the settling columns adjusted from Bagheri et al. \citep{Bagheri2014} (length of the scale bar is \SI{100}{\micro\metre}).}
  \label{particles_SEMmicroCT}
\end{figure}

Particles used in our experiments were separated in different sample sets based on their size: Sample Set I and Sample Set II. Sample Set I includes 100 irregular volcanic particles, 13 cylinders, 4 parallelepiped and one spherical particle with $155 \, \mu m \leq d_{eq} \leq 1.8 \, mm$. Size and shape of 12 selected irregular volcanic particles are fully characterized using the Scanning Electron Microscope micro Computed-Tomography (SEM micro-CT) \citep{Bagheri2014, P.VonlanthenJ.RauschR.A.KetchamB.PutlitzL.P.Baumgartner}, whereas the rest of sub-millimetric irregular particles are characterized using multiple-projection image analysis techniques (see Appendix A and Bagheri et al. \citep{Bagheri2014} for more details). Selected irregular particles in the Sample Set I are shown in Fig. \ref{particles_SEMmicroCT}. Volcanic particles are from Masaya (Nicaragua, Fontana Lapilli, 60 Ka), K\={\i}lauea (Hawaii, Mystery Unit of Keanakakoi formation, 1790 AD), Villarrica (Chile, Chaimilla unit, 3500 BP), Cotopaxi (Ecuador, layer 2, 290 years BP and layer 5, 1180 years BP), Llaima (Chile, 1957), Chait\'{e}n (Chile, 2008) and Stromboli (Italy, 2007) \citep{Bagheri2014} volcanoes.

\begin{figure*}
  \begin{center}
    \includegraphics[width=0.95\textwidth]{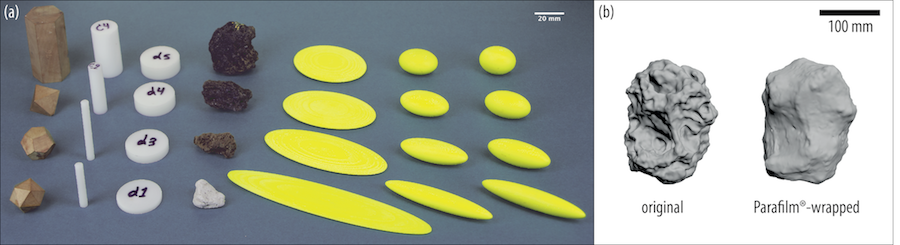}
  \end{center}
  \vspace{-1em}
  \caption[A selection of non-spherical particles of Sample Set II]{(a) A selection of non-spherical particles of Sample Set II tested in the wind tunnel experiments, (b) an irregular particle without and with Parafilm\textsuperscript{\textregistered} wrap.}
  \label{WT_particles}
\end{figure*}

Sample Set II includes 78 irregular volcanic particles, 21 ellipsoids, 12 cylinders, 8 disks and 21 regular shape particles with $10.9 \, mm \leq d_{eq} \leq 61.2 \, mm$ (Fig. \ref{WT_particles}a). In addition, 38 irregular volcanic particles were wrapped in Parafilm\textsuperscript{\textregistered} (a self-sealing, moldable and flexible wax film) in order to make their surface smooth without significantly changing their macroscopic shape characteristics (Fig. \ref{WT_particles}b). This provides insights into the influence of surface roughness on the drag coefficient. Volume and surface area of particles are measured with a NextEngine Inc. desktop 3D laser scanner with accuracy of $\approx 100 \, \mu m$ \citep{Bagheri2013a, Bagheri2014} and their mass were measured by a digital balance with accuracy of $0.001 \, gr$.

A list of all experimental data points used in this study, including those compiled from the literature, are summarized in Table \ref{TabParticles}. Form dimensions of all particles complied from the literature are recalculated based on the projection area protocol described in section \ref{section:shape} and shown in Table \ref{TabShape}.

\section{Methods}
\subsection{Stokes' regime: analytical solutions}
The analytical solution of Oberbeck \cite{Oberbeck1876} is solved numerically to obtain the drag coefficient of ellipsoids with both elongation $e$ and flatness $f$ between 0.01 and 1 (at 0.01 intervals), leading to $10^4$ data points. Oberbeck \citep{Oberbeck1876} showed that the ratio of the drag coefficient of an ellipsoid with the surface equation of   $x^2 / a^2 +y^2 / b^2+z^2 / c^2=1$ moving in the direction of $x$ axis (parallel to a), $C_{D, \, x}$, to the drag coefficient of a sphere with the same volume as the ellipsoid, $C_{D, \, sphere}$, at $Re \ll 1$ is equal to: 
\begin{equation}
  k_{S, \, x} \equiv \frac{C_{D, \, x}}{C_{D, \, sphere}}=\frac{8}{3} \, \frac{\sqrt[2/3]{a \, b \, c}}{\chi_0 + \alpha_0 \, a^3} \label{eq:kSOberbeck}
  \end{equation}
where $a$, $b$, $c$ are semi-axes of the ellipsoid and
\begin{equation}
\chi_0=a \, b \, c \, \int_0^\infty \! \frac{\mathrm{d}\lambda}{\Delta} \label{eq:ChiInteg}
  \end{equation}
 \begin{equation}
\alpha_0=a \, b \, c \, \int_0^\infty \! \frac{\mathrm{d}\lambda}{\left( a^2+\lambda \right)\Delta} \label{eq:AlphaInteg}
  \end{equation}
and 
 \begin{equation}
\Delta=\sqrt{\left( a^2+\lambda \right)\left( b^2+\lambda \right)\left( c^2+\lambda \right)} \label{eq:Delta}
  \end{equation}
  $k_{S,y}$ and $k_{S,z}$ can be obtained similarly for ellipsoids moving in parallel to $y$ and $z$ axes. Eqs. (\ref{eq:kSOberbeck} -- \ref{eq:Delta}) were solved numerically for each ellipsoid falling in $x$, $y$ and $z$ directions. Finally, average of $k_{S}$ for an ellipsoid moving in random orientations is calculated as \citep{Happel1983a,Clift2005}:
 \begin{equation}
k_S=3 \left(\frac{1}{k_{S,x}}+ \frac{1}{k_{S,y}}+\frac{1}{k_{S,z}}\right)^{-1} \label{eq:kSaverage}
  \end{equation}
\nomenclature[Gc]{$\chi_0$}{analytical expression defined in Eq. \ref{eq:ChiInteg}}
\nomenclature[Ga]{$\alpha_0$}{analytical expression defined in Eq. \ref{eq:AlphaInteg}}
\nomenclature[GD]{$\Delta$}{analytical expression defined in Eq. \ref{eq:Delta}}

\begin{small}
\begin{table*}

\footnotesize
  \centering
  \caption[Analytical and experimental databases used in this study.]{Analytical and experimental databases used in this study. \emph{No.} indicates number of experiments/datapoints considered in each category, $d_{eq}$ is the diameter of the volume-equivalent sphere, $\psi$ is the sphericity, $e$ is the elongation, $f$ is the flatness of the particle, and $\rho'$ is the particle-to-fluid density ratio. Literature data include spheres from Pettyjohn and Christiansen \cite{Pettyjohn1948}, Christiansen and Barker \cite{Christiansen1965a}, Schlichting \cite{Schlichting1968}, Roos and Willmarth \cite{Roos1971} and Achenbach \cite{Achenbach1972}; isometric particles (i.e. cube, cuboctahedron, octahedron, tetrahedron) from Pettyjohn and Christiansen \cite{Pettyjohn1948}, disks and cylinders from Willmarth et al. \cite{Willmarth1964}, Christiansen and Barker \cite{Christiansen1965a}, McKay et al. \cite{McKay1988}, Isaacs and Thodos \cite{Isaacs1967} and Clift et al. \cite{Clift2005}.}
  \vspace{-1em}
\begin{tabular}{l l cccccc} \toprule
 & shape &No.& $d_{eq} \, [mm]$& $\psi$ & $e$& $f$& $\rho'$
\\ \midrule
\multicolumn{8}{l}{\textbf{Stokes' regime\string:} $Re<0.1$} \\

\multicolumn{6}{l}{this work (analytical)} \\
& ellipsoid & $10^4$& -- & 0.02 -- 1 & 0.01 --1 & 0.01 -- 1 &  -- \\
\multicolumn{6}{l}{literature (analytical and experimental)}\\
 & isometric & 22& 1.7 -- 15.8 & 0.7 -- 0.9 &0.7 -- 0.9 &0.7 -- 1 & 1 -- 8\\
 & disk & 16 &  --  & 0.5 -- 0.9 & 0.7 -- 1 & 0.1 -- 1 &  -- \\
 & cylinder & 27 &  --  & 0.3 -- 0.9 & 0.02 -- 1 & 1 &  -- 
 \\ \midrule
\multicolumn{3}{l}{\textbf{Intermediate regime:} $0.1 \leq Re <1000$}\\
\multicolumn{6}{l}{this work (experimental, i.e. settling columns: Sample Set I)}\\ 
 & sphere & 5 & 1.45 & 1 & 1 &1& 2270\\
& cylinder &24 &0.63 -- 1.53  & 0.4 -- 0.8 & 0.03 -- 0.4 & 1 & 1400 \\
&prism & 4 & 0.47 -- 0.58 & 0.7 -- 0.8 & 0.7 -- 0.8 & 0.4 -- 0.6 & 1400 \\
&irregular & 196 & 0.15 -- 1.80 & 0.3 -- 0.9 & 0.3 -- 0.8 & 0.2 -- 1 & 2300\\
\multicolumn{6}{l}{literature (experimental)} \\
& sphere & 148 &  --  & 1 & 1 &1& 1 -- 15\\
&isometric&323&1.4 -- 15.8&0.7 -- 0.9&0.7 -- 0.9&0.7 -- 1&1 -- 11\\
&disk & 49 &0.8 -- 18.7 & 0.03 -- 0.8 & 0.9 -- 1 & 0.001 -- 0.5 & 1 -- 8\\
&cylinder & 7 & 4.5 -- 18.3 & 0.8 & 0.4 & 1 & 1 -- 3
 \\ \midrule
 \multicolumn{3}{l}{\textbf{Newton's regime:} $1000 \leq Re <3 \times 10^5$}\\
 \multicolumn{6}{l}{this work (experimental, i.e. vertical wind tunnel: Sample Set II)}\\
 &ellipsoid &120 &22.6 -- 23.2 &0.2 -- 1 &0.2 -- 0.8 &0.1 -- 1 &870\\
 &isometric &72 &17.7 -- 61.2 &0.8 -- 0.9 &0.7 -- 1 &0.7 -- 1 &150 -- 1000\\
 &disk &48 &16.2 -- 24.3 &0.5 -- 0.9 &0.7 -- 1 &0.1 -- 0.9 &1280\\
 &cylinder &72 &11.2 -- 35.9 &0.6 -- 0.9 &0.1 -- 0.7 &1 &560 -- 1300\\
 &Other reg. &48 &23.0 -- 39.0 &0.8 -- 0.9 &0.4 -- 0.7 &0.7 -- 1 &530 -- 750\\
 &Irr. rough &468 &10.9 -- 36.2 &0.5 -- 0.9 &0.5 -- 0.9 &0.4 -- 1 &175 -- 2130\\
 &Irr. smooth &228 &11.7 -- 37.8 &0.8 -- 0.9 &0.6 -- 0.9 &0.6 -- 1 &390 -- 910\\
 \multicolumn{6}{l}{literature (experimental)}\\
 &sphere &136 & --  &1 &1 &1 & -- \\
 &isometric &54 &2.9 -- 15.8 &0.7 -- 0.9 &0.7 -- 0.9 &0.7 -- 1 &2 -- 11\\
 &disk &40 &0.8 -- 21.3 &0.03 -- 0.9 &0.7 -- 1 &0.001 -- 8 &1 -- 10\\
 &cylinder &59 &4.5 -- 73.3 &0.7 -- 0.9 &0.2 -- 0.7 &1 &1 -- 2800
\\ \bottomrule
\end{tabular}
\label{TabParticles}
\end{table*}
\end{small}

\subsection{Intermediate regime: experiments in settling columns}  
\begin{figure}
  \begin{center}
    \includegraphics[width=0.47\textwidth]{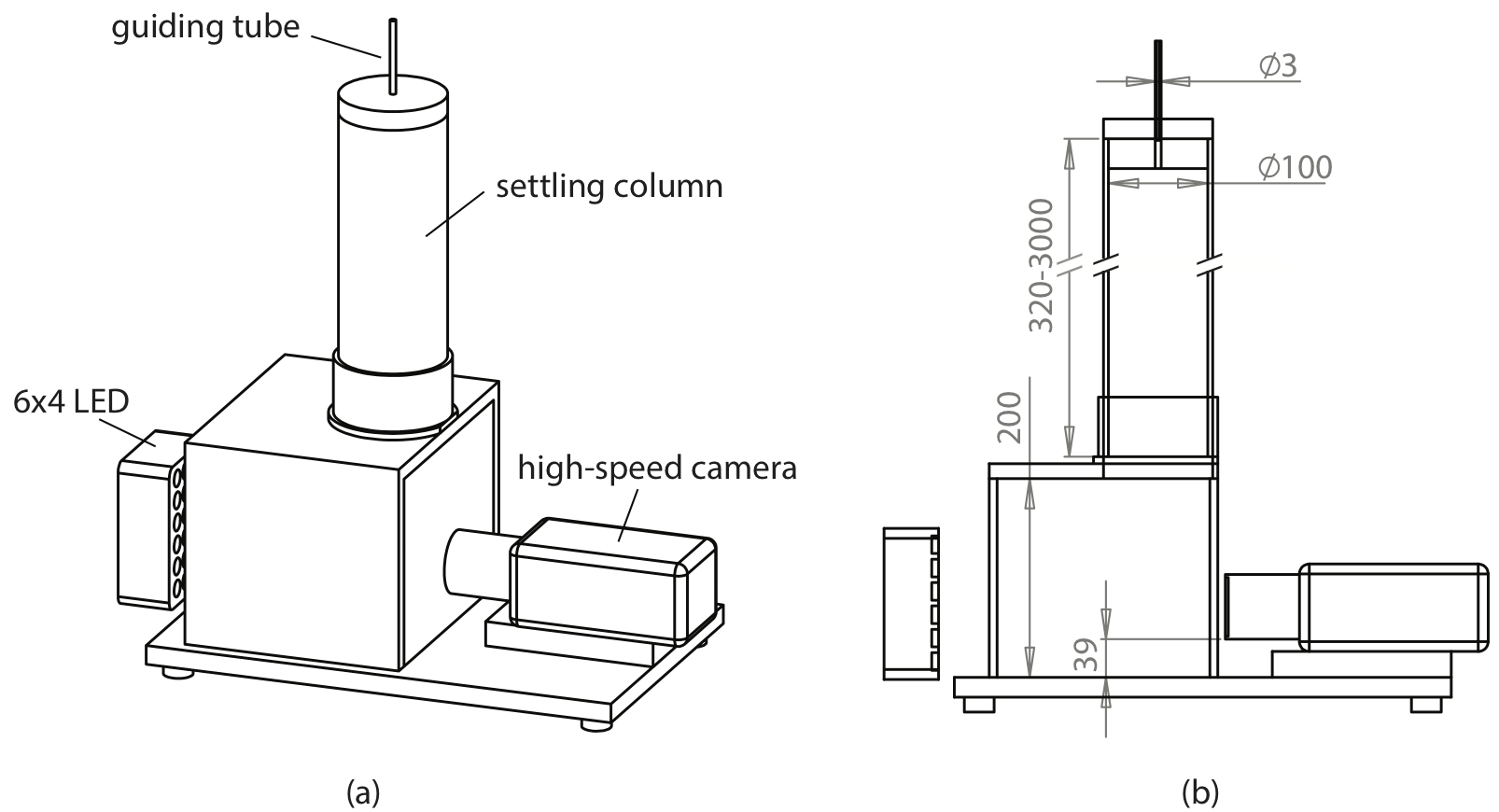}
  \end{center}
  \vspace{-2em}
  \caption[Schematic of the settling columns used in this study.]{Schematic of the settling columns used in this study ($9 \leq Re \leq 900$). (a) the perspective view, (b) the cross section (dimensions are in \SI{}{\milli\metre}).}
    \vspace{-1em}
  \label{SketchSettlingColumn}
\end{figure}

Settling columns of heights between 0.45 and $3.6 \, m$ are used for measuring the drag coefficient of particles of Sample Set I at the intermediate $Re$ ($0.1-1000$) (Fig. \ref{SketchSettlingColumn}). However, experiments could only cover the range $9 \leq Re \leq 900$ due to particle size and set-up characteristics. In each experiment run, a particle was released with zero initial velocity at the top of the settling column and it was filmed at $1600-2000 \, fps$ when it passed in front of a high-speed camera at the bottom of the column. A thin and short tube (guiding tube) is placed at the top of the settling column to keep the particle in the center after releasing. A high intensity 6x4 LED array and a holographic diffuser with transmission efficiency of $>85\%$ was used to backlight the camera field of view. The temperature difference between glass doors in the front and back of settling column was monitored to be $<1 \, ^{\circ} C$ during the experiments in order to prevent occurrence of natural convection inside the settling column. Effects of settling column walls on the measured velocity of falling particles are negligible since the ratio of particle cross-sectional area of the Sample Set I to that of settling columns (with diameter of $10 \, cm$) is very small. 
The high-speed camera was automatically triggered when the particle was in the field of view. By using a AF Micro-Nikkor $60 \, mm$ $f/2.8D$ lens it was possible to record high-speed movies with the pixel size of $20 \, \mu m$ and maximum field of view of $25.6 \, mm \times 16.0 \, mm$ ($1200\times 800$ pixels). Depending on the particle velocity and field of view, between 7 and 30 frames of falling particles were captured in each run. The resulting videos were then converted to 8 bit Tiff format images and analyzed by ImageJ software \citep{Schneider2012}. A Particle Tracking Velocimetry (PTV) code \citep{Bagheri2013a} was used to obtain particle velocity. 

The main error in measuring velocity of falling particles is due to the uncertainty in the particle centroid position. Considering the exposure time of videos (\SI{50}{\micro\second}) and falling velocity of particles (0.8 -- 7.6 \SI{}{\metre\per\second}), the uncertainty in the particle position $\delta y$ is between 41 -- 379 \SI{}{\micro\metre} (i.e. exposure time multiplied by the falling velocity). Thus, given that the vertical displacement of particles in the image $H$ is $800 \times \SI{20}{\micro\metre}$, the error on the measured velocity is between 0.3 -- 2.5\% ($=100\times \delta y/H$ ). Finally, to validate measurements, velocities of three glass spheres were measured and compared to previously published experimental data to validate the measurements. Comparisons showed that measurements have an acceptable average deviation of $5 \%$. 

In the settling column experiments the particle acceleration could not be calculated accurately since the field of view of the camera was relatively small. In order to make sure that particles reached their terminal velocity, they were tested at least in two column heights and the change in the measured velocity was monitored. The drag coefficient could be measured only for the 41 particles that reached their terminal velocity. Measurements for remaining particles, however, were used to benchmark the ability of the final drag coefficient model to predict particle velocity within a given falling distance. For benchmarking accuracy of models, we compare their relative errors with respect to reference values (i.e. analytical solutions or experimental measurements) as follows:

\nomenclature[Ae]{$error\%$}{relative error (see Eq. \ref{eq:error})}
\nomenclature[Sr]{$ref.$}{reference value}
\nomenclature[Sm]{$model$}{value obtained from empirical models}

 \begin{equation}
error(x)=\frac{| x_{ref.} - x_{model}| \times 100}{x_{ref.}} \label{eq:error}
  \end{equation}

\subsection{Newton's regime: experiments in a vertical wind tunnel}  
A $4 \, m$ high vertical wind tunnel \citep{Bagheri2013a} was used to measure the drag coefficient of Sample Set II particles. The vertical wind tunnel was built at the University of Geneva in collaboration with the fluid mechanics group (CMEFE) of the University of Applied Sciences Western Switzerland in Geneva (HES-SO//hepia). Particles were suspended in the upward airflow in the test section with an adjustable velocity of $5-27 \, \SI{}{\metre\per\second}$. Measurements in the wind tunnel on particles of sample Set II covered the range $8 \times 10^3 \leq Re \leq 6 \times 10^4$. The diverging design of the test section creates airflow with decreasing speed as the height of test section increases, which allows us to measure the variation of particle terminal velocity due to the change in their orientation. Particle motions in the test section were filmed with a high-speed camera and then were analyzed with the ImageJ software \citep{Schneider2012} and a PTV code to obtain mean and variation of particle drag coefficient, terminal velocity and projected area normal to airflow. For each particle, at least three experiments were conducted in different airflow speeds to make sure that the variability of the particle terminal velocity due to change in the particle orientation is captured. The reader is referred to Bagheri et al. \citep{Bagheri2013a} for more details on the design of the wind tunnel, the PTV code and experimental setup.

\section{Results}
We present a new model for the determination of the drag coefficient that is based on the Stokes and Newton drag corrections, i.e. $k_S$ and $k_N$. In fact, $k_S$ and $k_N$ are derived following Ganser \cite{Ganser1993} but accounting for shape descriptors that are more accurate and easier to determine than sphericity. First, we discuss the results for the Stokes' regime ($Re<0.1$) in order to parameterize $k_S$ and then the results for the Newton's regime ($10^3 \leq Re < 3 \times 10^5$) in order to parameterize $k_N$. Finally, we generalize the results for all $Re$, including the intermediate regime.

\subsection{Stokes' regime }
\subsubsection{Average \texorpdfstring{$C_D$}{TEXT} of particle in random orientations in the Stokes' regime}
\nomenclature[Gpo]{$\overline{\psi}_{\perp}$}{average of particle ${\psi}_{\perp} $ over 1000 random orientations}
\nomenclature[AA1]{$A_{Proj.}$}{projected area of the particle normal to the falling (or flow) direction, [\SI{}{\square\metre}]}
\nomenclature[SP]{$Proj.$}{Projected (area)}

\begin{figure}[!b]
  \begin{center}
    \includegraphics[width=0.47\textwidth]{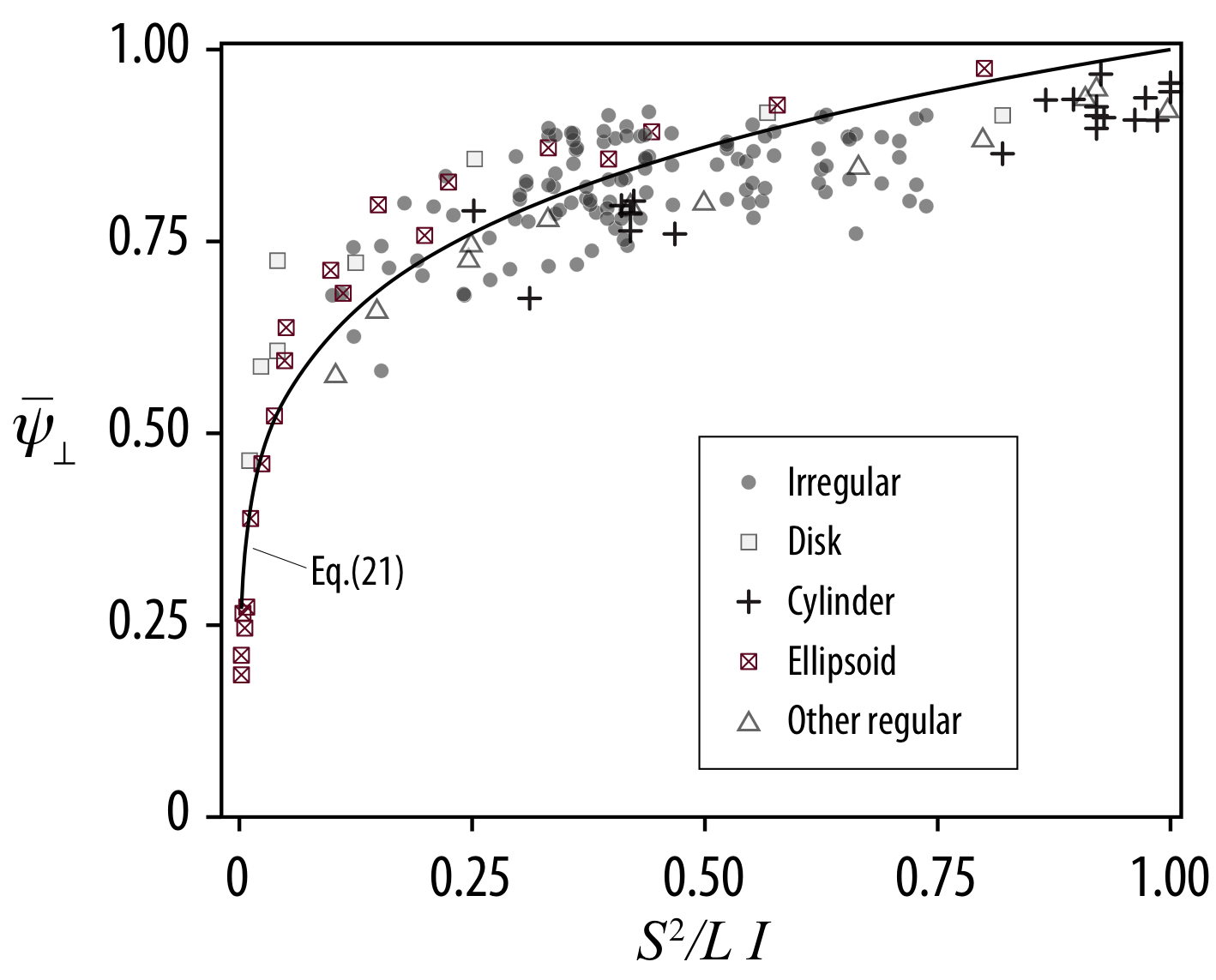}
  \end{center}
  \vspace{-2em}
  \caption[Dependency of the crosswise sphericity $\overline{\psi}_{\perp}$ averaged over random orientations for non-spherical particles of different shapes to the ratio of $S^2/L \, I$.]{Dependency of the crosswise sphericity $\overline{\psi}_{\perp}$ averaged over random orientations for non-spherical particles of different shapes to the ratio of $S^2/L \, I$. $\overline{\psi}_{\perp}$ is calculated by image analysis of projections obtained from 3D models of particles in Sample Set I and II in random orientations. \emph{Other regular} particles include cubes, pyramids, polyhedrons.}
  \label{SphCrossCorr} 
\end{figure}

In order to evaluate the perfoemence of Eqs. (\ref{LeithkS}) and (\ref{Holzer}), first we need to determine the particle crosswise sphericity, ${\psi}_{\perp}$ (assuming ${\psi}_{\parallel} \approx {\psi}_{\perp}$ in Eq. (\ref{Holzer}) as suggested by \cite{Holzer2008}). One way to achieve this, is to find a correlation between the crosswise sphericity averaged in many orientations, $\overline{\psi}_{\perp}$, and a simple shape description of the particle. Therefore, 1000 projections of 3D models of particles of different shapes in random orientations were created and their average projected area $\overline{A}_{Proj.}$ was measured (see \citep{Bagheri2013a,Bagheri2014} for more details). $\overline{\psi}_{\perp}$ was then calculated as the ratio of the projected area of an equivalent volume sphere ($\pi \, d_{eq}^2 /4$) to $\overline{A}_{Proj.}$. The correlation between $\overline{\psi}_{\perp}$ and various shape descriptors of particles (e.g. sphericity, flatness, elongation) was investigated, and it was found that $\overline{\psi}_{\perp}$ is best correlated with $S^2/L \, I$(Fig. \ref{SphCrossCorr}):
 \begin{equation}
\overline{\psi}_{\perp}=1.1 \, \left( S^2/ L \, I\right)^{0.177}-0.1 \label{eq:CrossSphCorr}
  \end{equation}
  
Eq. \ref{eq:CrossSphCorr} is associated with an average error of 13\% for ellipsoids and 7\% for all particle shapes. As mentioned earlier in section \ref{section:shape}, $S^2/L \, I$ is highly correlated with the particle flatness \citep{Bagheri2014}, which suggests that particle flatness is an important parameter for determining the particle projected area and $\overline{\psi}_{\perp}$.
 
\begin{figure}[]
  \begin{center}
    \includegraphics[width=0.47\textwidth]{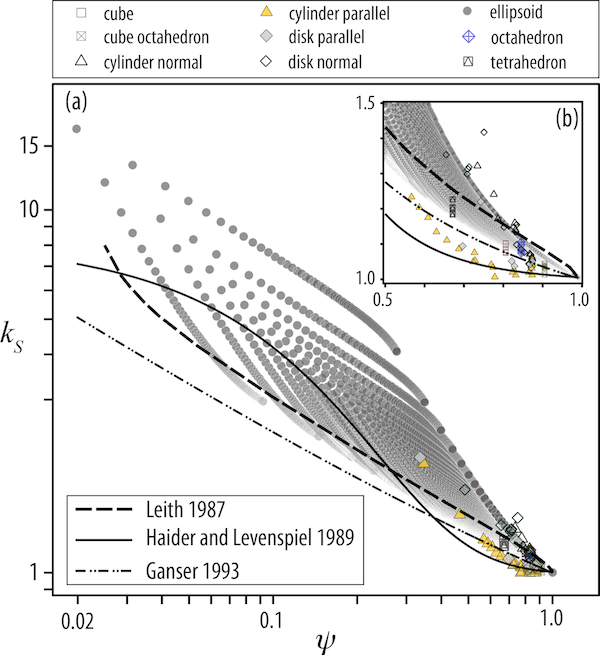}
  \end{center}
  \vspace{-2em}
  \caption[Stokes' drag correction $k_S$ ($C_D/C_{D, \, sphere}$) against sphericity for particles of various shapes calculated/measured in the Stokes' regime, $Re <0.1$.]{(a) Log plot showing the Stokes' drag correction $k_S$ ($C_D/C_{D, \, sphere}$) against sphericity for particles of various shapes calculated/measured in the Stokes' regime, $Re <0.1$ (see Table \ref{TabParticles}). (b) a zoom of plot (a) in linear scales. Cylinder and disks released with maximum projection area perpendicular to the falling direction called normal and those with minimum projection area normal to the falling direction called parallel. Data for non-ellipsoid shapes are from Pettyjohn and Christiansen \citep{Pettyjohn1948} and Clift et al. \citep{Clift2005}.}
    \vspace{-1em}
  \label{kS_SPH} 
\end{figure}

The accuracy of Eqs. (\ref{Hider} -- \ref{GanserkS}) and (\ref{Holzer}) (Table \ref{TabErrorStokes}) for estimating $k_S$ of particles calculated/measured in the Stokes' regime (see Table \ref{TabParticles}) is benchmarked. Fig. \ref{kS_SPH} shows that $k_S$ increases with decreasing sphericity $\psi$. For particles with $\psi>0.4$ the estimations of Ganser \cite{Ganser1993} (Eq. \ref{GanserkS}) and Leith \cite{Leith1987} (Eqs. \ref{LeithkS} and \ref{eq:CrossSphCorr}) are closer to the calculated $k_S$. In any case, from Fig. \ref{kS_SPH} it is evident that the sphericity $\psi$ is not a good candidate for estimating drag coefficient of non-spherical particles in the Stokes' regime, given the large spread in the data.

\begin{figure}[t]
  \begin{center}
    \includegraphics[width=0.48\textwidth]{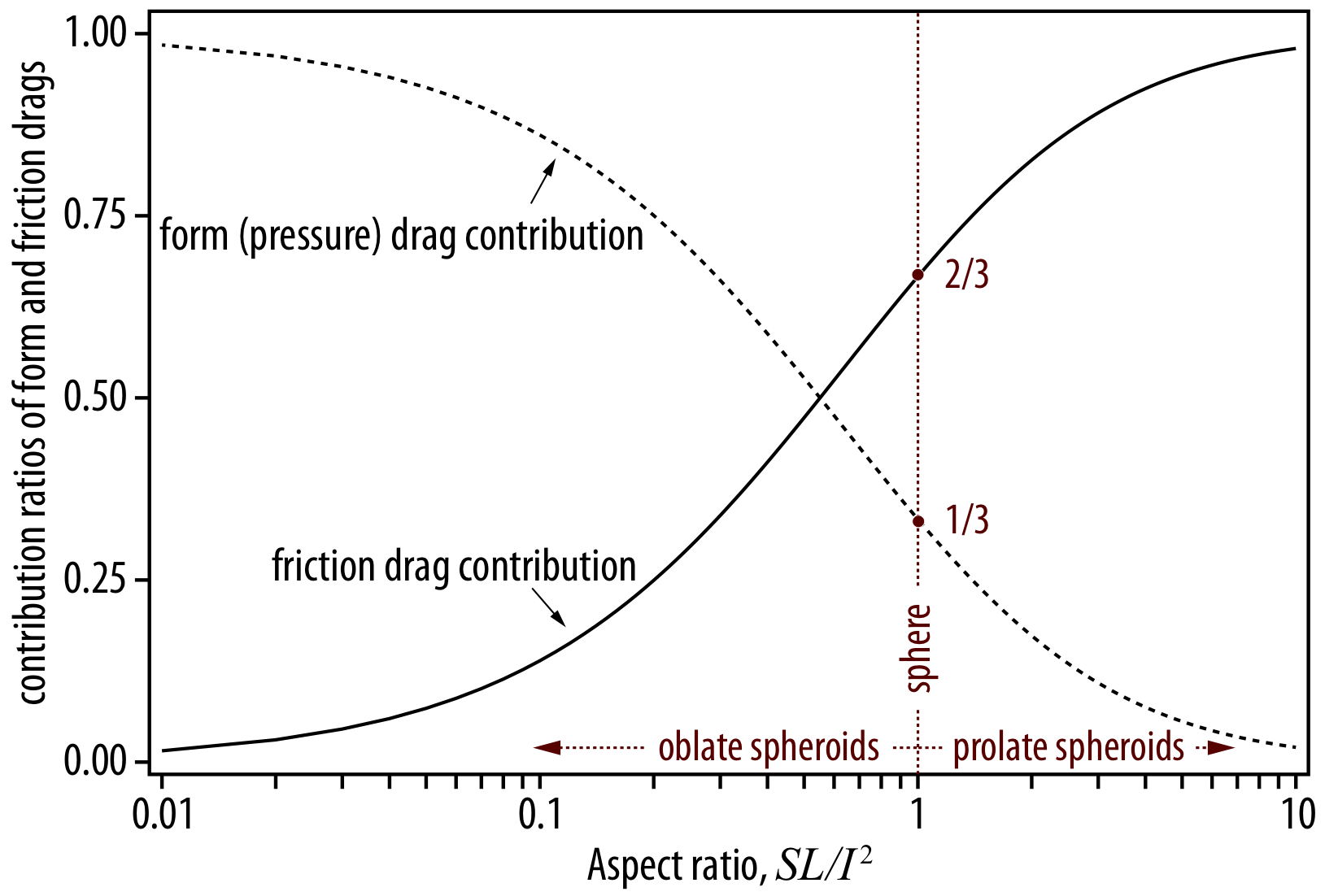}
  \end{center}
  \vspace{-2em}
  \caption{Contribution ratios of form and friction drags to the total drag exerted on oblate and prolate ellipsoids in the Stokes' regime versus ellipsoid aspect ratio. These ratios are calculated by analytical equations provided in Table 4.1 of Clift et al. \cite{Clift2005} for axisymmetric flow.}
    \vspace{-1em}
  \label{FrictionFormDragRatio} 
\end{figure}

The use of Eq. \ref{eq:CrossSphCorr} that takes into account particle orientation in the model of Leith \cite{Leith1987}, Eq. (\ref{LeithkS}), can fit the data better than the model of Ganser \cite{Ganser1993}. However, the improvement is not significant since these models assume that the contribution of form and friction drags are similar to those for sphere. In fact, the model of Leith \cite{Leith1987} is based on the fact that one third of the sphere drag in the Stokes' regime is due to the form drag (affected by the particle orientation) and two thirds of it is the friction drag (related to the particle surface area). These ratios, however, can significantly vary for non-spherical particles of different shapes. As an example, the contribution ratios for ellipsoids can vary significantly from those of the sphere (Fig. \ref{FrictionFormDragRatio}).  

A summary of error analyses for models shown in Table \ref{TabExistModel} is presented in Table \ref{TabErrorStokes}. The calculated $k_S$ based on the model of Haider and Levenspiel \cite{Haider1989} performs better for particles with $\psi<0.25$ compared to other models, but is still associated with large deviations up to 57.5\%.  

\begin{figure}[!b]
  \begin{center}
    \includegraphics[width=0.47\textwidth]{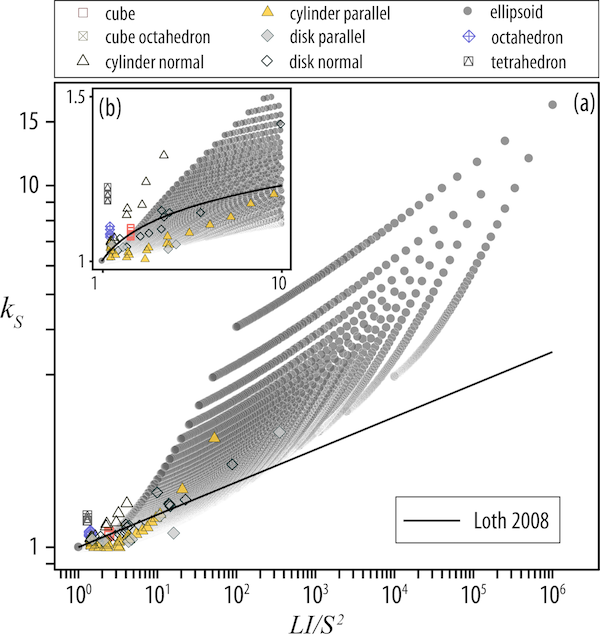}
  \end{center}
  \vspace{-2em}
  \caption[Stokes' drag correction $k_S$ against the shape descriptor introduced by Loth \cite{Loth2008}.]{(a) Log plot showing the Stokes' drag correction $k_S$ against the shape descriptor introduced by Loth \cite{Loth2008} for particles of various shapes moving in the Stokes' regime ($Re < 0.1$). (b) a zoom of plot (a) in linear scales. Data source is similar to Fig. \ref{kS_SPH}. }
  \label{LothkS} 
\end{figure}

Another shape descriptor suggested by Loth \cite{Loth2008} is a form factor defined as $L \, I \ S^2$ (Eq. \ref{eqLothkS} in Table \ref{TabErrorStokes}). Fig. \ref{LothkS} shows that, similar to the sphericity, $L \, I / S^2$ is not correlated well with $k_S$. In particular, it cannot discriminate isometric shapes, such as cuboctahedron, octahedron and tetrahedron, from each other. 

\begin{figure}[]
  \begin{center}
    \includegraphics[width=0.45\textwidth]{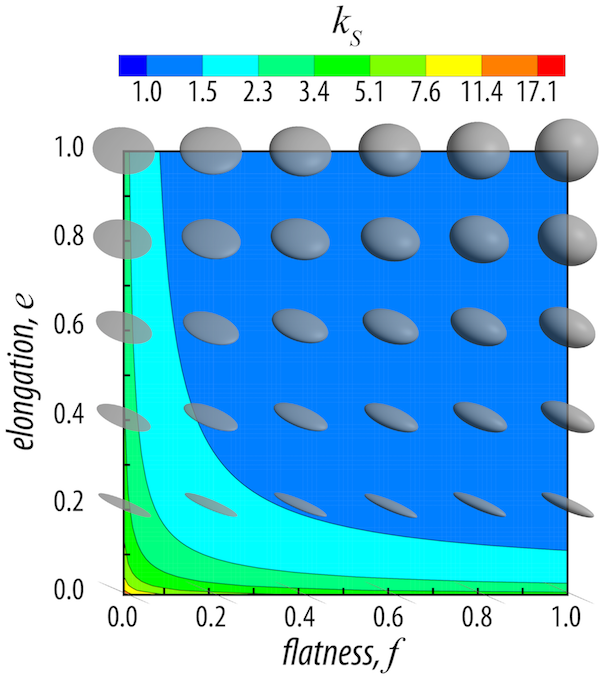}
  \end{center}
  \vspace{-2em}
  \caption{Impact of flatness $f$ and elongation $e$ on the particle Stokes' drag correction $k_S$.}
  \label{KS_vs_fl_el} 
\end{figure}

In the search for a better shape descriptor, we found that $k_S$ is almost equally sensitive to both elongation and flatness, with slightly higher sensitivity to elongation, as it is shown in Fig. \ref{KS_vs_fl_el}. Therefore, a simple form factor, such as $f \, e^{1.3}$ ($=S \, I^{0.3}/ L^{1.3}$), can correlate well with $k_S$ of ellipsoids. However, in order to avoid issues mentioned for the form factor of Loth \cite{Loth2008} (i.e. issues in discriminating isometric shapes), it is necessary to combine it with an additional parameter that is a function of characteristics of the particle other than form dimensions, such as $d_{eq}$. This parameter can be defined as $d_{eq}^3/L \, I \, S$ and if multiplied by the form factor found for ellipsoids, a new shape descriptors, which we define as Stokes form factor $F_S$, can be obtained: 
 \begin{equation}
F_S= f \, e^{1.3} \, \left( \frac{d_{eq}^3}{L \, I \, S}\right)= \frac{d_{eq}^3}{L^{2.3} \, I^{0.7}}  \label{eq:NewFS}
  \end{equation}
  
\nomenclature[Gpo]{$F_S$}{Stokes form factor, $f \, e^{1.3}$}

\begin{figure}[!b]
  \begin{center}
    \includegraphics[width=0.47\textwidth]{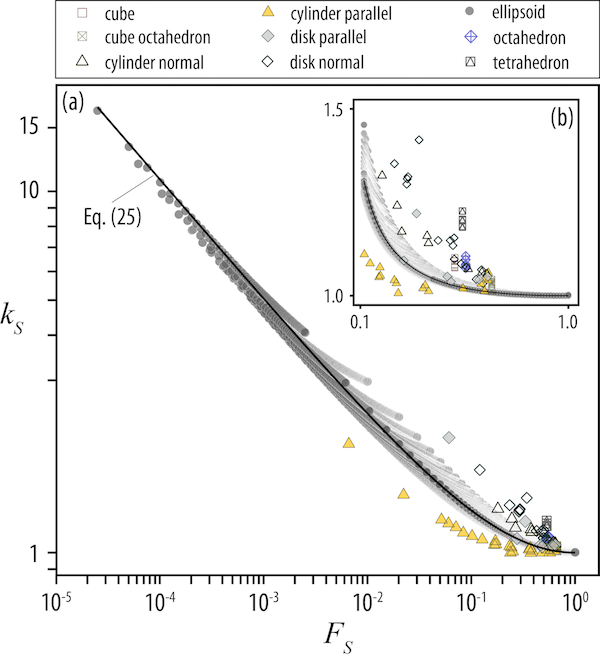}
  \end{center}
  \vspace{-2em}
  \caption[Stokes' drag correction $k_S$ against the new Stokes shape descriptor $F_S$.]{(a) Log plot showing the Stokes' drag correction $k_S$ against the new Stokes shape descriptor $F_S$ for particles of various shapes moving in the Stokes' regime ($Re<0.1$). (b) a zoom of plot (a) in linear scales. Data source are similar to Fig. \ref{kS_SPH}.}
    \vspace{-1em}
  \label{NewFS} 
\end{figure}
Eq. (\ref{eq:NewFS}) indicates that $F_S$ is comprised between 0 and 1; it is equal to 1 for a sphere and decreases as the particle shape becomes less spherical. It should be noted that for ellipsoids $d_{eq}^3=L \, I \, S$ and, therefore, $F_S$ reduces to $f \, e ^{1.3}$. Fig. \ref{NewFS} shows that $k_S$ correlates very well with $F_S$ for ellipsoids and other regular particles and a correlation can be found for estimating $k_S$ as a function of $F_S$:
 \begin{equation}
k_S=\frac{1}{2} \left(F_S^{1/3}+\frac{1}{F_S^{1/3}} \right)  \label{eq:NewkS}
  \end{equation}
Eq. (\ref{eq:NewkS}) is the most accurate and reliable equation with a mean error of 2.4\% and maximum error of 33.9\% (Table \ref{TabErrorStokes}). 

\begin{table*}[t]
  \centering
  \caption{Mean and maximum error of models presented in Table \ref{TabExistModel} in estimating the average Stokes' drag correction, Eq. (\ref{eq:kSaverage}), of $10^4$ ellipsoids. For models of Leith \cite{Leith1987} and H{\"o}lzer and Sommerfeld \cite{Holzer2008}, the average of crosswise sphericity in random orientations, Eq. (\ref{eq:CrossSphCorr}), is used for estimating the crosswise sphericity.}
  \vspace{-1em}
\begin{tabular}{l c c} \toprule
Correlation & \multicolumn{2}{c}{$error \%$} 
\\ \cmidrule{2-3}
& $mean$ & $max$ \\ \midrule
Haider and Levenspiel \cite{Haider1989}, Eq. (\ref{Hider})&
12.8 & 57.5 \\
Leith \cite{Leith1987} \& H{\"o}lzer and Sommerfeld \cite{Holzer2008}, Eqs. (\ref{LeithkS}) and (\ref{Holzer})&6.70&57.8\\
Ganser \cite{Ganser1993}, Eq. (\ref{GanserkS})&
10.4 & 69.7\\
Loth \cite{Loth2008}, Eq. (\ref{eqLothkS})&
10.3&79.3\\
Present, Eq. (\ref{eq:NewkS})&	2.44&33.9
\\ \bottomrule
\label{TabErrorStokes}
\end{tabular} 
\end{table*}

\subsubsection{Effects of particle orientation on \texorpdfstring{$C_D$}{TEXT} in the Stokes' regime}
\begin{figure}[!t]
  \begin{center}
    \includegraphics[width=0.48\textwidth]{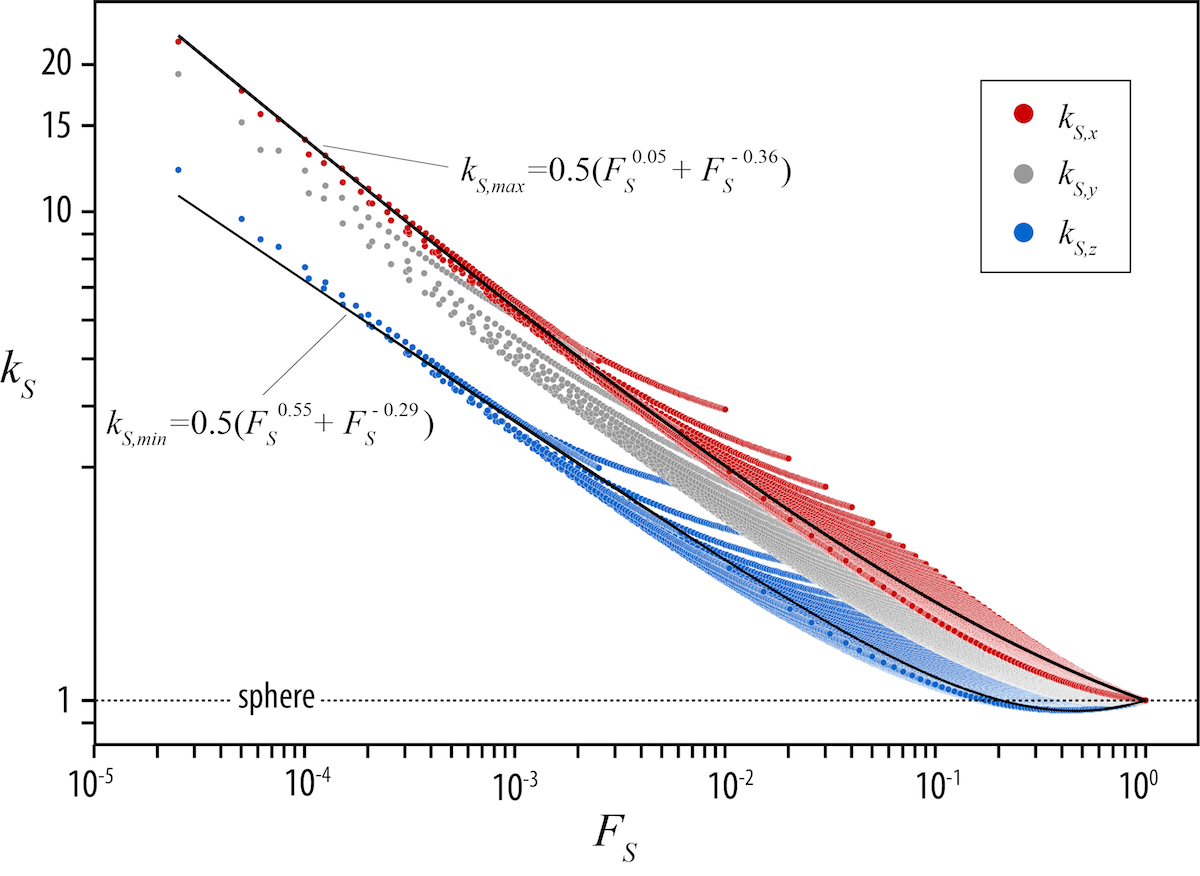}
  \end{center}
  \vspace{-2em}
  \caption[Calculated $k_S$ of ellipsoids falling in different orientations against $F_S$.]{Calculated $k_S$ of ellipsoids falling in different orientations against $F_S$. $k_{S, \, max}$ and $k_{S, \, min}$ are the Stokes' drag correction for ellipsoids that fall with their maximum and minimum projection areas normal to their falling paths, respectively.}
  \label{NewkSOrient} 
\end{figure}

As mentioned earlier (section \ref{sec_intro_orientation}), particle orientation is an important parameter that can significantly affect the drag. The effect of orientation of cylinders and disks on the drag coefficient can already be seen in Fig. \ref{NewFS}, which shows how the cylinders and disks falling with the largest area perpendicular to the flow (i.e. cylinder and disk normal) have higher drag compared to when they fall with the smallest projected area (i.e. cylinder and disk parallel). The drag coefficient for ellipsoids that settle parallel to one of the semi-axis, i.e. $k_{S, \, x}$, $k_{S, \, y}$, $k_{S, \, z}$, are calculated through Eqs. (\ref{eq:kSOberbeck} -- \ref{eq:Delta}) and shown in Fig. \ref{NewkSOrient}. The trend for $k_{S, \, x}$, $k_{S, \, y}$ and $k_{S, \, z}$ is similar to that of $k_S$, except that in some orientations it is possible that the ellipsoid experiences a drag lower than that of its volume-equivalent sphere (e.g. $k_{S, \, x}<1$). The minimum values for $k_{S, \, x}$, $k_{S, \, y}$ and $k_{S, \, z}$ are respectively 0.955, 0.988 and 0.998 that occur at $F_S$ of 0.417, 0.700 and 0.457, respectively. However, the average drag coefficient of ellipsoids in random orientations is always larger than that of the volume-equivalent sphere, i.e. $k_S>1$. The extremes of variation in the drag coefficient of an ellipsoid due to the change in its orientation can be predicted with a fit very similar to Eq. \ref{eq:NewkS}:
 \begin{equation}
k_S=\frac{1}{2} \left(F_S^{\alpha_1}+\frac{1}{F_S^{\beta_1}} \right)  \label{eq:NewkSOrient}
  \end{equation}
where $0.05<\alpha_1<0.55$ and $0.29<\beta_1<0.35$. The upper extreme curve $k_{S, \, max}$ occurs for $\alpha_1=0.55$ and $\beta_1=0.29$ in Eq. \ref{eq:NewkSOrient}, and $k_{S, \, min}$ occurs when  $\alpha_1=0.55$ and $\beta_1=0.29$ (Fig. \ref{NewkSOrient}). The average drag coefficient in random orientations, $k_S$, can be obtained simply by considering $\alpha_1=\beta_1=1/3$. 

\nomenclature[Sx]{$x,\, y,\,z$}{Cartesian coordinates}
\nomenclature[Ss]{$s$}{sphere}
\nomenclature[Sp]{$p$}{particle}
\nomenclature[Sf]{$f$}{fluid}
\nomenclature[SS]{$S$}{Stokes' regime: $Re\ll 1$}
\nomenclature[SS]{$N$}{Newton's regime: $1000<Re<3 \times 10^5$}
\nomenclature[Ga]{$\alpha_1$}{empirical variable used in Eq. (\ref{eq:NewkSOrient})}
\nomenclature[Gb]{$\beta_1$}{empirical variable used in Eq. (\ref{eq:NewkSOrient})}

\begin{figure}[!b]
  \begin{center}
    \includegraphics[width=0.47\textwidth]{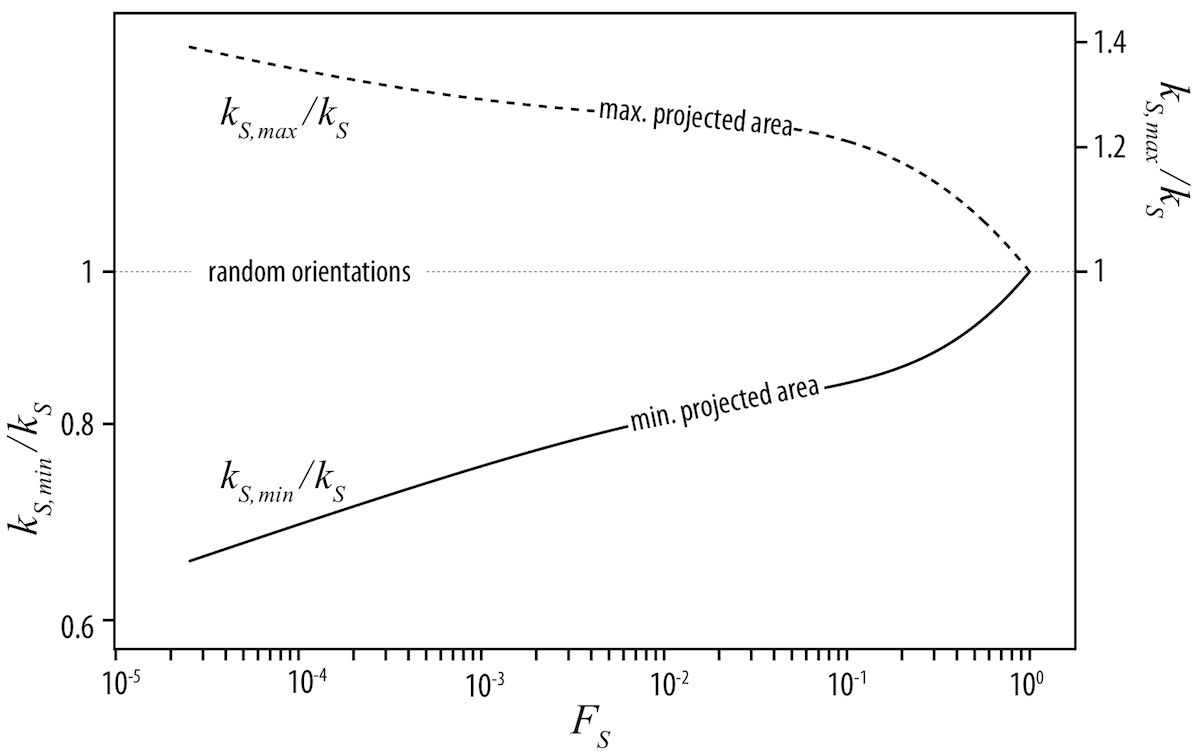}
  \end{center}
  \vspace{-2em}
  \caption[Effect of particle shape, $F_S$, on the sensitivity of ellipsoid drag to the change in orientation in the Stokes' regime.]{Effect of particle shape, $F_S$, on the sensitivity of ellipsoid drag to the change in orientation in the Stokes' regime. More the particle deviates from spherical shape (i.e. low $F_S$), more $k_{S, \, max}$ and $k_{S, \, min}$ deviate from $k_S$ that is obtained for randomly orientated ellipsoids. This shows that the effect of the orientation on the drag coeffcient is more significant for highly non-spherical particles.}
  \label{kSOrienationShape} 
\end{figure}

It is also important to know how the effect of particle orientation on the drag coefficient depends on particle shape. As the particle shape becomes less spherical the effect of particle orientation becomes more significant due to the increase of the ratio between maximum and minimum projection areas (Fig. \ref{kSOrienationShape}). It can also be noted that $k_{S, \, max}$ is on average 10\% (maximum of 20\%) higher than $k_S$, whereas $k_{S, \, min}$ is on average 13\% (maximum of 37\%) lower than $k_S$ for the particles considered here (Fig. \ref{kSOrienationShape}).

The accurate correlation for estimating $k_S$ from $F_S$, which is based on particle form dimensions and spherical equivalent diameter, is a great simplification in comparison to surface-area-dependent parameters, such as sphericity, in particular for irregular particles. However, when the spherical equivalent diameter cannot be measured directly, correlations presented by Bagheri et al. \cite{Bagheri2014} can be used that are based on form dimensions. Finally, $F_S$ can also be calculated by considering the term  $d_{eq}^3/L \, I \, S$ equal to one, in which case the particle shape will be approximated with an ellipsoid of a similar form (i.e. flatness and elongation).

\subsubsection{Effects of surface roughness and vesicularity on \texorpdfstring{$C_D$}{TEXT} in the Stokes' regime}

\begin{figure}[t]
  \begin{center}
    \includegraphics[width=0.47\textwidth]{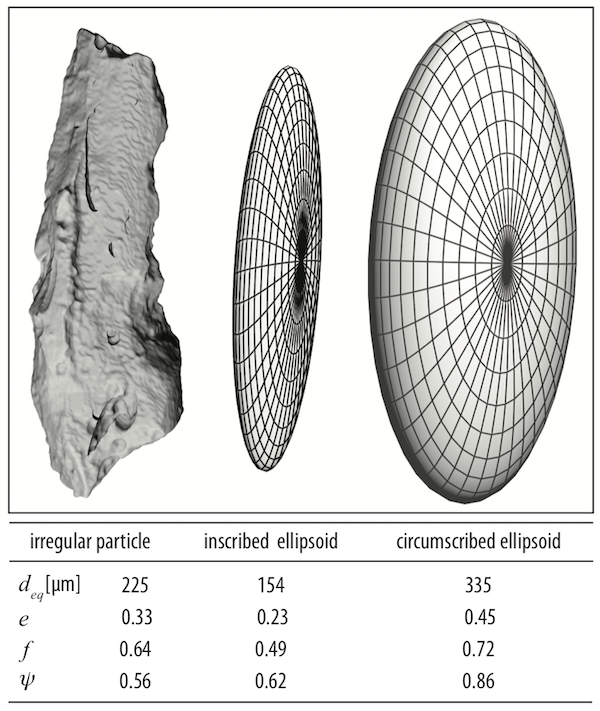}
  \end{center}
  \vspace{-2em}
  \caption[Inscribed and circumscribed ellipsoids found manually for the 3D model of an irregular volcanic particle.]{Inscribed and circumscribed ellipsoids found manually for the 3D model of an irregular volcanic particle. $f$, $e$ and $\psi$ are particle flatness, elongation and sphericity, respectively.}
  \label{kSHillsMethod} 
\end{figure}

Another fundamental question is: how irregularities in the particle shape, e.g. surface roughness, small-scale vesicularity, that cannot be captured by $F_S$, can affect the particle drag? To answer this question, we performed a test study by applying the theorem of Hill and Power \citep{Hill1956} to find the drag coefficient of an irregular particle (Fig. \ref{kSHillsMethod}). Assuming that the irregular particle shown in Fig. \ref{kSHillsMethod} is moving at $Re=0.01$ with constant relative velocity, $Re$ for inscribed and circumscribed ellipsoids will be $6.8 \times 10^{-3}$ and $1.5 \times 10^{-2}$, respectively, since their diameters are different and they should move with the relative velocity. As a result, by calculating $k_S$ of inscribed and circumscribed ellipsoids with Eq. \ref{eq:NewkS}, it can be found that $k_S$ for the irregular particle should be bounded between 0.97 and 1.64. However, we could improve the lower bound estimation furthermore by knowing that $k_S$ is always $\geq1$. Thus, Hill and Power \citep{Hill1956} principle suggests that $k_S=1.31$ for the irregular particle with a maximum uncertainty of 25\%. On the other hand, if we use Eq. \ref{eq:NewkS} directly, we would get $k_S=1.34$, which is within 2.5\% of deviation from the average of $k_S$ found by the method of Hill and Power \citep{Hill1956}. This indicates that small-scale irregularities and surface vesicularity do not significantly alter the drag coefficient.

Another insight provided by Hill and Power \citep{Hill1956} principle is that the sphericity is not an appropriate shape descriptor for estimating drag coefficient of irregular particles in the Stokes' regime. In fact, sphericity of the irregular particle is lower than sphericity of both inscribing and circumscribing ellipsoids. This implies that any correlation based on sphericity would predict higher drag for the irregular particle than both the inscribed and circumscribed ellipsoids.

\subsection{Newton's regime}
\subsubsection{Average of \texorpdfstring{$C_D$}{TEXT} for particles falling in the Newton's regime}
\begin{figure}[]
  \begin{center}
    \includegraphics[width=0.47\textwidth]{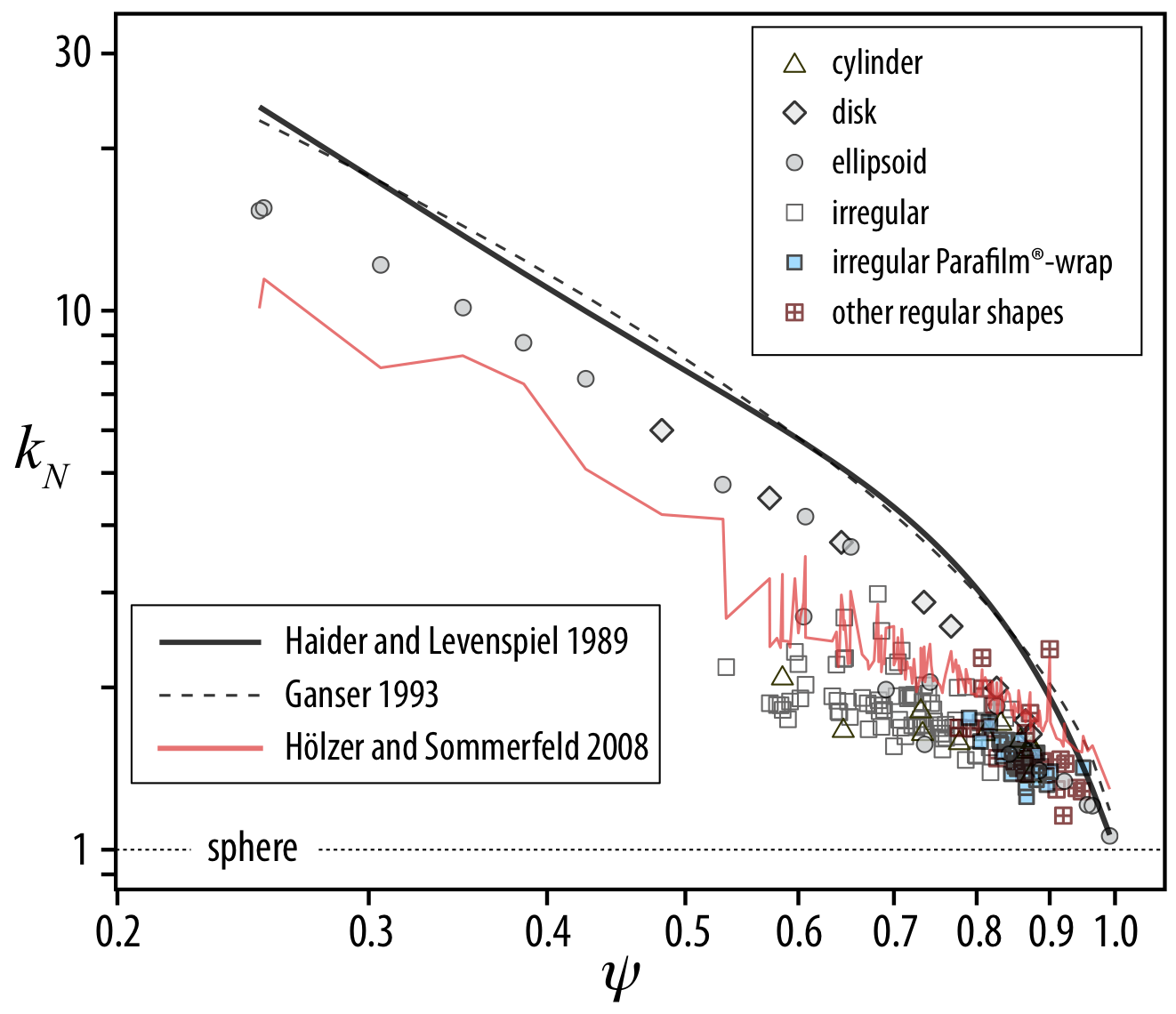}
  \end{center}
  \vspace{-2em}
  \caption{Newton's drag correction, $k_N$, of freely suspended non-spherical particles measured in the present study using the vertical wind tunnel against sphericity.  Estimations of models presented in Table \ref{TabExistModel} are also plotted.}
  \label{SphkN} 
\end{figure}

Non-spherical particles experience Newton's regime at different range of $Re$ depending on their shape. Here, the general range of $10^3 \leq Re \, k_N / k_S \leq 3\times 10^5$ is used to define the Newton's regime of any particle shape. $k_N$ for various non-spherical particles measured in our wind tunnel with $100 \leq \rho' \leq 2200$ are shown against sphericity in Fig. \ref{SphkN}. Although the trend shows that by decreasing the sphericity the drag coefficient increases, there is a considerable scatter at $\psi >0.5$ even for particles of regular shapes. For the sake of comparisons, estimations obtained from models of Haider and Levenspiel \cite{Haider1989}, Ganser \cite{Ganser1993} and H\"{o}lzer and Sommerfeld \cite{Holzer2008}(Eqs. \ref{Hider}, \ref{GanserkN} and \ref{Holzer}) that are based on measurements at $1 < \rho' <15$ are also shown in Fig. \ref{SphkN}. For the model of H\"{o}lzer and Sommerfeld \cite{Holzer2008} that accounts for particle orientation, the crosswise sphericity $\psi_{\perp}$ for each particle during the suspension in the wind tunnel is measured with computer vision algorithms \citep{Bagheri2013a}. 

Table \ref{TabErrorNewton} shows that models of Haider and Levenspiel \cite{Haider1989} and Ganser \cite{Ganser1993} are very close together and overestimate the drag coefficient of all particles with an average error of 90\% (max. error $\approx 240\%$). These large overestimations with respect to wind tunnel measurements is due to the fact that they are based on experiments at much lower density (low $\rho'$). On the other hand, the model of H\"{o}lzer and Sommerfeld \cite{Holzer2008} performs significantly better since it uses an additional variable to take into account the particle orientation, but it is still associated with a significant average error of 22\% (max. error $\approx \, 66\%$). In particular, it underestimates $k_N$ of regular particles and overestimates that of irregular particles.

\begin{figure}[]
  \begin{center}
    \includegraphics[width=0.45\textwidth]{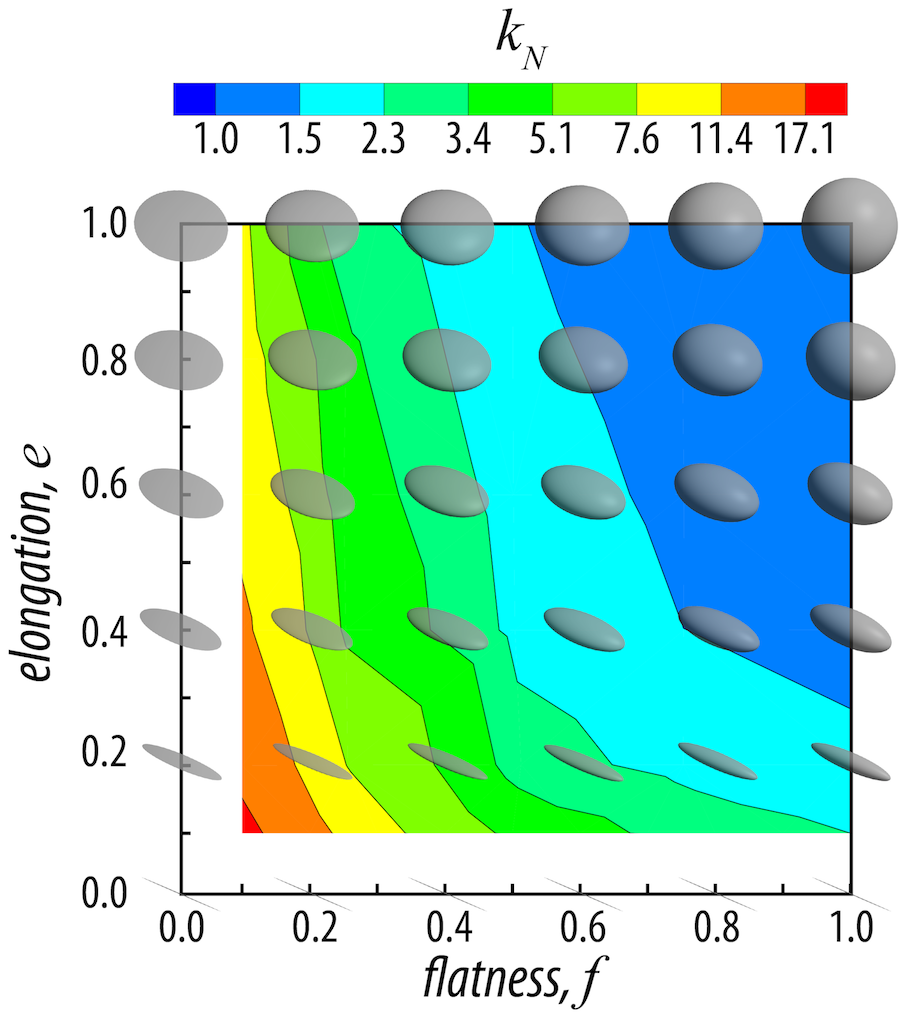}
  \end{center}
  \vspace{-2em}
  \caption{Impact of flatness $f$ and elongation $e$ on the Newton's drag correction $k_N$ of non-spherical particles measured in the present study using the vertical wind tunnel.}
    \vspace{-1em}
  \label{KN_vs_fl_el} 
\end{figure}

In order to find another shape descriptor that has a better correlation with $k_N$ than sphericity, various shape descriptors including flatness, elongation and circularity measures were tested; it was found that $k_N$ is more sensitive to flatness than to elongation (Fig. \ref{KN_vs_fl_el}). 
As a result, the following shape descriptor, here defined as the Newton shape descriptor $F_N$, was found:
 \begin{equation}
F_N=f^2 \, e \, \left( \frac{d_{eq}^3}{L \, I \, S}\right)= \frac{S \, d_{eq}^3}{L^{2} \, I^{2}}  \label{eq:NewFN}
  \end{equation}

\begin{figure}[]
  \begin{center}
    \includegraphics[width=0.47\textwidth]{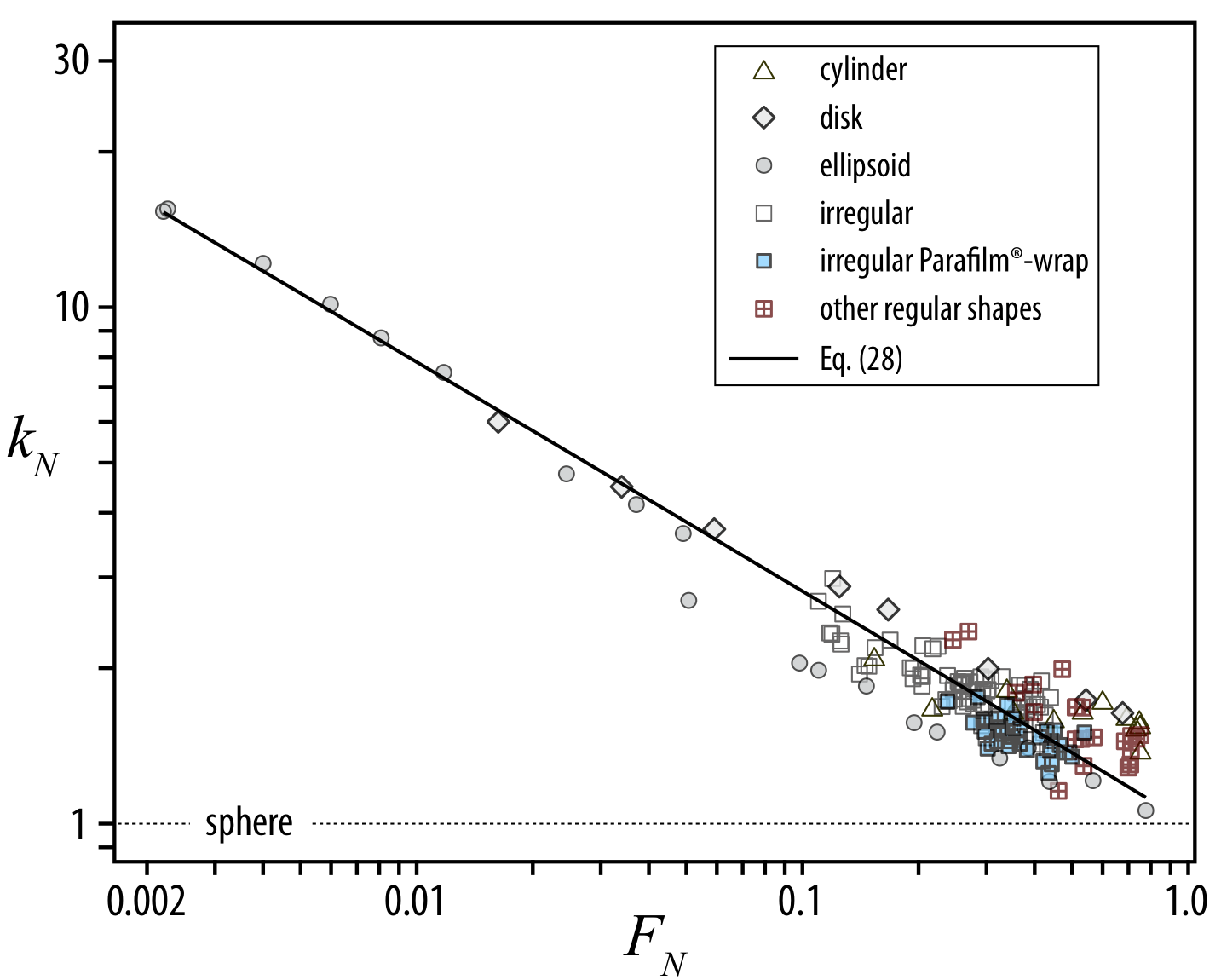}
  \end{center}
  \vspace{-1em}
  \caption{Newton's drag correction, $k_N$, of freely suspended non-spherical particles measured in the present study using the vertical wind tunnel versus the new Newton's shape descriptor $F_N$. Eq. (\ref{eq:NewkNAir}) found in this study for estimating $k_N$ is also shown on the plot. }
    \vspace{0em}
  \label{NewkNFNAir} 
\end{figure}

Note that $f^2 \, e$ ($=S^2/L \, I$) in Eq. (\ref{eq:NewFN}) is equivalent to the shape descriptor used in Eq. (\ref{eq:CrossSphCorr}) for estimating $\psi_{\perp}$ of particles in random orientations and, as mentioned earlier in section \ref{section:shape}, is the square of the so called \emph{Corey form factor} \citep{Corey1963}. Corey form factor is highly correlated with the particle flatness \citep{Bagheri2014} and has been used in several studies for estimating drag coefficient of particles \citep{Loth2008}. The term ${d_{eq}^3}/{L \, I \, S}$ in Eq. (\ref{eq:NewFN}) is used to avoid issues mentioned earlier for distinguishing isometric particles and is equal to 1 for ellipsoids. It can be seen in Fig. \ref{NewkNFNAir} that $k_N$ of particles measured in the wind tunnel is highly correlated with $F_N$ and a fit can be found as:
 \begin{equation}
 \log{\left( k_N \right)}=0.45 \, \left[-\log{\left( F_N \right)} \right]^{0.99} \quad \mbox{for } 150<\rho'<2130 \label{eq:NewkNAir}
  \end{equation}
  
As it is shown in Table \ref{TabErrorNewton}, average error of Eq. (\ref{eq:NewkNAir}) for estimating $k_N$ is about 10.9\% (max. error 43.6\%), which is considerably lower than errors found for existing models.

\begin{table*}[]
  \centering
  \caption[Mean and maximum error associated with the estimations of the drag coefficient of non-spherical particles (including particles of regular and irregular shapes) measured in the present study using the vertical wind tunnel ($150 \leq \rho' \leq 2130 $) based on selected correlations.]{Mean and maximum error associated with the estimations of the drag coefficient of non-spherical particles (including particles of regular and irregular shapes) measured in the present study using the vertical wind tunnel ($150 \leq \rho' \leq 2130 $) based on selected correlations. For a complete benchmark including all the particles studied in this work and literature see Table \ref{TabErrorNewtonAll}.}
  \vspace{-1em}
\begin{tabular}{l c c} \toprule
Correlation & \multicolumn{2}{c}{$error\%$} 
\\ \cmidrule{2-3}
& $mean$ & $max$ \\ \midrule
Haider and Levenspiel \cite{Haider1989}, Eq. (\ref{Hider})&
91.1&242\\
Ganser \cite{Ganser1993}, Eq. (\ref{GanserkN})&
89.4& 244\\ H{\"o}lzer and Sommerfeld \cite{Holzer2008}, Eq. (\ref{Holzer})&21.6& 66.3\\
Present, Eq. (\ref{eq:NewkNAir})&	10.9&43.6
\\ \bottomrule
\vspace{-2em}
\label{TabErrorNewton}
\end{tabular} 
\end{table*}

\subsubsection{Effects of surface roughness and vesicularity on \texorpdfstring{$C_D$}{TEXT} in the Newton's regime}
An important point that can be mentioned regarding $F_N$ is that it is not sensitive to the surface roughness and small-scale irregularities. However, as mentioned earlier, it is a known fact that for spheres and fixed cylinders in the Newton's regime, roughness can significantly decrease the drag force by shifting downwind the separation point of the boundary layer \citep{Achenbach2006}. To investigate the influence of roughness on the drag coefficient of irregular particles, 38 irregular particles were wrapped in Parafilm\textsuperscript{\textregistered} to create smooth surfaces for particles (see Fig. \ref{WT_particles}b). Wrapping particles with Parafilm\textsuperscript{\textregistered} increased both the particle mass, diameter, sphericity and $F_N$ for about 9\%, 6\%, 21\% and 19\%, respectively. 
\begin{figure}[!b]
  \begin{center}
    \includegraphics[width=0.48\textwidth]{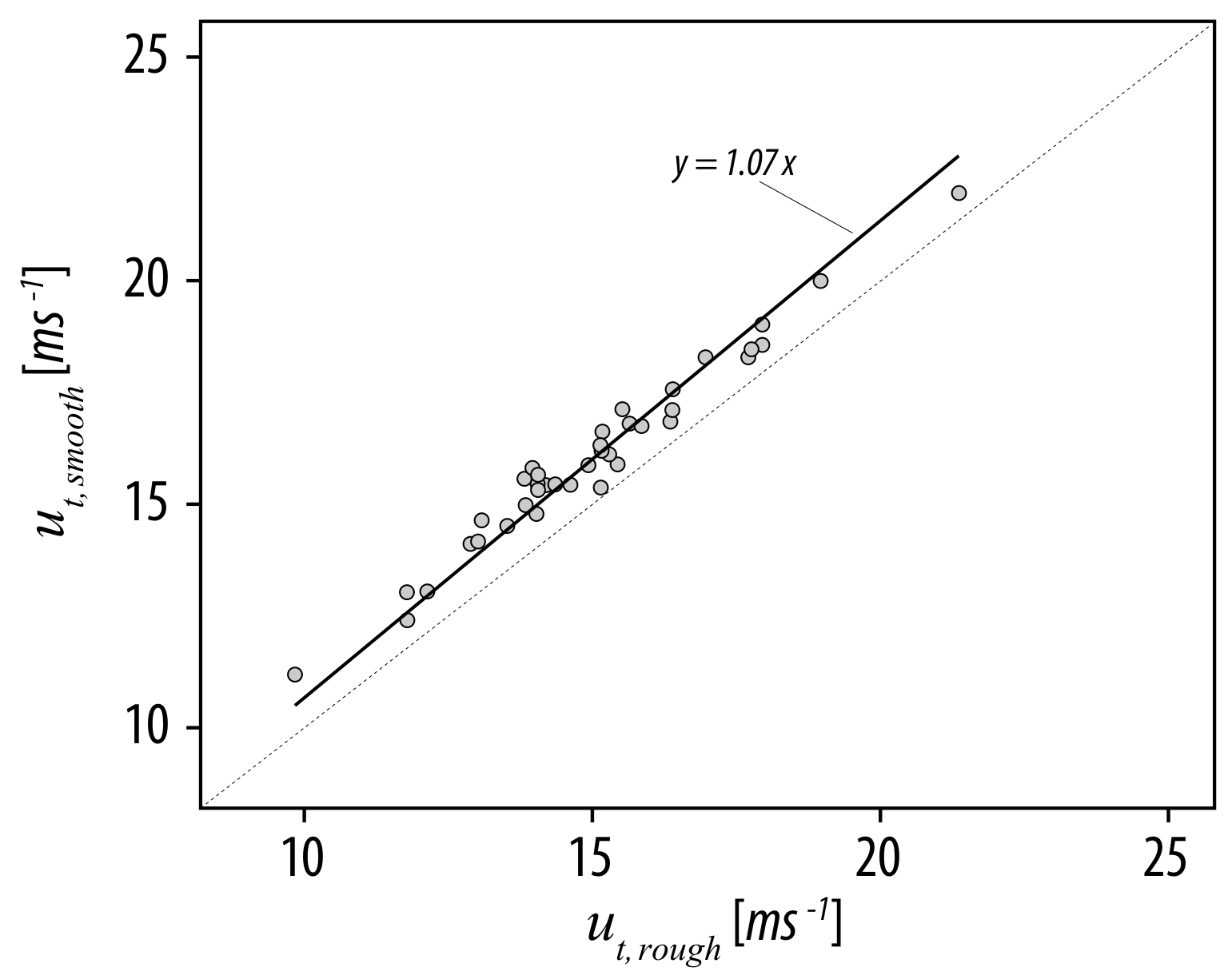}
  \end{center}
  \vspace{-2em}
  \caption{Comparison of terminal velocity, $u_t$, measured in the vertical wind tunnel for irregular particles without (rough) and with Parafilm\textsuperscript{\textregistered} wrap (smooth) ($7.9\times10^3<Re<4.5\times10^4$). }
  \label{RoughvsParafilm} 
\end{figure}
Fig. \ref{RoughvsParafilm} shows the terminal velocity for irregular particles with and without Parafilm\textsuperscript{\textregistered} wrap measured in the wind tunnel. It can be seen that the terminal velocity of particles wrapped in Parafilm\textsuperscript{\textregistered} increases by about 7\%, which is a sign a of a reduction in the drag coefficient. In fact, by wrapping particles with Parafilm\textsuperscript{\textregistered}, the drag coefficient decreases on average by about 19\%. Based on Eq. (\ref{eq:NewkNAir}), at least 8\% of this reduction can be explained by the increase in $F_N$. The rest can be due to changes in shape characteristics that cannot be explained by $F_N$ and Eq. (\ref{eq:NewkNAir}).

In any case, this decrease cannot be due to the shift in the separation point for boundary layer, since, if this was the case, the reduction in the drag coefficient should have been much larger (e.g. $\sim 75 \%$ reduction for sphere). In conclusion, the effect of surface roughness and vesicularity on the drag coefficient of irregular particles, at least for those measured here and at $8 \times 10^3 \leq Re \leq 6 \times 10^4$, is about $~10\%$. Such an effect is negligible compared to the effect of particle shape, such as  $F_N$ (i.e. $~48\%-77\%$).

\subsubsection{Effects of particle orientation and density ratio on \texorpdfstring{$C_D$}{TEXT} in the Newton's regime} \label{sssec:density_newton}
\begin{figure}[]
  \begin{center}
    \includegraphics[width=0.48\textwidth]{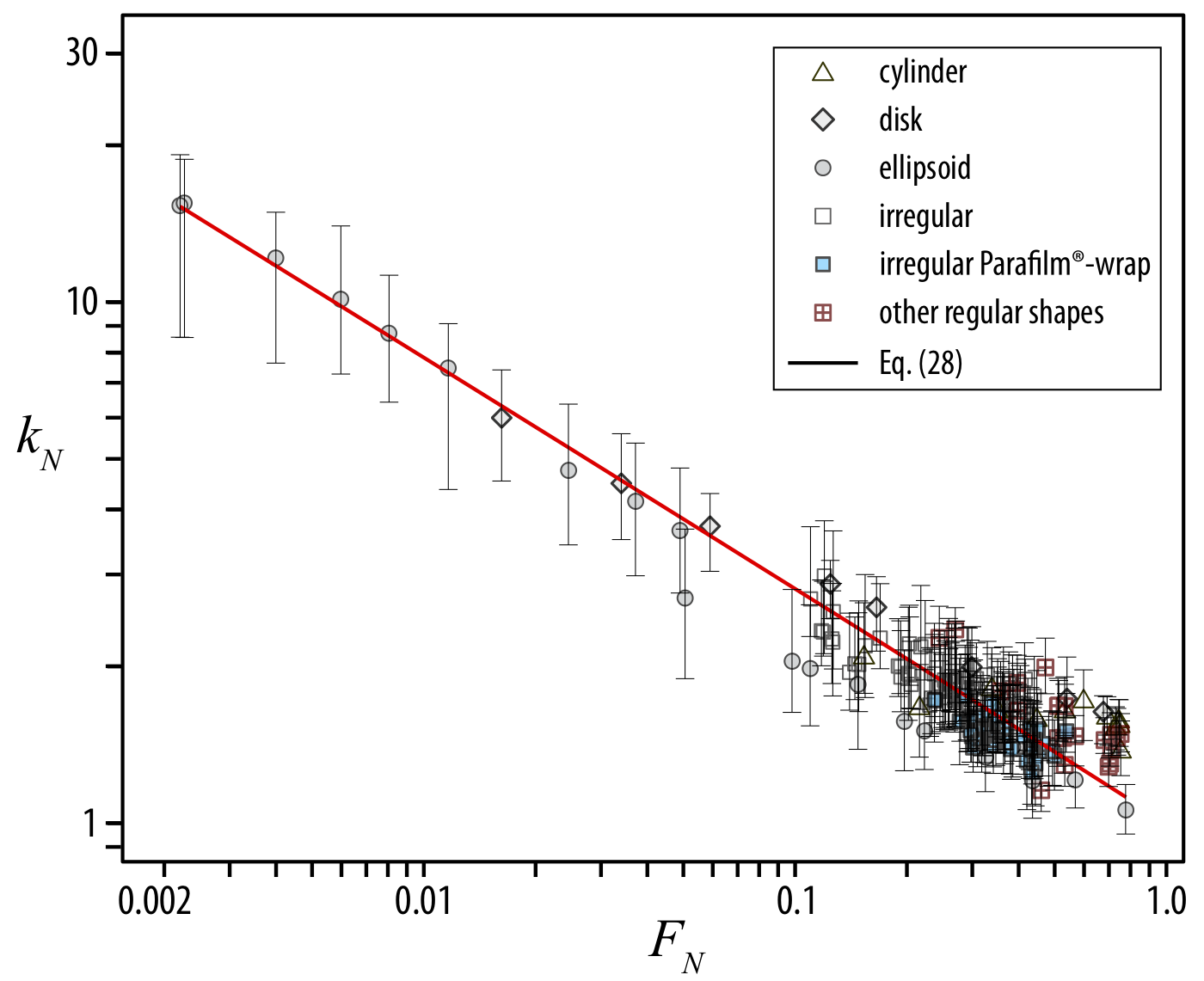}
  \end{center}
  \vspace{-2em}
  \caption{Variability of the Newton's drag correction, $k_N$, of non-spherical particles measured in the present study using the vertical wind tunnel. This variability is due to the change in the orientation of particles under free suspension conditions. Note that this plot is valid for particles falling in gases since it is based on the measurements at $150 \leq \rho' \leq 2130 $. }
  \vspace{-1em}
  \label{Fig15_Varaibility} 
\end{figure}

Fig. \ref{Fig15_Varaibility} shows the variability of the Newton's drag correction, $k_N$, of non-spherical particles measured in the vertical wind tunnel. The variability in the particle drag is due to the fact that the orientation of non-spherical particles is not fixed during free suspension (or free fall). To capture this variability, results of at least three experiments conducted at different wind speeds are merged together for each particle. As it can be seen from Fig. \ref{Fig15_Varaibility}, the drag coefficient (hence the terminal velocity) of non-spherical particles is not constant and it is better described by a range of values. It is possible that the variability in the drag coefficient is broader than those we could measured in the vertical wind tunnel since highly flat/elongated particles could not be suspended in extreme orientations (i.e. maximum and minimum projection area) long enough to perform the measurements.

\begin{figure}[]
  \begin{center}
    \includegraphics[width=0.48\textwidth]{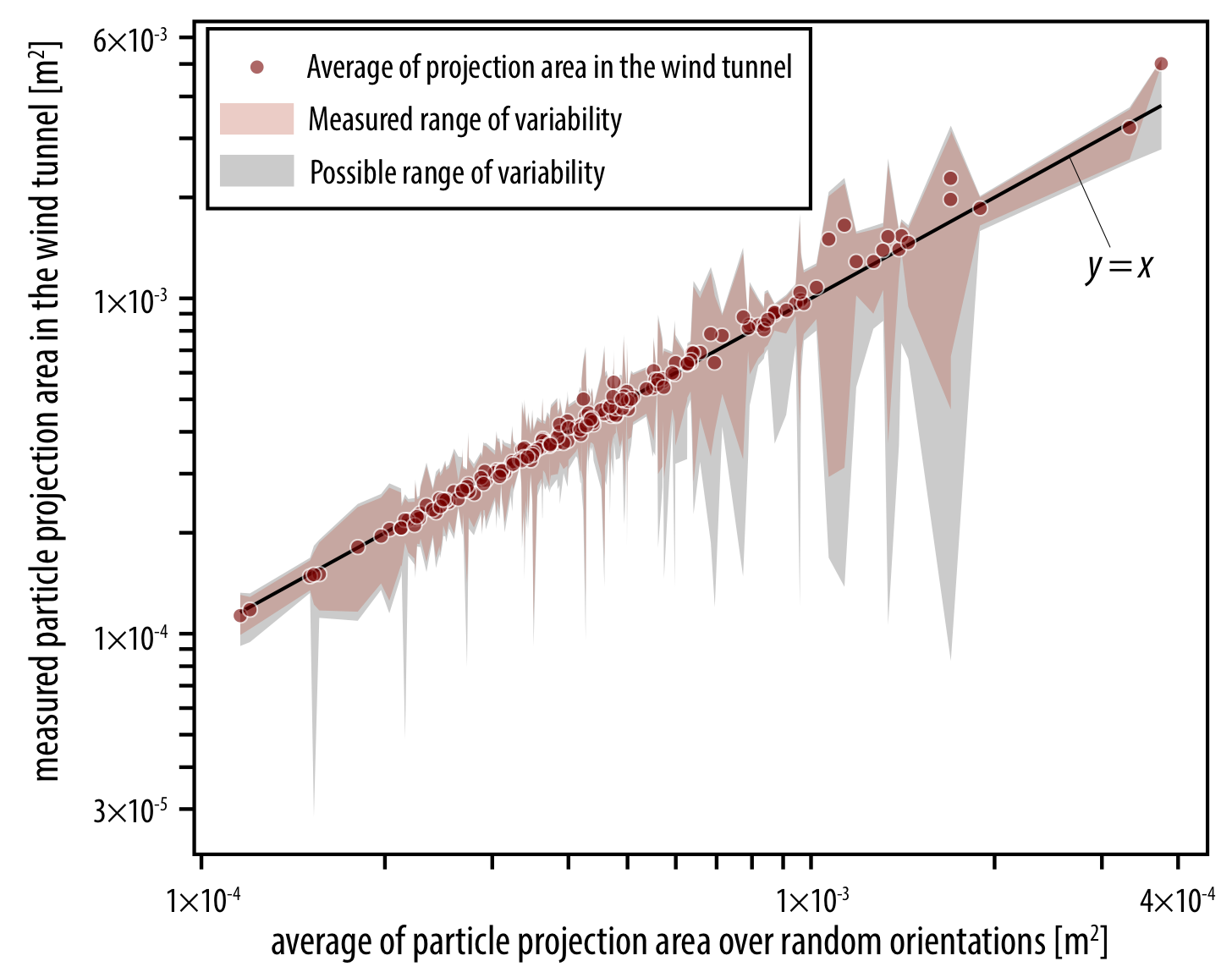}
  \end{center}
  \vspace{-2em}
  \caption[Average of particle projection area normal to the direction of flow measured in the vertical wind tunnel versus average of particle projection area over 1000 random orientations.]{Average of particle projection area normal to the direction of flow measured in the wind tunnel versus average of particle projection area over 1000 random orientations. Variation of projection area is both measured based on wind tunnel experiments (red shading) and calculated from the particle 3D model (gray shading).}
  \label{WTProjectedArea} 
\end{figure}

Additional data resulted from our wind tunnel study is the average of particle projection area normal to the airflow during suspension in the wind tunnel (Fig. \ref{WTProjectedArea}). Interestingly, the average of particle projection area in the wind tunnel is very close to the average of projected areas of particles in random orientations, which is closer to their maximum projected area rather than to their minimum. This suggests that for a freely falling particle at high $\rho'$, the \emph{preferred orientation} is very close to the average of its random orientations.

In the Newton's regime, as mentioned earlier (sec. \ref{sec_intro_orientation}), the orientation of freely falling particles is a function of particle-to-fluid density ratio, $\rho'$. The effect of particle orientation on the drag at high $\rho'$ is already presented in Fig. \ref{Fig15_Varaibility}. We also investigated the effect of $\rho'$ for particles measured in the vertical wind tunnel only, but no correlation could be found between $k_N$ and $\rho'$. This indicates that when $\rho'> 100$ the drag coefficient is no more affected by $\rho'$. However, in order to find a general correlation for estimating the drag coefficients of freely falling particles valid at any $\rho'$, more data of particles with low $\rho'$ should also be considered. This can be achieved by adding the available results in the literature for the drag of freely falling particles measured in liquids . 

\begin{figure}[]
  \begin{center}
    \includegraphics[width=0.47\textwidth]{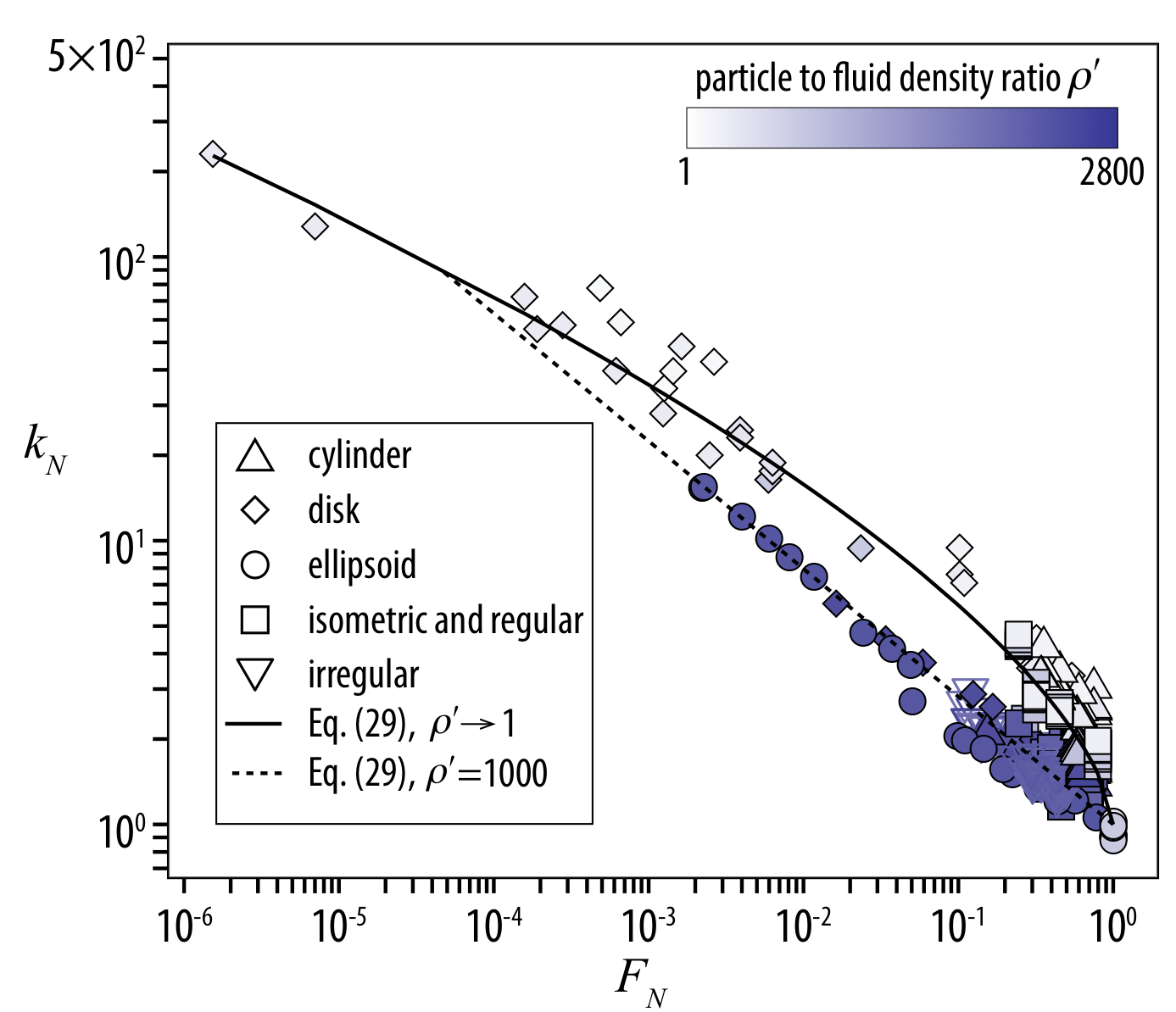}
  \end{center}
  \vspace{-2em}
  \caption[Newton's drag correction $k_N$ for freely falling non-spherical particles versus $F_N$ at different particle-to-fluid density ratios $\rho'$.]{Newton's drag correction $k_N$ for freely falling non-spherical particles versus $F_N$ at different particle-to-fluid density ratios $\rho'$ for our wind tunnel experiments (Fig. \ref{NewkNFNAir}) and published data from Pettyjohn and Christiansen\cite{Pettyjohn1948}, Willmarth et al. \cite{Willmarth1964}, Christiansen and Barker \cite{Christiansen1965a}, Isaacs and Thodos \cite{Isaacs1967} and McKay et al. \cite{McKay1988} (see Table \ref{TabParticles}). }
  \vspace{-1em}
  \label{kNDensityEffect} 
\end{figure}

In Fig. \ref{kNDensityEffect}, $k_N$ for measurements made in both gases (i.e. present wind tunnel data, same as in Fig. \ref{NewkNFNAir}) and liquids (i.e. published, see figure caption and Table \ref{TabParticles}) is plotted against $F_N$. It is evident that at any given $F_N$, $k_N$ of particles with higher $\rho'$ is lower. Since particle shape is fixed, the only explanation is that by decreasing $\rho'$, particles tend to have higher projection areas perpendicular to the falling direction and hence their drag coefficient increases. By taking $\rho'$ into account and using non-linear regressions, a general correlation for obtaining $k_N$ based on $F_N$ and $\rho'$ can be found that is valid for freely falling particles at any $\rho'>1$:
 \begin{equation}
 \log{\left( k_N \right)}=\alpha_2 \, \left[-\log{ \left( F_N \right)} \right]^{\beta_2} \quad \mbox{for } \rho'>1 \label{eq:NewkNGeneral}
  \end{equation}
where $\alpha_2$ and $\beta_2$ are sigmoidal functions of $\rho'$
 \begin{equation}
 \mathcolorbox{yellow}{\alpha_2=0.45+\frac{10}{\exp{\left( 2.5 \, \log{\rho'}\right)+30 }}} 
 \label{eq:alpha2}
  \end{equation}
 \begin{equation}
 \mathcolorbox{yellow}{\beta_2=1-\frac{37}{\exp{\left( 3 \, \log{\rho'}\right)+100}} }
 \label{eq:beta2}
  \end{equation}

\nomenclature[Ga]{$\alpha_2$}{empirical expression defined in Eq. (\ref{eq:alpha2})}
\nomenclature[Gb]{$\beta_2$}{empirical expression defined in Eq. (\ref{eq:beta2})}

A summary of error analyses of selected correlations on estimating $k_N$ of all data points in the Newton's regime is presented in Table \ref{TabErrorNewtonAll}, which shows that Eqs.(\ref{eq:NewkNGeneral} -- \ref{eq:beta2}) are associated with a remarkable average error of 14.3\% (max. error of 51.2\%). Unfortunately, the model of H\"{o}lzer and Sommerfeld \cite{Holzer2008} could not be benchmarked here, since particle orientation was not known for data points from the literature.

\begin{table*}[]
  \centering
  \caption{Mean and maximum error associated with the estimations of the drag coefficient of non-spherical particles in the Newton's regime measured in various liquids (compiled  from the literature) and air (present study), see Table \ref{TabParticles}.}
    \vspace{-1em}
\begin{tabular}{l c c} \toprule
Correlation & \multicolumn{2}{c}{$error$\%} 
\\ \cmidrule{2-3}
& $mean$ & $max$ \\ \midrule
Haider and Levenspiel \cite{Haider1989}, Eq. (\ref{Hider})&
54.9&242\\
Ganser \cite{Ganser1993}, Eq. (\ref{GanserkN})&
53.9& 244\\
This work, Eq. (\ref{eq:NewkNGeneral})&	14.3&51.2
\\ \bottomrule
\vspace{-2em}
\label{TabErrorNewtonAll}
\end{tabular} 
\end{table*}

\begin{figure}[!b]
  \begin{center}
    \includegraphics[width=0.48\textwidth]{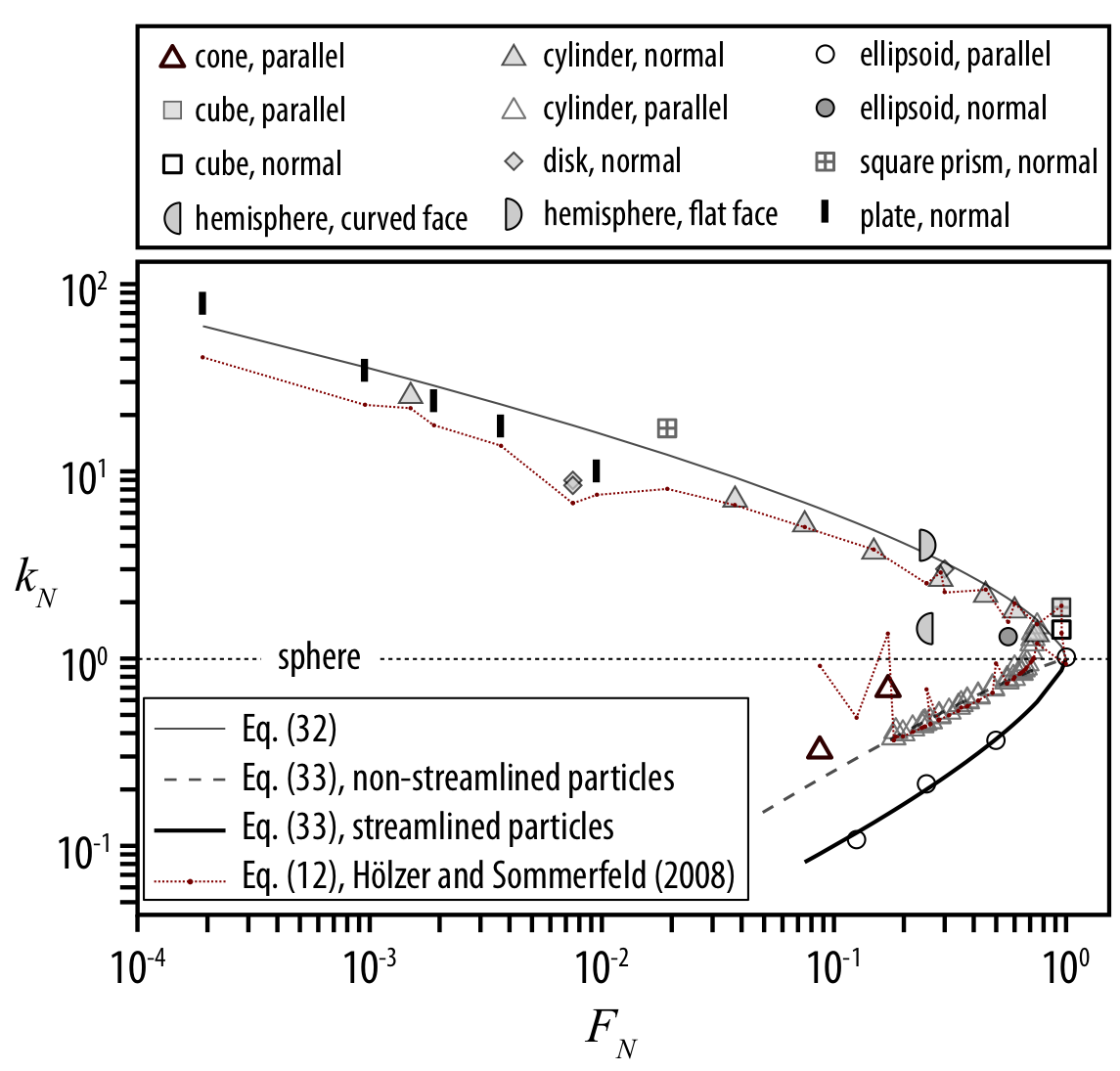}
  \end{center}
  \vspace{-2em}
  \caption{Newton's drag correction $k_N$ for various non-spherical particles measured experimentally in fixed orientations against $F_N$ from the data of Hoerner \citep{Hoerner1965}, White \citep{White1998} and Higuchi et al. \citep{Higuchi2008}.}
  \label{kNFixedOrientation} 
\end{figure}

Eqs.(\ref{eq:NewkNGeneral} -- \ref{eq:beta2}) take into account the effects of preferred orientations of particles on the drag coefficient through $\rho'$, however, not all possible orientations might happen when particles freely fall in a fluid. In fact, highly non-spherical particles (i.e. low $F_N$) might have very different drag coefficients in their extreme orientations. To explore this, the dependency of $k_N$ on $F_N$ for various non-spherical particles at fixed orientations is plotted in Fig. \ref{kNFixedOrientation} (from published data). Most particles are divided in two groups depending on their orientation relative to the flow: \emph{normal} (maximum projection area perpendicular to the flow direction) and \emph{parallel} (minimum projection area perpendicular to the flow direction). The remaining particles (half spheres) are described based on the orientation of the hemispheres with respect to the direction of the flow. In general, the drag coefficient is always higher than that of the volume-equivalent sphere when they are fixed normal to the flow ($k_N>1$) and is smaller than that of the sphere when they are fixed parallel to the flow ($k_N<1$). The exception is cylindrical particles fixed parallel to the flow, where $k_N$ is smaller than unity only when $F_N<0.7$. 

Fig. \ref{kNFixedOrientation} shows that the drag coefficient in a fixed orientation is dependent not only on the particle shape but also on the direction of the flow. As an example, for a hemisphere when the flow impinges the curved face, $k_N$ is 1.4, which largely increases to 4.0 when the flat face is in the front. In this case, while particle projection area normal and parallel to flow is constant (i.e. constant $\psi_{\perp}$), the drag coefficient can change up to 185\%. In addition, the differences between $k_N$ for streamlined (e.g. ellipsoid) and flat-nose particles (e.g. cylinder) fixed parallel to the flow are significantly different, while their values of $F_N$ and $\psi_{\perp}$ are very close together. Thus, no unique correlation as a function of $F_N$, $\psi$, $\psi_{\perp}$ or any other shape/orientation descriptor can be found for estimating $k_N$ for all shapes. At most, general correlations can be found that can roughly constrain the extremes of variation in $k_N$ at different orientations.

In Fig. \ref{kNFixedOrientation} a curve based on Eq. (\ref{eq:NewkNGeneral}) for $\rho' \rightarrow 1$ is shown (same as solid line in Fig. \ref{kNDensityEffect}), which, interestingly, is very close to $k_N$ for particles fixed normal to the flow. This indicates that at the limit of $\rho' \rightarrow 1$ particles fall with their maximum projection area perpendicular to their falling direction. Therefore, the maximum drag, $k_{N, \, max}$, occurs for particles oriented normal to the flow and can be estimated by inserting $\rho' \rightarrow 1$ in Eq. (\ref{eq:NewkNGeneral}):
 \begin{equation}
 \log{\left( k_{N, \, max} \right)}=0.77 \, \left[-\log{\left( F_N \right)} \right]^{0.63} \label{eq:kNmax}
  \end{equation}
For particles in the parallel orientation, for which the drag coefficient are the lowest at a given $F_N$, the simplest way for estimating $k_{N, \, min}$ is to define two separate correlations for non-streamline and streamline particles:
\begin{equation}
\begin{split}
\log{\left( k_{N, \, min} \right)}= \left\{ 
\begin{array}{l l}
  -0.6 \, \left[-\log{\left( F_N \right)} \right]^{1.17} & \\ \quad\quad \mbox{for non-streamline, } F_N > 0.1\\
  - \, \left[-\log{\left( F_N \right)} \right]^{0.48} & \\ \quad\quad \mbox{for streamline, } F_N > 0.1 \\ \end{array} \right.  \label{eq:kNmin}
  \end{split}
\end{equation}
It should be noted that Eq. (\ref{eq:kNmin}) results in $k_{N, \, min}<1$ for all particles, which is not the case for cylinders with $F_N>0.7$, but it is the only solution if we want to avoid complex correlations or using orientation-dependent parameters. In addition, Eq. (\ref{eq:kNmin}) is valid only at $F_N>0.1$ since no data at lower values of $F_N$ were available to check its validity. Eq. (\ref{eq:kNmin}) is associated with average error of 21\% (max. error of 152\%) for estimating drag coefficient end members of various particle shapes. Estimations of model of H\"{o}lzer and Sommerfeld \cite{Holzer2008} are also plotted in Fig. \ref{kNFixedOrientation}, which shows that this is an accurate model with an average error of 17\% (max. error of 184\%) for all particles except for parallel ellipsoids. The error associated with parallel ellipsoids is large (up to 347\%), since the model significantly overestimates the drag coefficient of parallel ellipsoids (the error is even higher than that associated with the estimation of the drag coefficient of parallel cylinders).

\subsection{The new general drag coefficient model}
\begin{figure*}[t]
  \begin{center}
    \includegraphics[width=0.84\textwidth]{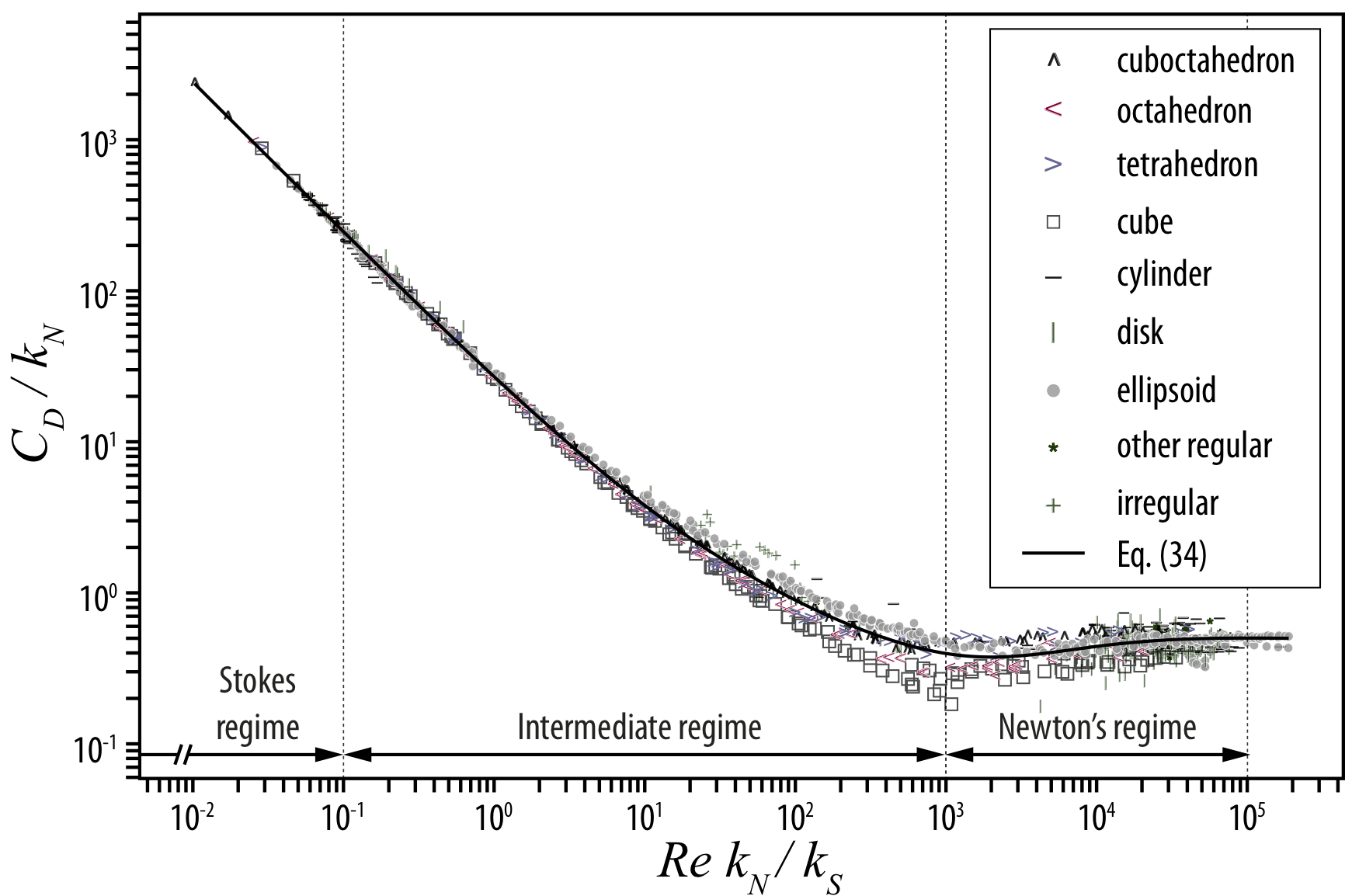}
  \end{center}
  \vspace{-1em}
  \caption[Dependency of normalized drag coefficient of freely falling particles on normalized Reynolds number.]{Dependency of normalized drag coefficient of freely falling particles on normalized Reynolds number. Data are from present study, Pettyjohn and Christiansen \citep{Pettyjohn1948}, Willmarth et al. \citep{Willmarth1964}, Christiansen and Barker \citep{Christiansen1965a},Isaacs and Thodos \citep{Isaacs1967}, Schlichting \citep{Schlichting1968}, Roos and Willmarth \citep{Roos1971}, Achenbach \citep{Achenbach1972}, Clift et al. \citep{Clift2005} and McKay et al. \citep{McKay1988}. (see Table \ref{TabParticles}).}
  \vspace{-1em}
  \label{AllCDStarReStar} 
\end{figure*}

\vspace{0em}
Based on the dimensional analysis of Ganser \cite{Ganser1993}, the drag coefficient of non-spherical particles for any subcritical Reynolds number can be predicted as a function of Stokes' $k_S$ and Newton's $k_N$ drag corrections. In particular, by normalizing the drag coefficient $C_D$ and particle Reynolds number $Re$ as $C_D/k_N$ and $Re \, k_N/k_S$ \citep{Ganser1993}, repectively, all data points obtained for freely falling particles show a similar trend as it is illustrated in Fig. \ref{AllCDStarReStar}. Finally, a general correlation for estimating the normalized drag coefficient based on normalized Reynolds number can be found that is valid for any particle shape:
 \begin{equation} 
 \begin{split}
\frac{C_D}{k_N}=\frac{\displaystyle 24 \, k_S}{\displaystyle Re \, k_N} \left( 1+0.125 \left( Re \, k_N / k_S \right) ^{2/3} \right)\quad\quad\quad\quad \\ \quad\quad\quad +\frac{\displaystyle 0.46}{\displaystyle 1+5330/{\displaystyle \left( Re \, k_N / k_S \right)}}
\end{split}
 \label{eq:GeneralModel}
  \end{equation}

It is important to note that the fitting constants in Eq. (\ref{eq:GeneralModel}) are different compared to those of Ganser \cite{Ganser1993}. In addition, $k_S$ and $k_N$ are based on different shape descriptors. Eq. (\ref{eq:GeneralModel}) is associated with an average error of 5.3\% for all data points (12.1\% if $10^4$ ellipsoids in the Stokes' regime are excluded). A detailed error analysis of Eq. (\ref{eq:GeneralModel}) along with comparison with other models is shown in Table \ref{TabErrorFinal}. It can be seen how Eq. (\ref{eq:GeneralModel}) has the lowest mean of relative error for estimating drag coefficient of non-spherical particles that is half of that for the models of Haider and Levenspiel \cite{Haider1989} and Ganser \cite{Ganser1993}. If, for the sake of simplicity, we approximate non-spherical particles to ellipsoids by neglecting the term $d_{eq}^3 / L\,I\,S$ for calculating $F_S$ and $F_N$ (i.e. $F_S =f\, e^{1.3}$, $F_N=f^2\,e$), the average error of Eq. (\ref{eq:GeneralModel}), for the particle considered in Table \ref{TabErrorFinal}, slightly increases to 10.7\% (maximum of 87.9\%).
 
 \begin{table*}[]
 \vspace{1em}
  \centering
  \caption[Mean and maximum error associated with the estimations of the drag coefficient of non-spherical particles.]{Mean and maximum error associated with the estimations of the drag coefficient of all non-spherical particles freely falling at $Re<3\times 10^5$, including data points compiled from the literature and those obtained in the present study (see Table \ref{TabParticles}). It should be noted that the error analysis presented here was performed only for 500 of the $10^4$ ellipsoids calculated in this study for the Stokes' regime in order to have a uniform distribution of data points at different $Re$.}
    \vspace{-1em}
\begin{tabular}{l c c} \toprule
Correlation & \multicolumn{2}{c}{$error\%$} 
\\ \cmidrule{2-3}
& $mean$ & $max$ \\ \midrule
Haider and Levenspiel \cite{Haider1989}, Eq. (\ref{Hider})&
19.4&244.0\\
Ganser \cite{Ganser1993}, Eqs. (\ref{GanserCd}), (\ref{GanserkS}) and (\ref{GanserkN})&
20.0& 247.6\\
This work, Eqs. (\ref{eq:NewkS}), (\ref{eq:NewkNGeneral}) and (\ref{eq:GeneralModel}) &	9.8&73.4
\\ \bottomrule
\label{TabErrorFinal}
\end{tabular} 
\end{table*}

If we take a closer look at Fig. \ref{AllCDStarReStar}, a scatter in the data at intermediate Reynolds numbers can be observed. \citet{Loth2008} suggested that this scatter in the intermediate regime is due to the effect of particle orientation, which for some specific shapes (e.g. sphere, broadside falling cylinder) results in circular cross sections in the direction of the flow while for other shapes (e.g. broadside falling disk and cubes) results in sharp cross sections. He argued that this scatter can be explained considering that the separation point of the boundary layer for particles with circular cross sections is dependent on $Re$, whereas for the others the separation point remains almost fixed after initiation at low $Re$. A solution for this problem is to find separate fits for estimating the drag coefficient of particles of circular and non-circular sections \citep{Loth2008}. However, here we have decided not to present any correlation other than Eq. (\ref{eq:GeneralModel}), since in any case the associated error is low (i.e. 5.9\%) and the gain in the added accuracy is not worth the extra complications.

 \begin{figure*}[]
  \begin{center}
    \includegraphics[width=0.95\textwidth]{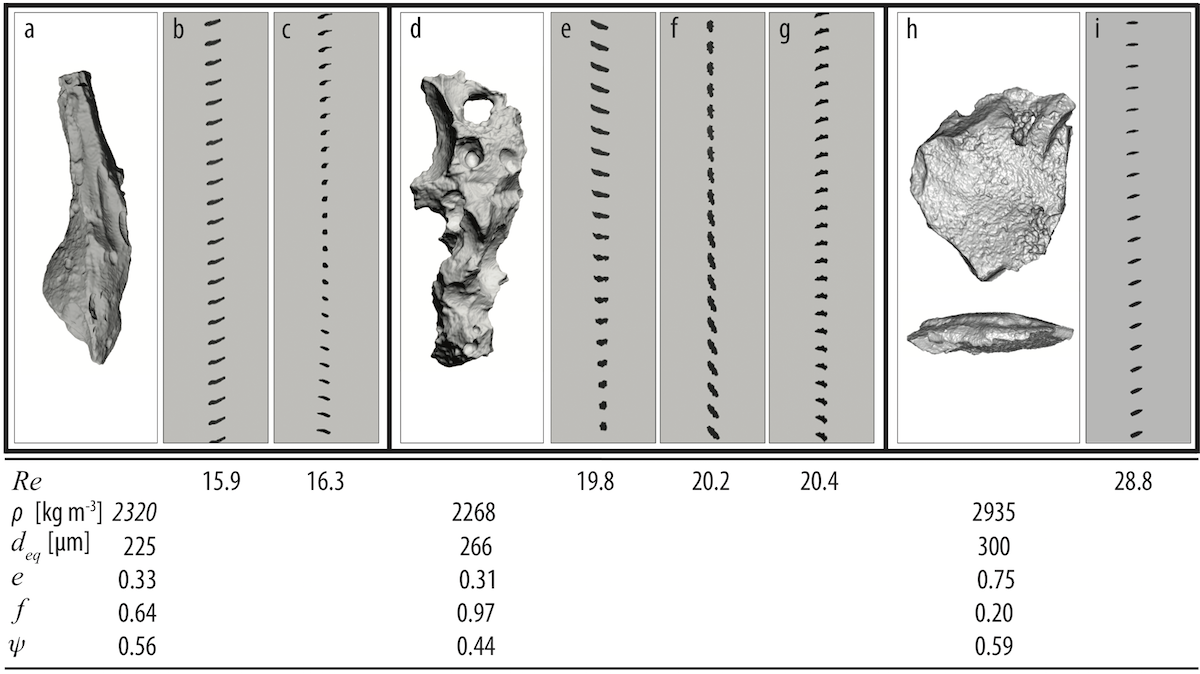}
  \end{center}
  \vspace{-1em}
  \caption[Falling pattern of irregular particles in settling columns. The 3D model of the falling particle is shown in the left side of high-speed image sequences.]{Falling pattern of irregular particles in settling columns. The 3D model of the falling particle is shown in the left side of high-speed image sequences. High-speed image sequences shown in b, e, f are from experiments carried out in the short settling column with the falling distance of $\approx 0.45 \, m$, image height of $ 15.6 \, mm$ and recording speed of $1600 fps$; and in c, g and i are from experiments performed in the intermediate settling column with falling distance of $\approx 1.13 \, m$, image height of $ 14.9 \, mm$ and recording speed of $2000 fps$.}
  \label{FallingPattern} 
\end{figure*}

\begin{figure}[!b]
  \begin{center}
    \includegraphics[width=0.35\textwidth]{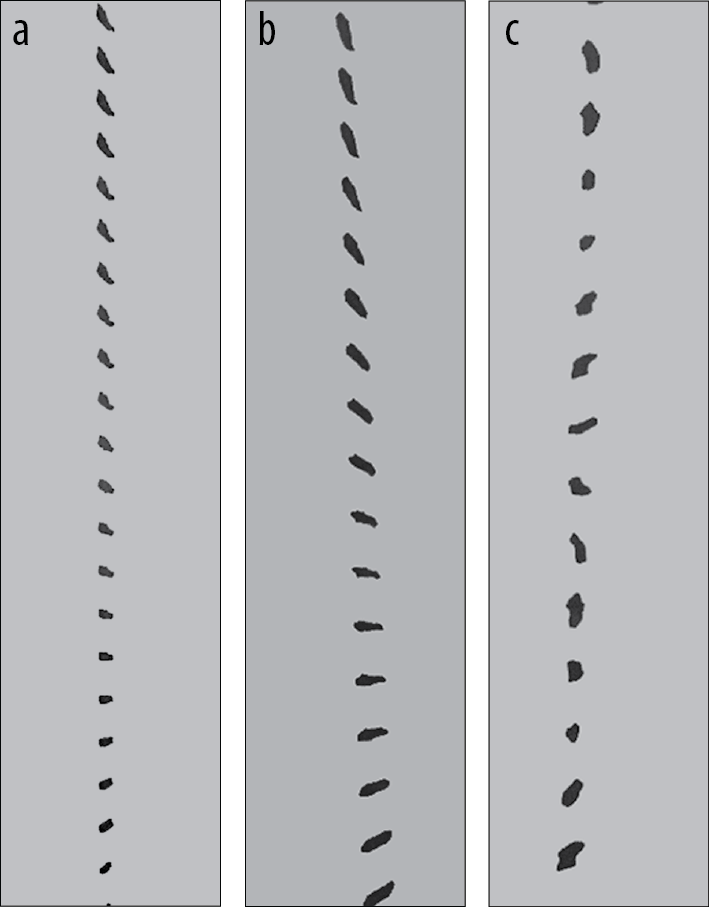}
  \end{center}
  \vspace{-1em}
  \caption[Irregular particles in settling columns falling at Reynolds number of (a) 120, (b) 190 and (c) 250.]{Irregular particles in settling columns falling at Reynolds number of (a) 120, (b) 190 and (c) 250. These experiments are conducted in the long settling column with falling distance of $\approx 3.6 \, m$. Time interval between each snapshot of the particle is $0.625 \, ms$ ($1600 fps$) and image height is $40.7 \, mm$.}
  \label{FallingPatternHighRe} 
\end{figure}
Results obtained in this study do not allow us to characterize secondary motion of irregular particles at intermediate range of Reynolds number systematically, simply because the field of view through our high-speed camera is too small ($ 24 \times 16 \, mm$). Nevertheless, given that in contrast to previous studies our experiments are conducted in air-filled settling columns, some interesting aspects can be analysed by inspecting the high-speed videos of falling particles. Fig. \ref{FallingPattern} shows that the falling orientation at $15 < Re<30$ for irregular particles, is either steady (Fig. \ref{FallingPattern}a), steady while rotating around a vertical axis (Fig. \ref{FallingPattern}b) or associated with some oscillations (Fig. \ref{FallingPattern}e -- i). Additionally, particles shown in Fig. \ref{FallingPattern}a -- c and \ref{FallingPattern}h -- i fall with orientations close to their maximum projection area normal to their falling paths, whereas the projected area of the particle in Fig. \ref{FallingPattern}d -- g is variable (oscillation frequency $\approx 8 \, Hz$). In any case, we did not observe any oscillation for particles with $Re<18$, which is much lower compared to $Re$ found for steady fall of cylinders (i.e. $Re<80-300$) and disks (i.e. $Re<100$) in liquids. This suggests that for irregular particles falling in quiescent gases when Brownian motion is not important and at $0.05<Re<18$, $k_{S, \, max}$ and $k_{N, \, max}$ can provide better estimations of the drag coefficient through Eq. (\ref{eq:GeneralModel}) than $k_S$ and $k_N$, given that particles fall with their maximum projected area normal to the flow.

At higher $Re$, the frequency of oscillation for irregular particles increases significantly and in some cases can lead to strong lateral deviations as it is shown in Fig. \ref{FallingPatternHighRe}. However, even at high $Re$ some particles have been observed to fall with a steady orientation and, hence, a general conclusion cannot be made.

\section{Caveats of the new model}
\begin{figure}[!htb]
  \begin{center}
    \includegraphics[width=0.48\textwidth]{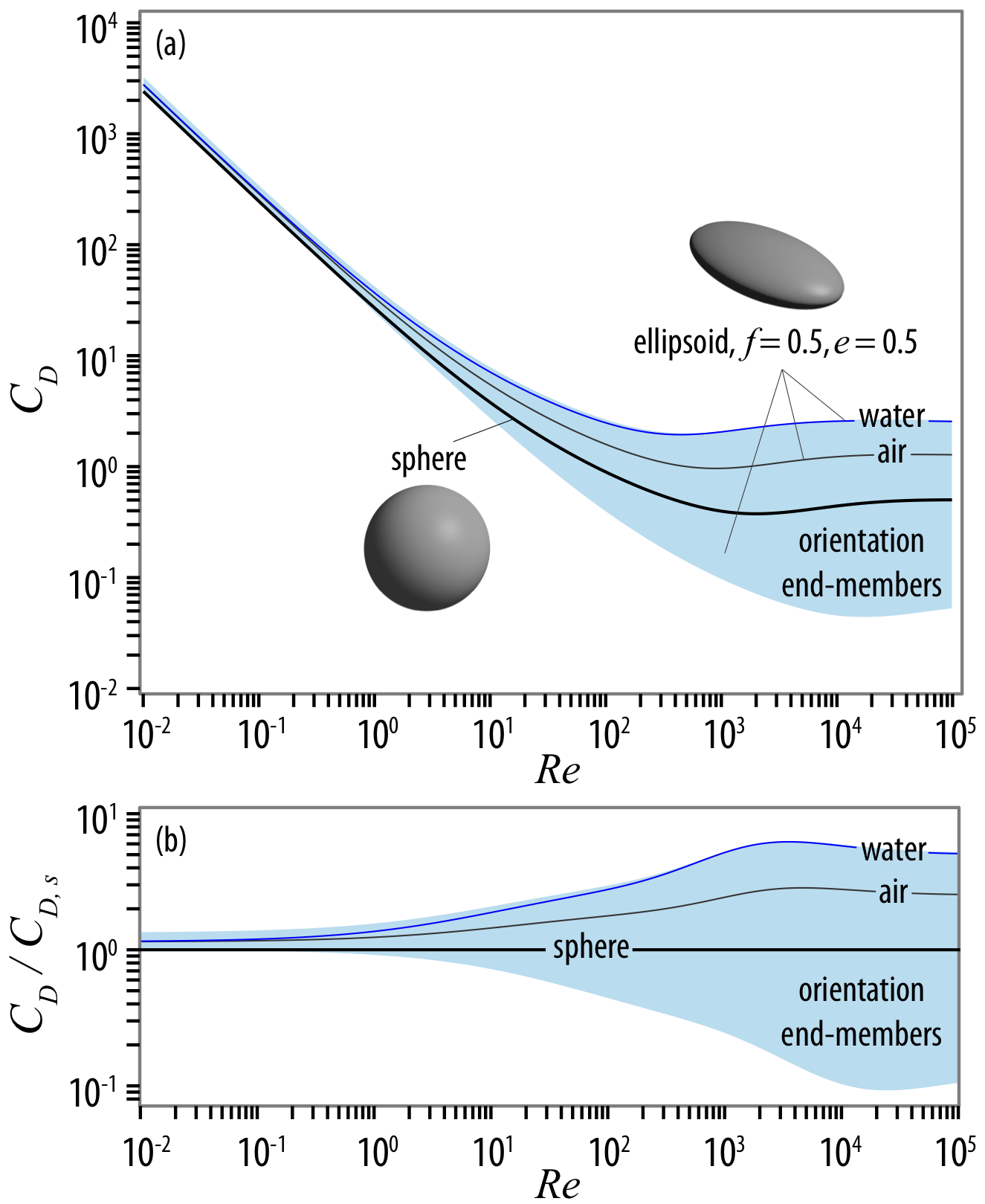}
  \end{center}
  \vspace{-2em}
  \caption[Effect of orientation on the drag coefficient of an ellipsoid with flatness and elongation of 0.5.]{Effect of orientation on the drag coefficient of an ellipsoid with flatness and elongation of 0.5 and density of $\SI{2000}{\kilogram\per\cubic\metre}$ estimated by Eq. \ref{eq:GeneralModel} using Eqs. (\ref{eq:NewkSOrient}, \ref{eq:kNmin} and \ref{eq:kNmax}). (a) Ellipsoid drag coefficient against Reynolds number; (b) same as (a) with the ellipsoid drag coefficient normalized by the sphere drag coefficient. For the sake of comparison the average drag coefficient for free fall in water and air is also shown. }
  \vspace{-1em}
  \label{Final_orientation} 
\end{figure}

Although a large number of data points in a wide range of $Re$ are used to derive the general model for particle drag coefficient and other correlations in this study, it is important to discuss the main assumptions and limitations of our approach. One of the crucial assumptions for obtaining the general drag coefficient model, Eq. (\ref{eq:GeneralModel}), is that the drag coefficient of a particle with a given shape, density ratio and orientation is solely a function of $Re$, $k_S$ and $k_N$. This can be questionable in some particular cases (e.g. the observed spread in the data in Fig. \ref{AllCDStarReStar} at intermediate $Re$). An example is shown in Fig. \ref{Final_orientation}, which plots the drag coefficient of an ellipsoid with $f=e=0.5$ and density of \SI{2000}{\kilogram\per\cubic\metre} falling in water and air predicted by Eq. (\ref{eq:GeneralModel}) using Eqs. (\ref{eq:NewkS}), (\ref{eq:NewkNGeneral}). It can be seen that effect of the density ratio starts to be noticeable at $Re>1$, while it was expected to be an influencing parameter at higher $Re$ (at least not before $Re$ of $18$, see section \ref{sssec:density_newton}).

\begin{figure}[!htb]
  \begin{center}
    \includegraphics[width=0.46\textwidth]{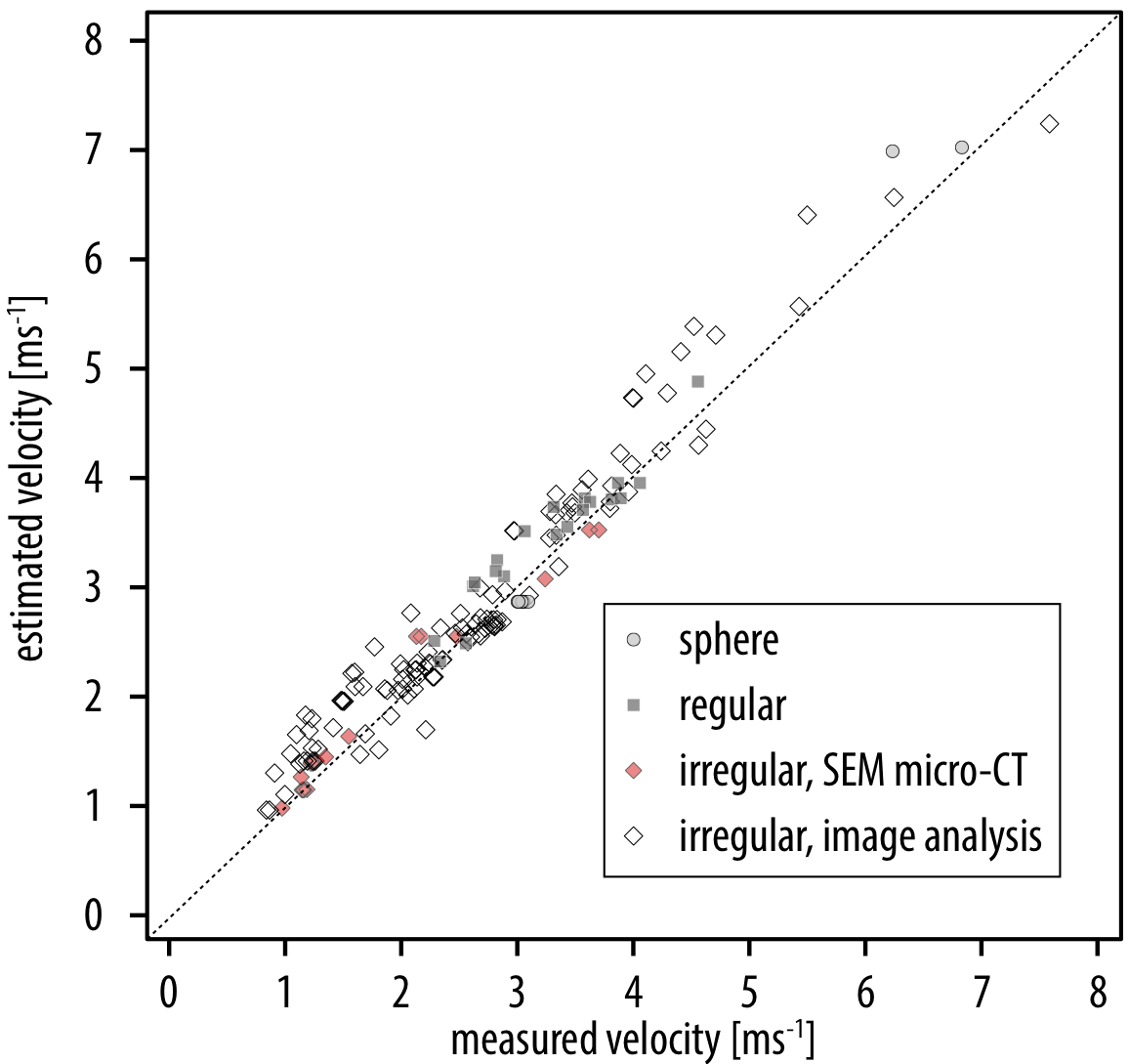}
  \end{center}
  \vspace{-2em}
  \caption[Falling velocity of particles measured in settling columns ($9 \leq Re \leq 300$) against velocity estimated through the general drag coefficient model.]{Falling velocity of particles measured in settling columns ($9 \leq Re \leq 300$) against velocity estimated through the general drag coefficient model, Eq. (\ref{eq:GeneralModel}). Regular particles include cylinders and prisms (see Table \ref{TabParticles}). Characteristics of irregular particles (e.g. volume, form dimensions) are quantified either by SEM micro-CT or by performing image analyses on 2 -- 3 projections of the particle.}
      \vspace{-1em}
  \label{FallingVelocity} 
\end{figure}

This premature influence of density ratio at low $Re$ can lead to an artificial underestimation of the drag coefficient. In order to check this issue, falling velocities of particles measured in settling columns ($9<Re<900$) are compared to those predicted by Eq. (\ref{eq:GeneralModel}). As it is shown in Fig. \ref{FallingVelocity}, the terminal velocity of particles in this range of $Re$ seems to be slightly overestimated by Eq. (\ref{eq:GeneralModel}). However, the average error for all particles is 12.5\% and it is even lower for irregular particles that are better characterized by SEM micro-CT and regular particles (i.e. 7.5\%). So, we can conclude that the overestimation of terminal velocity (i.e. the underestimation of the drag coefficient) does not affect the overall estimation error of Eq. (\ref{eq:GeneralModel}). Finally, given that all correlations derived in this work are empirical, it is important to apply them within the range of their validity.

\section{Discussion and conclusions}
The drag coefficient of non-spherical particles of regular and irregular shapes at subcritical $Re$ ($Re<3 \times 10^5$) was investigated through analytical and experimental methods. Effects of particle shape, surface roughness, orientation and particle-to-fluid density ratio $\rho'$ on the drag coefficient were discussed in detail. Two new shape descriptors, namely Stokes $F_S$ and Newton shape descriptor $F_N$, were introduced that are based on particle flatness $f$, elongation $e$ and spherical equivalent diameter $d_{eq}$. Compared to the sphericity that is the most used shape descriptor in the literature, the new shape descriptors are significantly easier to measure, are not a function of measurement scale and are better correlated with the drag coefficient. Based on our results the following conclusions can be drawn:
\begin{table*}[!h]
  \centering
  \caption[The general correlation for estimating the average drag coefficient, $C_D$, of freely falling solid non-spherical particles in liquids or gases.]{The general correlation for estimating the average drag coefficient, $C_D$, of freely falling solid non-spherical particles in liquids or gases. $f$ and $e$ are particle flatness and elongation defined as the ratio of $S/I$ and $I/L$, respectively; where $L$, $I$ and $S$ are particle form dimensions and defined as the longest, intermediate and shortest lengths of the particle, receptively. $d_{eq}$ is the diameter of a volume-equivalent sphere, $Re$ is the particle Reynolds number defined in Eq. (\ref{eq:Re}) and $\rho'$ is the particle-to-fluid density ratio. By neglecting the term $d_{eq}^3 / L\,I\,S$ for calculating $F_S$ and $F_N$, shape of non-spherical particles will be approximated to ellipsoids of similar flatness and elongation.}
    \vspace{-1em}
\begin{tabular}{l} \toprule
$C_D=\frac{\displaystyle 24 \, k_S}{\displaystyle Re} \left( 1+0.125 \left( Re \, k_N / k_S \right) ^{2/3} \right) + \frac{\displaystyle 0.46 \, k_N}{\displaystyle 1+5330/{\displaystyle \left( Re \, k_N / k_S \right)}} $ \\
where\\
$k_S=\left(F_S^{1/3}+{F_S^{-1/3}} \right)/2$\\
$\mathcolorbox{yellow}{k_N =10^{\displaystyle \, \alpha_2 \, \left[-\log{ \left( F_N \right)} \right]^{\displaystyle \beta_2}}}$\\
$\mathcolorbox{yellow}{\alpha_2=0.45+10/{\left(\exp{\left( 2.5 \, \log{\rho'}\right)+30 }\right)}}$\\
 $\mathcolorbox{yellow}{\beta_2=1-37/{\left(\exp{\left( 3 \, \log{\rho'}\right)+100}\right)}}$
 \\
and\\
$F_S=f \, e^{1.3} \, \left( \frac{d_{eq}^3}{L \, I \, S}\right)$, or simpler but less accurate: $F_S=f \, e^{1.3}$\\
$F_N =f^2 \, e \, \left( \frac{d_{eq}^3}{L \, I \, S}\right)$, or simpler but less accurate: $F_N =f^2 \, e$
\\ \bottomrule
    \vspace{-1em}
\label{TabSummaryGenModel}
\end{tabular} 
\end{table*}

\begin{itemize}
\item A new general drag coefficient correlation is presented, Eq. \ref{eq:GeneralModel} that is summerized in Table \ref{TabSummaryGenModel} and Fig. \ref{AllCDStarReStar}. This correlation can be used to predict the average drag coefficient of particles falling in fluids (gases and liquids). The main assumption is that the particle orientation in the Stokes' regime is random and in the Newton regime is a function of particle-to-fluid density ratio $\rho'$. The average error of the new general drag coefficient correlation for predicting the drag coefficient of non-spherical particles presented in this study and literature  is $\sim 10\%$ (Table \ref{TabErrorFinal}).

\item  If, we approximate non-spherical particles to ellipsoids by neglecting the term $d_{eq}^3 / L\,I\,S$ for calculating $F_S$ and $F_N$ (i.e. $F_S =f\, e^{1.3}$, $F_N=f^2\,e$), the average error of Eq. (\ref{eq:GeneralModel}), for the particle considered in Table \ref{TabErrorFinal}, slightly increases to $~11\%$.

\item Effect of particle orientation on the drag coefficient is significant, in particular at high $Re$ (Figs. \ref{NewkSOrient} and \ref{kNFixedOrientation}). By using Eqs. (\ref{eq:NewkSOrient}), (\ref{eq:kNmax}) and (\ref{eq:kNmin}) within Eq. (\ref{eq:GeneralModel}) end-members of the particle drag coefficient due to change in the orientation can be found (Fig. \ref{Final_orientation}a). These end-members at high $Re$, however, are valid for specific orientations of particles that might occur rarely as the particle falls.  
\item Out of all parameters describing particle shape, it is the particle form that has the greatest impact on the drag coefficient as opposed to the surface-related characteristics, such as sphericity.
\item In the Stokes' regime ($Re<0.1$), the drag coefficient is slightly more sensitive to changes in the elongation than in the flatness, i.e. $F_S \propto f \, e^{1.3}$, whereas in the Newton's regime ($1000 \leq Re \leq 3 \times 10^5$), the impact of the flatness on the drag coefficient is much more significant than the impact of the elongation, i.e. $F_N \propto f^2 \, e$.
\item The average drag coefficient of a non-spherical particle falling in a gas or a liquid is always higher than the drag coefficient of its volume-equivalent spheres, i.e. $k_S, \, k_N >1$ (Figs. \ref{NewFS} and \ref{kNDensityEffect}). However, in some specific orientations, the particle drag can be even lower than its volume-equivalent sphere (e.g. minimum projected area normal to the motion path) (Fig. \ref{Final_orientation}). 
\item The impact of both shape and orientation on the drag coefficient of non-spherical particles increases with $Re$ (Fig. \ref{Final_orientation}b).
\item Effects of surface vesicularity and roughness on the drag coefficient of freely falling non-spherical particles was found to be $<25\%$ at $Re\ll1$ (or $\ll 25\%$ for fine-scale surface roughness, see Fig. \ref{kSHillsMethod}) and $<10\%$ at $7.9\times10^3<Re<4.5\times10^4$ (see Fig. \ref{RoughvsParafilm}). 
\item In the Newton's regime ($1000 \leq Re<3 \times 10^5$), particle secondary motions and orientation are functions of the particle-to-fluid density ratio $\rho'$ (Fig. \ref{kNDensityEffect}). Particles falling in liquids (low $\rho'$) have orientations close to their maximum projected area  normal to their falling path, while those falling in gases (high $\rho'$) have random orientations and projection areas lower than their maximum. As a result, a solid particle of a given shape will experience higher drag when it falls in a liquid compared to when it falls in a gas. 
\end{itemize}

\section*{Acknowledgement}
This project was funded by the Swiss National Science Foundation (SNSF, Grant No. 200020-125024). Authors are grateful to I. Manzella for insightful discussions on the wind tunnel set-up and particle shape characterization, P. Pontelandolfo and P. Haas for their help and constructive discussions on the wind tunnel set-up and calibration, L. Dominguez for her help on particle image analysis, P. Vonlanthen for his support on particle SEM micro-CT, J. Phillips for brainstorming ideas at the design stage of the wind tunnel and F. Arlaud for his help to design and construct the settling columns.

\bibliography{final_clean}

\begin{thebibliography}{53}
\providecommand{\natexlab}[1]{#1}
\providecommand{\url}[1]{\texttt{#1}}
\expandafter\ifx\csname urlstyle\endcsname\relax
  \providecommand{\doi}[1]{doi: #1}\else
  \providecommand{\doi}{doi: \begingroup \urlstyle{rm}\Url}\fi

\bibitem[Achenbach(1972)]{Achenbach1972}
Elmar Achenbach.
\newblock {Experiments on the flow past spheres at very high Reynolds numbers}.
\newblock \emph{Journal of Fluid Mechanics}, 54\penalty0 (03):\penalty0
  565--575, March 1972.

\bibitem[Achenbach(1974)]{Achenbach2006}
Elmar Achenbach.
\newblock {The effects of surface roughness and tunnel blockage on the flow
  past spheres}.
\newblock \emph{Journal of Fluid Mechanics}, 65\penalty0 (01):\penalty0 113,
  March 1974.

\bibitem[Albertson(1953)]{Albertson1953}
Maurice~L. Albertson.
\newblock {Effect of Shape on the Fall Velocity of Gravel Particles}.
\newblock In John~S. McNown and M.~C. Boyer, editors, \emph{Proceedings of the
  Fifth Hydraulics Conference}, page 308, Iowa City, 1953. State University of
  Iowa.

\bibitem[Alfano et~al.(2011)Alfano, Bonadonna, Delmelle, and
  Costantini]{Alfano2011c}
Fabrizio Alfano, Costanza Bonadonna, Pierre Delmelle, and Licia Costantini.
\newblock {Insights on tephra settling velocity from morphological
  observations}.
\newblock \emph{Journal of Volcanology and Geothermal Research}, 208\penalty0
  (3-4):\penalty0 86--98, December 2011.

\bibitem[Baba and Komar(1981)]{Baba1981}
Jumpei Baba and P.D. Komar.
\newblock {Measurements and analysis of setting velocities of natural quartz
  sand grains}.
\newblock \emph{Journal of Sedimentary Research}, 51\penalty0 (2):\penalty0
  631, 1981.

\bibitem[Bagheri et~al.(2013)Bagheri, Bonadonna, Manzella, Pontelandolfo, and
  Haas]{Bagheri2013a}
G.~H. Bagheri, C.~Bonadonna, I.~Manzella, P.~Pontelandolfo, and P.~Haas.
\newblock {Dedicated vertical wind tunnel for the study of sedimentation of
  non-spherical particles}.
\newblock \emph{Review of Scientific Instruments}, 84\penalty0 (5):\penalty0
  054501, 2013.

\bibitem[Bagheri et~al.(2015)Bagheri, Bonadonna, Manzella, and
  Vonlanthen]{Bagheri2014}
G.H. Bagheri, C.~Bonadonna, I.~Manzella, and P.~Vonlanthen.
\newblock {On the characterization of size and shape of irregular particles}.
\newblock \emph{Powder Technology}, 270:\penalty0 141--153, January 2015.

\bibitem[Blott and Pye(2007)]{Blott2007}
SIMON~J. Blott and KENNETH Pye.
\newblock {Particle shape: a review and new methods of characterization and
  classification}.
\newblock \emph{Sedimentology}, 55:\penalty0 31--63, September 2007.

\bibitem[Brosse and Ern(2013)]{Brosse2013}
Nicolas Brosse and Patricia Ern.
\newblock {The motion of an axisymmetric body falling in a tube at moderate
  Reynolds numbers}.
\newblock \emph{Journal of Fluid Mechanics}, 714:\penalty0 238--257, January
  2013.

\bibitem[Cheng(1997)]{Cheng1997}
N.S. Cheng.
\newblock {Simplified settling velocity formula for sediment particle}.
\newblock \emph{Journal of hydraulic engineering}, 123\penalty0
  (February):\penalty0 149, 1997.

\bibitem[Chhabra et~al.(1999)Chhabra, Agarwal, and Sinha]{Chhabra1999a}
R.P. Chhabra, L.~Agarwal, and N.K. Sinha.
\newblock {Drag on non-spherical particles: an evaluation of available
  methods}.
\newblock \emph{Powder Technology}, 101\penalty0 (3):\penalty0 288--295, March
  1999.

\bibitem[Chow and Adams(2011)]{Chow2011}
Aaron~C. Chow and E~ERIC Adams.
\newblock {Prediction of Drag Coefficient and Secondary Motion of Free-Falling
  Rigid Cylindrical Particles with and without Curvature at Moderate Reynolds
  Number}.
\newblock \emph{Journal of Hydraulic Engineering}, 137\penalty0 (11):\penalty0
  1406--1414, November 2011.

\bibitem[Christiansen and Barker(1965)]{Christiansen1965a}
EB~B. Christiansen and Dee~H. Barker.
\newblock {The effect of shape and density on the free settling of particles at
  high Reynolds numbers}.
\newblock \emph{AIChE Journal}, 11\penalty0 (1):\penalty0 145--151, January
  1965.

\bibitem[Clift and Gauvin(1971)]{Clift1971}
R.~Clift and W.~H. Gauvin.
\newblock {Motion of entrained particles in gas streams}.
\newblock \emph{The Canadian Journal of Chemical Engineering}, 49\penalty0
  (4):\penalty0 439--448, August 1971.

\bibitem[Clift et~al.(2005)Clift, Grace, and Weber]{Clift2005}
R.~Clift, J.~R. Grace, and M.~E. Weber.
\newblock \emph{{Bubbles, Drops, and Particles}}.
\newblock Dover Publications, Mineola, New York, 2005.
\newblock ISBN 0486445801.

\bibitem[Corey(1963)]{Corey1963}
Arthur~Thomas Corey.
\newblock \emph{{Influence of shape on the fall velocity of sand grains}}.
\newblock Audio Visual Service, Colorado State University., 1963.

\bibitem[Cox(1965)]{Cox1965}
R.~G. Cox.
\newblock {The steady motion of a particle of arbitrary shape at small Reynolds
  numbers}.
\newblock \emph{Journal of Fluid Mechanics}, 23\penalty0 (04):\penalty0
  625--643, 1965.

\bibitem[Dellino et~al.(2005)Dellino, Mele, Bonasia, Braia, {La Volpe}, and
  Sulpizio]{Dellino2005}
Pierfrancesco Dellino, Daniela Mele, Rosanna Bonasia, Giuseppe Braia, Luigi {La
  Volpe}, and Roberto Sulpizio.
\newblock {The analysis of the influence of pumice shape on its terminal
  velocity}.
\newblock \emph{Geophysical Research Letters}, 32\penalty0 (21):\penalty0
  L21306, 2005.

\bibitem[Ganser(1993)]{Ganser1993}
G.H. Ganser.
\newblock {A rational approach to drag prediction of spherical and nonspherical
  particles}.
\newblock \emph{Powder Technology}, 77\penalty0 (2):\penalty0 143--152, 1993.

\bibitem[G\"{o}g\"{u}s et~al.(2001)G\"{o}g\"{u}s, İpek\c{c}i̇, and
  K\"{o}kpinar]{Gogus2001}
M.~G\"{o}g\"{u}s, ON~İpek\c{c}i̇, and MA~K\"{o}kpinar.
\newblock {Effect of particle shape on fall velocity of angular particles}.
\newblock \emph{Journal of Hydraulic Engineering}, 127\penalty0 (10):\penalty0
  860, 2001.

\bibitem[Haider and Levenspiel(1989)]{Haider1989}
A.~Haider and O.~Levenspiel.
\newblock {Drag coefficient and terminal velocity of spherical and nonspherical
  particles}.
\newblock \emph{Powder Technology}, 58\penalty0 (1):\penalty0 63--70, May 1989.

\bibitem[Happel and Brenner(1983)]{Happel1983a}
John Happel and Howard Brenner.
\newblock \emph{{Low Reynolds number hydrodynamics: with special applications
  to particulate media}}, volume~1.
\newblock Springer Science \& Business Media, December 1983.
\newblock ISBN 9024728770.

\bibitem[Higuchi et~al.(2008)Higuchi, Sawada, and Kato]{Higuchi2008}
Hiroshi Higuchi, Hideo Sawada, and Hiroyuki Kato.
\newblock {Sting-free measurements on a magnetically supported right circular
  cylinder aligned with the free stream}.
\newblock \emph{Journal of Fluid Mechanics}, 596:\penalty0 49--72, January
  2008.

\bibitem[Hill and Power(1956)]{Hill1956}
R~Hill and G~Power.
\newblock {Extremum Principles For Slow Viscous Flow And The Approximate
  Calculation Of Drag}.
\newblock \emph{The Quarterly Journal of Mechanics and Applied Mathematics},
  9\penalty0 (3):\penalty0 313--319, 1956.

\bibitem[Hoerner(1965)]{Hoerner1965}
SF~Hoerner.
\newblock \emph{{Fluid-dynamic drag: practical information on aerodynamic drag
  and hydrodynamic resistance}}.
\newblock Hoerner Fluid Dynamics Midland Park, NJ, 1965.

\bibitem[H\"{o}lzer and Sommerfeld(2008)]{Holzer2008}
Andreas H\"{o}lzer and Martin Sommerfeld.
\newblock {New simple correlation formula for the drag coefficient of
  non-spherical particles}.
\newblock \emph{Powder Technology}, 184\penalty0 (3):\penalty0 361--365, June
  2008.

\bibitem[Isaacs and Thodos(1967)]{Isaacs1967}
Jack~L. Isaacs and George Thodos.
\newblock {The free-settling of solid cylindrical particles in the turbulent
  regime}.
\newblock \emph{The Canadian Journal of Chemical Engineering}, 45\penalty0
  (3):\penalty0 150--155, June 1967.

\bibitem[Jayaweera and Mason(1965)]{Jayaweera1965}
K.~O. L.~F. Jayaweera and B.~J. Mason.
\newblock {The behaviour of freely falling cylinders and cones in a viscous
  fluid}.
\newblock \emph{Journal of Fluid Mechanics}, 22\penalty0 (04):\penalty0 709,
  March 1965.

\bibitem[Komar and Reimers(1978)]{Komar1978}
P.D. Komar and CE~Reimers.
\newblock {Grain shape effects on settling rates}.
\newblock \emph{The Journal of Geology}, 86\penalty0 (2):\penalty0 193--209,
  1978.

\bibitem[Leith(1987)]{Leith1987}
David Leith.
\newblock {Drag on Nonspherical Objects}.
\newblock \emph{Aerosol Science and Technology}, 6\penalty0 (2):\penalty0
  153--161, 1987.

\bibitem[List and Schemenauer(1971)]{List1971}
Roland List and Robert~S. Schemenauer.
\newblock {Free-Fall Behavior of Planar Snow Crystals, Conical Graupel and
  Small Hail}.
\newblock \emph{Journal of the Atmospheric Sciences}, 28\penalty0 (1):\penalty0
  110--115, January 1971.

\bibitem[Loth(2008)]{Loth2008}
E.~Loth.
\newblock {Drag of non-spherical solid particles of regular and irregular
  shape}.
\newblock \emph{Powder Technology}, 182\penalty0 (3):\penalty0 342--353, March
  2008.

\bibitem[Mand\o and Rosendahl(2010)]{Mando2010}
Matthias Mand\o and Lasse Rosendahl.
\newblock {On the motion of non-spherical particles at high Reynolds number}.
\newblock \emph{Powder Technology}, 202\penalty0 (1-3):\penalty0 1--13, August
  2010.

\bibitem[Marchildon et~al.(1964)Marchildon, Clamen, and Gauvin]{Marchildon1964}
E.~K. Marchildon, A.~Clamen, and W.~H. Gauvin.
\newblock {Drag and oscillatory motion of freely falling cylindrical
  particles}.
\newblock \emph{The Canadian Journal of Chemical Engineering}, 42\penalty0
  (4):\penalty0 178--182, August 1964.

\bibitem[Marchildon and Gauvin(1979)]{Marchildon1979}
EK~Marchildon and WH~Gauvin.
\newblock {Effects of acceleration, deceleration and particle shape on
  single-particle drag coefficients in still air}.
\newblock \emph{AIChE Journal}, 25\penalty0 (6):\penalty0 938--948, 1979.

\bibitem[McKay et~al.(1988)McKay, Murphy, and Hillis]{McKay1988}
G.~McKay, R.~W. Murphy, and M.~Hillis.
\newblock {Settling characteristics of discs and cylinders}.
\newblock \emph{Chemical Engineering Research and Design}, 16\penalty0
  (1):\penalty0 107--112, 1988.

\bibitem[McNown and Malaika(1950)]{McNown1950}
J.S. McNown and J.~Malaika.
\newblock {Effects of particle shape on settling velocity at low Reynolds
  numbers}.
\newblock \emph{Trans. Am. Geophys. Union}, 31:\penalty0 74--82, 1950.

\bibitem[Nakamura and Tomonari(1982)]{Nakamura2006}
Y.~Nakamura and Y.~Tomonari.
\newblock {The effects of surface roughness on the flow past circular cylinders
  at high Reynolds numbers}.
\newblock \emph{Journal of Fluid Mechanics}, 123:\penalty0 363--378, April
  1982.

\bibitem[Oberbeck(1876)]{Oberbeck1876}
Anton Oberbeck.
\newblock {Ueber station\"{a}re Fl\"{u}ssigkeitsbewegungen mit
  Ber\"{u}cksichtigung der inneren Reibung}.
\newblock \emph{Journal f\"{u}r die reine und angewandte Mathematik},
  81:\penalty0 62--80, 1876.

\bibitem[Pettyjohn and Christiansen(1948)]{Pettyjohn1948}
ES~Pettyjohn and EB~Christiansen.
\newblock {Effect Of Particle Shape On Free- Settling Rates Of Isometric
  Particles}.
\newblock \emph{Chemical Engineering Progress}, 44\penalty0 (2):\penalty0
  157--172, 1948.

\bibitem[Roos and Willmarth(1971)]{Roos1971}
F.~W. Roos and W.~W. Willmarth.
\newblock {Some experimental results on sphere and disk drag}.
\newblock \emph{AIAA Journal}, 9\penalty0 (2):\penalty0 285--291, February
  1971.

\bibitem[Schlighting(1968)]{Schlichting1968}
H~Schlighting.
\newblock \emph{{Boundary-Layer Theory}}, volume 539.
\newblock McGraw-Hill New York, 6th editio edition, 1968.

\bibitem[Schneider et~al.(2012)Schneider, Rasband, and Eliceiri]{Schneider2012}
Caroline~A Schneider, Wayne~S Rasband, and Kevin~W Eliceiri.
\newblock {NIH Image to ImageJ: 25 years of image analysis}.
\newblock \emph{Nat Meth}, 9\penalty0 (7):\penalty0 671--675, July 2012.

\bibitem[Sneed and Folk(1958)]{Sneed1958}
E.D. Sneed and R.L. Folk.
\newblock {Pebbles in the lower Colorado River, Texas a study in particle
  morphogenesis}.
\newblock \emph{The Journal of Geology}, 66\penalty0 (2):\penalty0 114--150,
  1958.

\bibitem[Stokes(1851)]{GeorgeGabriel1851}
George~Gabriel Stokes.
\newblock \emph{{On the Effect of the Internal Friction of Fluids on the Motion
  of Pendulums}}, volume~9.
\newblock 1851.

\bibitem[Stringham et~al.(1969)Stringham, Simons, and Guy]{Stringham1969a}
GE~E Stringham, D.B.~B Simons, and H.P.~P Guy.
\newblock {The behavior of large particles falling in quiescent liquids}.
\newblock \emph{GEOL SURV PROF PAP 562-C, PP C 1-C 36, 1969. 36 P, 27 FIG, 7
  TAB, 23 REF.}, pages 1--36, 1969.

\bibitem[Tran-Cong et~al.(2004)Tran-Cong, Gay, and Michaelides]{Tran-Cong2004}
Sabine Tran-Cong, Michael Gay, and Efstathios~E. Michaelides.
\newblock {Drag coefficients of irregularly shaped particles}.
\newblock \emph{Powder Technology}, 139\penalty0 (1):\penalty0 21--32, January
  2004.

\bibitem[Vonlanthen et~al.(2015)Vonlanthen, Rausch, Ketcham, Putlitz,
  Baumgartner, and
  Grob\'{e}ty]{P.VonlanthenJ.RauschR.A.KetchamB.PutlitzL.P.Baumgartner}
Pierre Vonlanthen, Juanita Rausch, Richard~A. Ketcham, Benita Putlitz, Lukas~P.
  Baumgartner, and Bernard Grob\'{e}ty.
\newblock {High-resolution 3D analyses of the shape and internal constituents
  of small volcanic ash particles: The contribution of SEM micro-computed
  tomography (SEM micro-CT)}.
\newblock \emph{Journal of Volcanology and Geothermal Research}, 293:\penalty0
  1--12, February 2015.

\bibitem[Wadell(1933)]{Wadell1933}
H.~Wadell.
\newblock {Sphericity and roundness of rock particles}.
\newblock \emph{The Journal of Geology}, 41\penalty0 (3):\penalty0 310--331,
  1933.

\bibitem[White(1998)]{White1998}
Frank~M. White.
\newblock \emph{{Fluid Mechanics}}.
\newblock Mcgraw-Hill College, 1998.
\newblock ISBN 0072281928.

\bibitem[Wieselsberger(1922)]{Wieselsberger1922}
C~Wieselsberger.
\newblock {Further information on the laws of fluid resistance}.
\newblock \emph{Physikalische Zeitschrift}, 23:\penalty0 219--244, 1922.

\bibitem[Willmarth et~al.(1964)Willmarth, Hawk, and Harvey]{Willmarth1964}
William~W. Willmarth, Norman~E. Hawk, and Robert~L. Harvey.
\newblock {Steady and Unsteady Motions and Wakes of Freely Falling Disks}.
\newblock \emph{Physics of Fluids}, 7\penalty0 (2):\penalty0 197, 1964.

\bibitem[Wilson and Huang(1979)]{Wilson1979}
L~Wilson and T.C. Huang.
\newblock {The influence of shape on the atmospheric settling velocity of
  volcanic ash particles}.
\newblock \emph{Earth and Planetary Science Letters}, 44\penalty0 (2):\penalty0
  311--324, August 1979.

\end{thebibliography}

\renewcommand{\theequation}{A.\arabic{equation}}    
  \setcounter{equation}{0}  
  \section*{Appendix A: Image analysis and density measurements}  
Given that the size of particles tested in the settling columns is very small ($\SI{0.15}{\milli\metre}<d_{eq}<\SI{1.8}{\milli\metre}$), the most accurate method for characterizing size and shape was to reconstruct 3D models of particles by SEM micro-CT \citep{P.VonlanthenJ.RauschR.A.KetchamB.PutlitzL.P.Baumgartner}. However, this is a time consuming method and could not be used for all of irregular particles. Therefore, 3D models of only 12 irregular particles were obtained by using a SEM micro-CT (with a resolution in the order of 1--3 \SI{}{\micro\metre}) and the remaining particles were characterized by image analysis. For each particle without SEM micro-CT data, two to three images (i.e. projections) in different particle orientations (including minimum and maximum projection area) were obtained manually with a binocular microscope. The particle orientation under the microscope was changed and a sticky paper was used to keep the particle in the desired orientation. Using two or three projections for size and shape characterization of irregular particles based on image analysis was proven to be the best compromise between the accuracy and number of considered projections \citep{Bagheri2014}.  Particle projections were analyzed by ImageJ software \citep{Schneider2012} to obtain particle form dimensions (i.e. $L$, $I$, $S$) and projection area $A_p$, perimeter $P$, circle equivalent diameter $d_{2D}$( $=\sqrt{4 \, A_p / \pi}$), diameter of the largest inscribed circle $D_i$ and the smallest circumscribed circle $D_c$. The process of particle characterization based on image analysis and its comparison against 3D measurements are discussed in more details by Bagheri et al. \citep{Bagheri2014}. Here, we use the following equations to obtain sphericity $\psi$, spherical equivalent diameter $d_{eq}$ and surface area $\SA_{p}$ based on image analysis \citep{Bagheri2014}:
\begin{equation}
\psi = \left\{ 
\begin{array}{l l}
  \sqrt{\overline{D_i}/ \overline{D_c}} & \quad \mbox{non-vesicular surface}\\
  4 \, \pi \, \overline{A_p}/ \overline{P}^2 & \quad \mbox{vesicular surface}\\ \end{array} \right.  \label{eq:Sph2D}
\end{equation}
\begin{equation}
  d_{eq} =d_{2D}/1.022 \, \psi^{-0.29}  \label{eq:deq2D}
\end{equation}
\begin{equation}
  \SA_{p}=\pi \, d_{eq}^2 / \psi  \label{eq:SA2D}
\end{equation}
where overbars indicate the arithmetic average of variables obtained from multiple projections. Estimations of Eqs. (\ref{eq:Sph2D}--\ref{eq:SA2D}) are associated with average errors of 1.9--4.6\% compared to measurements obtained by a 3D laser scanner and SEM micro-CT on 127 irregular particles \citep{Bagheri2014}. A water pycnometer with nominal volume of $50.48 \, cc$ is used for measuring density of particles. The density of each irregular particle is considered to be equal to the density measured by the water pycnometer for a few grams ($1-24 \,\SI{}{\gram}$) of a sample of the same origin and sieve-size as the particle. Through this procedure the internal porosity of sample particles will be close to that of the selected particle particles and the density measurements will be more reliable.

\end{document}